\documentclass[10pt,journal,compsoc]{IEEEtran}
\AtBeginDocument{%
	\providecommand\BibTeX{{%
			\normalfont B\kern-0.5em{\scshape i\kern-0.25em b}\kern-0.8em\TeX}}}

\usepackage{setspace}
\usepackage{longtable}
\usepackage[figuresright]{rotating}
\usepackage{epsfig,endnotes,color}
\usepackage{graphicx}
\usepackage{amsmath}
\usepackage{caption}
\usepackage{graphicx}
\usepackage{url}

\usepackage{CJKutf8}
\usepackage{subfigure}
\usepackage{lscape, latexsym, amssymb, algorithmic, multirow}
\usepackage[linesnumbered, vlined, ruled]{algorithm2e}
\usepackage{multirow}
\usepackage{mathtools, bbm, color}
\usepackage{booktabs}
\usepackage{amsthm,mathrsfs,amsfonts,dsfont}
\usepackage{enumerate}
\usepackage{hhline}
\usepackage{enumitem}
\usepackage{tabu}
\usepackage{colortbl}
\usepackage{enumerate}
\usepackage[graphicx]{realboxes}

\usepackage{footnote}
\usepackage{color}
\usepackage{colortbl}
\usepackage{pifont}
\usepackage{bbding}
\usepackage{threeparttable}
\usepackage{rotating}
\def\BibTeX{{\rm B\kern-.05em{\sc i\kern-.025em b}\kern-.08em
    T\kern-.1667em\lower.7ex\hbox{E}\kern-.125emX}}

\newcommand{\bin}[1]{\textcolor{black}{#1}}
\newcommand{\system}{{\sc FirmSec}\xspace}

\begin{document}
\title{One Bad Apple Spoils the Barrel: Understanding the Security Risks Introduced by Third-Party Components in IoT Firmware}

\IEEEtitleabstractindextext{%
\begin{abstract}

Currently, the development of IoT firmware heavily depends on third-party components (TPCs) to improve development efficiency. Nevertheless, TPCs are not secure, and the vulnerabilities in TPCs will influence the security of IoT firmware. Existing works pay less attention to the vulnerabilities caused by TPCs, and we still lack a comprehensive understanding of the security impact of TPC vulnerability against firmware. To fill in the knowledge gap, we design and implement \system, which leverages syntactical features and control-flow graph features to detect the TPCs in firmware, and then recognizes the corresponding vulnerabilities. Based on \system, we present the first large-scale analysis of the security risks raised by TPCs on $34,136$ firmware images. We successfully detect $584$ TPCs and identify $128,757$ vulnerabilities caused by $429$ CVEs. Our in-depth analysis reveals the diversity of security risks in firmware and discovers some well-known vulnerabilities are still rooted in firmware. Besides, we explore the geographical distribution of vulnerable devices and confirm that the security situation of devices in different regions varies. Our analysis also indicates that vulnerabilities caused by TPCs in firmware keep growing with the boom of the IoT ecosystem. Further analysis shows $2,478$ commercial firmware images have potentially violated GPL/AGPL licensing terms. 

\end{abstract}

}

\author{Binbin Zhao,
        Shouling Ji, Jiacheng Xu,
        Yuan Tian, Qiuyang Wei, Qinying Wang, Chenyang Lyu, Xuhong Zhang, Changting Lin, Jingzheng Wu, Raheem Beyah
\IEEEcompsocitemizethanks{\IEEEcompsocthanksitem B. Zhao, S. Ji, J. Xu, Q. Wei, Q. Wang, C. Lyu, X. Zhang are with the College of Computer Science and Technology at Zhejiang University, Hangzhou, Zhejiang, 310027, China; B. Zhao and S. Ji are also with the  School of Electrical and Computer Engineering, Georgia Institute of Technology, Atlanta,
GA, 30332. E-mail: binbin.zhao@gatech.edu, \{sji, stitch, weiqiuyang, wangqinying, puppet, zhangxuhong\}@zju.edu.cn. S. Ji is the corresponding author.
\IEEEcompsocthanksitem Y. Tian is with  the Electrical and Computer Engineering at University of California, Los Angeles, CA, 90095. Email: yuant@ucla.edu.
\IEEEcompsocthanksitem C. Lin is with the Binjiang Institute of Zhejiang University. Email: linchangting@zju.edu.cn.

\IEEEcompsocthanksitem J. Wu is with the  Institute  of  Software,  Chinese Academy  of  Sciences,  Beijing. Email: jingzheng08@iscas.ac.cn.

\IEEEcompsocthanksitem R. Beyah is with the College of Engineering, Georgia Institute of Technology, Atlanta,
GA, 30332. \protect Email: rbeyah@coe.gatech.edu. 

* This work was partially conducted  when B. Zhao was at Zhejiang University.

}% <-this % stops a space
}

\maketitle
\vspace{-0.1cm}
\section{Introduction}
\iffalse

Motivations: IoT firmware has so many security problems but are not well studied. Show real examples of security bugs in IoT Firmware iamges. 

Research questions: understand and measure the real-world security issues in the IoT firmware images (including the different angles we have discussed.)

Challenges: present the challenges for such a large scale analysis 

Solutions: briefly introduce the method and explain the intuitions and reasons for these design choices.

Results: summarize the results first (directly answer the research questions). We can also provide interesting cases as a brief case study here.

Contributions: 1) open-source tool/dataset (need to check if we want to make the dataset open to public for now); 2) measurement results of the large-scale analysis 
\fi

% \yuan{You can condense this paragraph.}
The Internet of Things (IoT) has become ubiquitous and offers great convenience to our daily lives~\cite{gubbi2013internet,li2015the}.
From the traditional IoT devices, e.g., router and printer, to the smart homes, e.g., smart light and smart plugin, IoT devices have an ever-growing presence in modern life. 
According to a recent report~\cite{gartner}, Gartner forecasts that the number of IoT devices will triple from 2020 to 2030.
Inevitably, the booming of  IoT devices also raises the public’s concern about their security  risks~\cite{suo2012security, zhao2020large, DBLP:conf/uss/WangJT0ZKLLDLB21}  and several real-world attacks further aggravate this panic. For example, Mirai has compromised millions of IoT devices including  IP  cameras,  DVRs,  and routers, to form a botnet~\cite{stone-gross2009your, bayer2009view} and launch DDoS attacks against various  online  services~\cite{antonakakis2017understanding,kolias2017ddos}.

Firmware is an integrated software package that booted in IoT devices which serve the essential roles. Nowadays, firmware widely adopts TPCs, e.g., \textit{BusyBox}~\cite{busybox} and \textit{OpenSSL}~\cite{openssl}, to accelerate and simplify the development process. However, the wide usage of TPCs is a double-edged sword since a significant number of TPCs have known vulnerabilities that will open up many new attack surfaces to IoT firmware. 
% TPCs have many , and the vulnerabilities in TPCs will bring serious security risks to firmware.
% Currently, the development of firmware depends heavily on TPCs, . Nevertheless, the wide usage of TPCs is a double-edged sword. Though utilizing  TPCs can reduce development efforts and costs, the vulnerabilities discovered in TPCs will turn back influence the security of IoT devices.  
% In addition, a single vulnerability caused by TPCs may cause a butterfly effect in the IoT ecosystem which can be easily spread across to thousands or even millions of IoT devices from multiple vendors. 
For example, the Heartbleed vulnerability~\cite{durumeric2014matter} in \textit{OpenSSL} 
% and the ShellShock vulnerability~\cite{shellshock} in the \textit{Unix Bash shell} 
has greatly affected at least millions of IoT devices. To make things worse, vendors may reuse a set of the same TPCs in different kinds of firmware, which accelerates the spread of potential vulnerabilities. Therefore, it is vital to recognize the vulnerable TPCs used in firmware.
A single vulnerability caused by TPCs may cause a butterfly effect in the IoT ecosystem.

% All the IoT devices are equipped with a consolidated software package, often called \textit{firmware}, which is usually presented as \textit{firmware image}. Like the traditional software that used in PC, the \textit{firmware image} also
% widely adopts the TPCs for development. Nevertheless, Nevertheless, the vulnerabilities discovered in the TPCs will turn back influence the security of IoT devices.

% includes vulnerabilities that may be caused by misusing of TPCs or bugs. During the last decade, there are a great number of new vulnerabilities are discovered in a wide range of IoT devices~\cite{iotvulnerability}. 
Although a series of research~\cite{chen2016towards, costin2014large, shoshitaishvili2015firmalice,xu2017neural, corteggiani2018inception, costin2016automated, 182944, david2018firmup, feng2016scalable,feng2020P2IM} has adopted static or dynamic approaches to evaluate firmware security, they are still limited since they pay less attention to the vulnerabilities caused by TPCs in firmware, lack the consideration of non-Linux based firmware, and/or are unscalable on large-scale firmware security analysis. 
Specifically, first, they lack the analysis of N-days vulnerabilities introduced by TPCs, which may result in more serious problems than unknown vulnerabilities in reality~\cite{overlookedVul}.  Second, most of them are limited to analyze the Linux-based firmware, but short of the analysis of monolithic firmware, which is widely adopted in new ubiquitous, low-power embedded systems, e.g., smart homes.
% non-Linux based firmware, e.g., smart homes~\cite{chan2009smart}, which is becoming the largest part of the IoT ecosystem~\cite{stojkoska2017review}.
Last but not least, their scalability is a challenge when given large-scale firmware tests.  For instance, several approaches require real IoT devices in analysis or lots of manual work for configuration, which greatly limits their scalability.
% In addition, these approaches take quite a lot of time to deal with a single firmware image since most firmware should be handled case by case. 
% They are suitable to conduct an in-depth analysis of firmware, but not for discovering the TPCs caused vulnerabilities in firmware images at large-scale.
Therefore, we still lack a comprehensive understanding of the usage of TPCs in multiple kinds of firmware and the vulnerabilities introduced by them, with a scalable and practical method.

\vspace{-0.3cm}
\subsection{Challenges}
% \yuan{\color{red}It overlaps with former paragrahs, need to merge the content.}
% Though previous studies~\cite{chen2016towards, costin2014large, shoshitaishvili2015firmalice, xu2017neural, corteggiani2018inception, costin2016automated, 182944, david2018firmup, feng2016scalable} have already proposed techniques to address the critical problems of detecting the vulnerabilities from IoT devices, most of their experiences are not suitable for the current booming IoT ecosystems. We still encounter several challenges of performing large-scale analysis on firmware images in the following aspects.

To obtain a comprehensive overview of the TPCs used in firmware and their corresponding vulnerabilities, 
% and design a practical and scalable system to address this problem, 
we have the following key challenges.

\noindent\textbf{Firmware Dataset Construction.} To enable our study, the first challenge is to construct a large-scale and comprehensive firmware dataset covering different kinds of firmware from various vendors. Thus we can obtain convincing results of the current security issues of the firmware.
%To enable our study, the first challenge is to collect a large-scale real firmware dataset. The dataset should cover a great number of different kinds of firmware images from different vendors. Thus we can obtain comprehensive and convincing results of the current security issues of the firmware.
Nevertheless, there is no publicly accessible firmware dataset for research.
Besides, more and more vendors begin to prohibit the public from downloading firmware and adopt anti-scraping techniques~\cite{wang2018anti-crawler} against the hostile web crawler, which in turn greatly increases the data collection difficulty.

\noindent\textbf{Firmware Processing.}  There are two major challenges in firmware processing. (1) Extract the contained objects, which have essential information for TPC detection, from Linux-based firmware. Though existing tools, e.g., \textit{binwalk}~\cite{binwalk}, can be used to unpack firmware, they have limitations on dealing with the firmware which has adopted the latest or customized filesystems. 
% For instance, \textit{binwalk} fails to handle the
% uncommon filesystems, e.g., Minix filesystem and YAFFS filesystem~\cite{YAFFS}. Moreover, the integrated extraction tools in \textit{binwalk}, e.g., \textit{unsquashfs} and \textit{jffsdump}, are unable to process the customized filesystems which have specific implementations addded by vendors. 
% Thus, we still need to implement a new tool to unpack different kinds of firmware. 
(2) Deal with the monolithic firmware, which is widely used in lower-power embedded systems, e.g., smart homes. Monolithic firmware usually lacks a traditional operating system or metadata, e.g., RAM/ROM start address, needed for analysis. Besides, its code, libraries, and data are highly intermixed.  Regarding these challenges, yet, existing tools, e.g., \textit{IDA}, cannot deal with the monolithic firmware without extra configuration.

\noindent\textbf{TPC Detection and Vulnerability Identification.} We have two major challenges in TPC detection and vulnerability identification. (1) Detect the TPCs at version-level. To accurately identify the corresponding vulnerabilities of TPCs used in firmware, we need to detect them at version-level rather than only at TPC-level since the vulnerabilities of different versions of TPCs might be different. Nevertheless, it is hard to distinguish the same TPCs at version-level since the code difference of different versions might be tiny. Besides, without source code, we can only obtain limited features from firmware for TPC identification.  Previous 
% state-of-the-art 
works for firmware analysis~\cite{david2018firmup, feng2016scalable, ding2019asm2vec} cannot meet our requirements since they do not support detecting the TPCs at version-level in firmware. (2) Construct a TPC database. We need a comprehensive and easily usable TPC database that indicates the possible TPCs used in firmware and the vulnerabilities of each version of TPCs. To the best of our knowledge, there is no previous work has built such a TPC database for IoT firmware. It is a challenge to collect as many TPCs as possible and map the vulnerabilities to the TPC versions. 

\vspace{-0.3cm}
\subsection{Methodology}
% To address the above challenges, we propose \system to conduct a large-scale analysis of the security of IoT firmware.
In this paper, we dedicate to providing a comprehensive understanding of the usage of TPCs in firmware and the potential security risks caused by TPCs. Toward this, we develop a scalable and automated framework \system to conduct a large-scale analysis of IoT firmware. Our analysis method is as follows.

% \yuan{\color{red}What are your research questions?}
\textbf{First}, to solve the challenge of constructing a firmware dataset, we collect firmware images from public sources and private sources, e.g., official website and private firmware repository.
% We then develop a web crawler \yuan{\color{red}Why is your search and crawler better than previous researchers? Need to highlight your innovations. This applies to several other places as well.} for collecting firmware images and custom it for each site with respect to the robots exclusion standard. Our crawler supports solving the common anti-scraping techniques, e.g., IP address restriction and CAPTCHA, which greatly increases the effectiveness of collecting firmware images.
% In addition, we get access to a private firmware repository from a well-known company which contains tens of thousands of firmware images. 
% From the above four sources, we obtain a total of $34,129$ firmware images which include $11,079$ publicly accessible firmware images and $23,050$ private firmware images. 
% In comparison with previous works, our dataset has a much larger scale and covers a wider range of different kinds of firmware images.
% \yuan{\color{red}Instead of mentioning the challenges again, we should focus on describing our solution. What are the novel points in your firmware unpacking framework? Also need to explain why you design it this way.}
\textbf{Second}, we customize several plugins for existing tools, e.g., \textit{binwalk} and \textit{IDA}, to address the problem in firmware processing. The customized tools support unpacking and disassembling both Linux-based firmware that utilizes popular and uncommon filesystems, and monolithic firmware.
\textbf{Third}, 
we propose a new detection strategy to identify the TPCs used in firmware. The main idea behind our strategy is to leverage two kinds of features, syntactical features, and control-flow graph features, that are extracted from the TPC and firmware.
% a coarse-grained method and a fine-grained method to analyze the firmware at large-scale. For coarse-grained analysis, we first propose a method to identify the TPCs at version level in firmware with $91.02\%$ precision.
We then search the corresponding vulnerabilities based on versions check~\cite{backes2016reliable, zhang2019libid, zhan2021atvhunter}, which relies on our TPC database.
% For fine-grained analysis, we adopt a state-of-art \textit{Gemini}~\cite{xu2017neural} to identify the specific vulnerabilities, e.g., OpenSSL Heartbleed vulnerability, in firmware. Our fine-grained analysis supports detecting 12 specific vulnerabilities with higher than $90\%$ accuracy.
% implement a firmware analysis engine that serves as the core of \system. It first identifies the TPCs used in firmware. 
% % We maintain a large-scale TPCs list and design a new matching strategy which solves the low true positive rate of matching complex patterns by traditional methods. Our matching strategy consists of three aspects which will be discussed in Section~\ref{section: system design}.
% Then, the analysis engine will search the vulnerabilities of the detected TPCs according to our customized vulnerability database that builds on CVE Detail~\cite{cvedetail}. 
% The vulnerability database covers detailed vulnerability information of TPCs in our list.
Based on the results, \system will generate a report for each firmware image which indicates its potential risks. Moreover, the report will provide a series of suggestions for fixing the vulnerabilities.

To provide more in-depth insights, we conduct further studies to explore the security status quo of the IoT ecosystem from four aspects. First, we evaluate the vulnerability of the firmware of different kinds and from different vendors. 
Second, we investigate the usage trending of  TPCs in firmware in different periods and the corresponding vulnerabilities they caused. 
Next, we explore the geographical difference of the security of IoT devices.
% The results show that most vulnerable IoT devices are mainly located in several regions, e.g., South Korea and China.
% we identify the corresponding IoT devices with vulnerable firmware images and then adopt 
% We find the number of N-days vulnerabilities increases dramatically over time while the TPCs do not change a lot.
% the TPCs adopted  
Then, we explore the outdated TPC problem in firmware.
% We reveal that the version of the TPCs is on average five years older than the latest version when the firmware  was published.
Then, we study the changes of TPCs during the firmware update.
Finally, we analyze the TPC GPL/AGPL license violations in firmware.

% and explore the geographical difference in the security of IoT devices. Also, we split the firmware images by their release time. We then figure out the changes in the usage of TPCs and their corresponding vulnerabilities in firmware images in the past decade. At the same time, we also check the release time of the TPCs used in firmware images. Based on this, we conclude that the version of the TPCs in the firmware (that we analyzed) is on average five years older than the newest version of the libraries when the firmware was published.

% our tool integrates a series of customized script that can assist \textit{binwalk} to extract the objects from specific kinds of IoT devices.

% Third, we convert the extracted objects into strings and match the TPCs by using named entity recognition. After extraction, more 

\vspace{-2mm}
\subsection{Contributions}
We summarize our main contributions below. 

% \begin{itemize}
% \item  
$\bullet$
We construct so far the largest firmware dataset, which includes $11,086$ publicly accessible firmware images and $23,050$ private firmware images. It contains 35 kinds of firmware, with most of them rarely studied in previous works. To facilitate future 
% firmware security 
research, we open-source this dataset at https://github.com/BBge/FirmSecDataset.
% , which can serve as a benchmark for further studies, e.g., cross-platform vulnerability detection.
 
% \item
$\bullet$
We propose \system, a scalable and automated framework to analyze the TPCs used in firmware and identify the corresponding vulnerabilities. \system has $91.03\%$ of precision and $92.26\%$ of recall  in identifying the TPCs  at version-level in firmware, which significantly outperforms state-of-the-art works from academia and commercial systems from industry, e.g., \textit{Gemini}~\cite{xu2017neural}, \textit{BAT}~\cite{hemel2011finding}, \textit{OSSPolice}~\cite{duan2017identifying}, \textit{binaré}~\cite{binare}, \textit{360 FirmwareTotal}~\cite{360iot}, and \textit{Alibaba FSS}~\cite{aliiot}.

% \system also manages to detect more TPCs and vulnerabilities within less time compared to industry systems\cite{aliiot},\cite{360iot},\cite{binare}.

% \textit{binaré}~\cite{binare}, 
% \textit{360 FirmwareTotal}~\cite{360iot}, and \textit{Alibaba FSS}~\cite{aliiot}. 
% and supports detecting  12  specific  vulnerabilities  with  higher  than 90\% accuracy. Besides, \system performs better than one academic incubated system~\textit{binaré}~\cite{costin2014large,binare} and two industry-leading systems developed by 360~\cite{360iot} and Alibaba~\cite{aliiot}. 
% % It can detect an average of $11.7$ TPCs and $104.2$ vulnerabilities per firmware while the corresponding number of 360 and Alibaba are $4.8$ and $62.3$, $3$ and $55.5$, respectively. Moreover, \system spends an average of $4.2s$ analyzing each firmware while 360 spends $60s$ and Alibaba spends $174.7s$. 
% We also will open-source \system to facilitate future research.

% \item 
$\bullet$
We conduct the first large-scale analysis of the vulnerable TPC problem in firmware. We identify $584$ TPCs and detect $429$ CVEs in $34,136$ firmware images. According to the results, we reveal the widespread usage of vulnerable and outdated TPCs in IoT firmware.
% , which have fallen behind by an average of five years. 
Besides, we notice that firmware contains more and more vulnerabilities caused by TPCs over time, and vendors hardly update vulnerable TPCs when the firmware is updated. 
Moreover, we confirm the geographical difference in the security of IoT devices where several regions, e.g., South Korea and China, are in a more severe situation than other regions. Finally, we discover $2,478$ firmware images potentially violate GPL/AGPL licensing terms, which may lead to lawsuits.

\section{Design and Implementation}
\label{section: system design}
% With \system, manual efforts are only required at the first time setting up the framework.
% Then, we introduce the firmware dataset we constructed.

% %%Begin: Figure: the Architecture of XXX
% \begin{figure*}
% % \centering
% \includegraphics[width=\textwidth]{Figures/overview.pdf}
% \caption{Architecture of \system}
% \label{Figure: the Architecture of XXX}
% \end{figure*}
% %%End: Figure: Vulnerable Rate Trending of Route

\begin{figure}
\setlength{\abovecaptionskip}{0.02cm}
\centering
\includegraphics[scale=0.20]{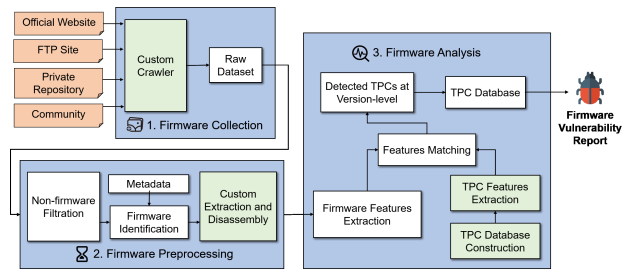}
\caption{Architecture of \system. Green submodules require some manual work during the initialization of \system, while other submodules can be automatically executed.}
\label{Figure: the Architecture of XXX}
\vspace{-7mm}
\end{figure}

\system is designed to automatically identify the TPCs used in firmware and detect the corresponding vulnerabilities.
% \subsection{System Overview}
As shown in Figure~\ref{Figure: the Architecture of XXX}, \system mainly contains three components: \emph{firmware collection}, \emph{firmware preprocessing} and \emph{firmware analysis}.
% \yuan{\color{red}This paragrah is not clear. Please organize in the following: what each module do, and how are they connected. We only need to explain the modules very briefly here and highlight your innovations.}
 The firmware collection module is mainly designed to collect firmware from different sources.
% : 1) official website, 2) FTP site, 3) community,  and 4) private firmware repository. 
Next, the firmware preprocessing module will process the collected firmware in three steps: 1) filter out the non-firmware files from the raw dataset; 2) identify the necessary information of firmware;
% e.g., category, architecture, filesystems, and so on; 
and 3) unpack and disassemble firmware.
Finally, the firmware analysis module will detect the TPCs at version-level in firmware according to the syntactical features and control-flow graph features extracted from the TPC and firmware. The corresponding vulnerabilities will be recognized through versions check according to our TPC database. 
The implementation behind each component is discussed in the following subsections.
% \yuan{\color{red}need figure for each component}
% \begin{enumerate}[label=(\roman*)]
% 	\item \textit{Firmware Collection}: The first step of our system XXX is to collect a great number of firmware. Our firmware are mainly collected from four sources: 1) Official Websites. Vendors usually allow users to download firmware on their official websites. 2) FTP sites. 3) External Websites.  4) Private Firmware Repository. Tuya provides us tens of thousands of private firmware since we have a great partnership. This part is our main source of firmware.
% 	We develop a custom web crawler to obtain the firmware from the above sources.
%     \item \textit{Firmware Extraction}: In the second step, we write a custom \textit{Python} script based on \textit{binwalk}~\cite{binwalk} API to unpack firmware. \textit{Binwalk} is a well-known unpacking tool developed by Craig Heffner~\cite{binwalk} which can extract the root file systems of firmware.
%     \item \textit{Firmware Analysis Engine}:
% \end{enumerate}

\vspace{-0.3cm}
\subsection{Firmware Collection}
% \yuan{\color{red}This part seems to be just engineering effort, we can be more brief about this.}

The firmware collection module is responsible for collecting firmware to construct a large-scale raw dataset. 
% Chen et al.~\cite{chen2016towards} proposed web crawlers that can obtain firmware from $42$ vendors. 
% % We first want to adopt their web crawler to collect data. 
% However, most of the individual spider of their web crawlers are failed since the vendors' websites are updated and a portion of download links are expired.  Due to safety concerns, more and more vendors refuse to provide publicly accessible firmware, which heavily increases the difficulty of collecting firmware. Moreover, previous works 
% mainly collect firmware of traditional IoT devices, e.g., router and camera. They 
% do not provide useful guides on collecting smart homes.
% In addition, the development of anti-scraping techniques, e.g., CAPTCHA~\cite{ahn2003captcha} and IP Address Restrictions, also bring a lot of trouble. 
% To address these issues, we develop a customized web crawler with respect to the robots exclusion standard. 
% \textbf{First}, 
To solve the problem of limited firmware resources, we implement a web crawler to collect firmware from three sources: 1) Official website;  Several vendors provide firmware download links on their official websites. 
2) FTP site; We discover a list of ftp sites that store the old version firmware which can be freely download.
3) Community. We collect a part of firmware images of several vendors, e.g., Xiaomi, from the community, including the related forums and GitHub repositories;
% For example, Giese et al.~\cite{dustcloud} created a GitHub repository to share the firmware for research purpose. 
4) Private firmware repository. We obtain the permission to access the private firmware repository of a world-leading company that mainly focuses on smart homes. In this paper, we use TSmart to represent this company.
% Tuya. Tuya is a well-known IoT vendor that mainly focuses on smart homes.
TSmart's firmware repository contains tens of thousands of firmware images of smart homes manufactured by hundreds of vendors, e.g., Philips, and never be studied.
\textbf{Second}, we implement a web crawler based on \textit{Scrapy}~\cite{Scray} which supports collecting firmware from $31$ sites of 13 vendors, such as TP-Link, D-Link, Trendnet, etc. 

\vspace{-0.2cm}
\subsection{Firmware Preprocessing}
In this part, we mainly preprocess the collected firmware by three steps: firmware filtration, firmware identification, and firmware extraction and disassembly.

\noindent\textbf{Firmware Filtration.} The first step is to exclude the non-firmware files from the raw dataset.
Analyzing the non-firmware files in our raw dataset will produce unexpected results that influence the credibility of our final results. Therefore, we need to exclude the non-firmware files from the collected firmware. 
First, we filter the obvious non-firmware files, e.g., .txt files, through suffix matching. Second, we adopt Binary Analysis Next Generation (BANG)~\cite{bang}, which supports recognizing 136 kinds of files, to get rid of other non-firmware files, e.g., Android Dex. We apply it to the remaining collected files and remove the non-firmware files based on the returned file types.
%  
% % Though we obtain a great number of firmware images through the firmware collection module, a great number of them may be non-firmware files. 
%  We first  Next, we  file types of the remaining files, which  BANG supports identifying 136 kinds of files, e.g., 
% will return
% Thus, it is essential to filter the non-firmware files before we conduct further analysis. 
% To filter the non-firmware files, we first 
% construct a whitelist that includes firmware formats, such as \textit{.bin} and \textit{.hex}. The whitelist is constructed as follows. We first select the binary files from our raw dataset and record their formats. Then, we query these formats in Google to confirm whether they are valid formats of IoT firmware. Next, we add the confirmed formats to our whitelist. We regard the binary files whose formats are not included in the whitelist as non-firmware files.

%%Begin: Figure: the Architecture of XXX
%%End: Figure: Vulnerable Rate Trending of Route

\noindent\textbf{Firmware Identification.} 
The second step is to identify the necessary information of firmware, e.g., operating systems, architectures, and filesystems, which is crucial for further analysis.
Though  non-firmware files have been excluded from the dataset, there are still a large number of firmware images whose categories are still unknown. Besides, we also lack the information about their operating systems, architectures, filesystems and so on, which is  crucial for further analysis. 
% We obtain the missed information of firmware based on two steps. First, we
adopt two methods to identify the category of unknown firmware. (1) We 
extract the information from the metadata files, which are crawled with the firmware. The metadata file describes the firmware, including its corresponding device model, category, and so on. The metadata file is usually in the same archive with firmware.
(2) We implement a script to query the filename of the unknown firmware on the Internet and crawl the top-3 returned sites' content, which may contain the relevant information, e.g., the category of firmware. We finally combine the information
obtained from the above two methods to identify the category of unknown firmware.
Second, we adopt \textit{binwalk} to scan all firmware images to obtain their operating systems, architectures, and filesystems. 
\noindent\textbf{Firmware Extraction.} 
% \noindent$\bullet$\textit{Firmware extraction.}
The contained objects in firmware have many syntactical features, which are vital for further TPC detection. Besides, the linked libraries extracted from firmware are the main sources for creating the TPC database. Therefore, it is necessary to conduct firmware extraction before we perform further analysis. Nevertheless, existing tools more or less have issues, e.g., incomplete extraction,
% and endless loop,
when extracting the firmware which adopts the latest or customized filesystems. The main challenge here is to extract the filesystems in firmware thoroughly without recursively extracting compressed data. To address this challenge, we equip \textit{binwalk} with a series of plugins.  First, we analyze the filesystems used in the collected firmware and find the original \textit{binwalk} mainly has issues on dealing with three popular systems:  SquashFS, JFFS2, YAFFS, and UBIFS. Then, we utilize \textit{sasquatch}~\cite{sasquatchgithub} and \textit{jefferson}~\cite{jffs2github} to replace the \textit{unsquashfs} and \textit{jffsdump} in \textit{binwalk}. These plugins support many vendor-specific SquashFS and JFFS2 implementations. Next, we implement a new plugin based on \textit{yaffshiv}~\cite{yaffshiv} for \textit{binwalk} to unpack YAFFS, which supports the brute-forcing of YAFFS build parameters, e.g., page size, and extracting the contained objects even after the filesystem has run many rounds. Finally, we develop a plugin based on \textit{UBI Reader}~\cite{ubireader} for  \textit{binwalk} to unpack UBIFS.

\noindent\textbf{Firmware Disassembly.} 
% \noindent$\bullet$\textit{Firmware disassembly.}  
The control-flow graph (CFG) in firmware contains the essential information to identify the TPCs in firmware. To obtain the CFGs, we need to disassemble the firmware first. However, existing disassemble tools are unable to handle the monolithic firmware without extra configuration. 
For instance, both native \textit{IDA} and \textit{GHIDRA}~\cite{ghidra} cannot deal with many monolithic firmware images since they adopt uncommon processors and discard some necessary information, e.g., the RAM/ROM start address,  required for analysis. The main challenge here is to recover the missing information based on the known limited information of the monolithic firmware. 
To solve this challenge, we first analyze the processors used in monolithic firmware to recover the missing information and then implement a series of customized plugins for \textit{IDA} to load and disassemble the monolithic firmware. More specifically, first, we analyze the processor types, e.g., ESP8266~\cite{schwartz2016internet}, used in the collected monolithic firmware. Second, we collect the corresponding reference manual or datasheet for each processor. The reference manual and datasheet can provide us three kinds of useful information: the core, memory map, and interrupt vector table of the processor. We can figure out the exact instruction sets to disassemble the firmware according to the core, and retrieve the RAM/ROM start address used for loading firmware from the memory map. Moreover, we can obtain the firmware start address from the interrupt vector table.
% According to the core, we can figure out Based on the memory map, 
Finally, we implement $7$ plugins, which correspond to $7$ kinds of processors used in monolithic firmware, for \textit{IDA} based on the recovered information.
The plugins enable \textit{IDA} to load and disassemble the monolithic firmware automatically.

\label{section: TPCs from TSmart.}
\begin{table}[h]
\caption{TPCs from TSmart.}
\centering
\setlength\tabcolsep{8pt}
\footnotesize
\label{Table: common TPCs}
\begin{tabular}{cccc}
\toprule[1.5pt]
% \multicolumn{4}{c}{TPC Name}                          \\ \midrule[1pt]
Dosfstools     & ALSA         & U-Boot    & Boa            \\
iPerf          & BusyBox      & uClibc    & Dropbear       \\
IPTables       & Hostapd      & ECC       & sudo           \\
MTD-Utils      & libnl        & apr\_sha1 & Wpa supplicant \\
udhcp          & libpcap      & OpenSSL   & FreeRTOS       \\
Wireless Tools & libuv        & PCRE      & lwIP           \\
ZBar           & Mbed TLS     & SimCList  & libmqtt        \\
cJSON          & opencore-amr & jq        & Espressif      \\
librtmp        & Opus         & libusb    & Speex          \\
zlib           & RTMPDump     & Jansson   & Fdk-acc        \\ \bottomrule[1.5pt]
\end{tabular}
\end{table}

\vspace{-0.3cm}
% \subsection{\textcolor{red}{Firmware Analysis}}
\subsection{Firmware Analysis}
% \textbf{#TODO}
It is a common practice 
% academia~\cite{backes2016reliable, zhang2019libid, zhan2021atvhunter} and industry~\cite{binare,360iot,aliiot} 
to use the versions check~\cite{backes2016reliable, zhang2019libid, zhan2021atvhunter} to identify the corresponding vulnerabilities of TPCs. In \system, we also adopt this strategy to analyze firmware. The firmware analysis module includes three submodules: 1) \emph{TPC database construction}, 2) \emph{TPC detection}, and 3) \emph{vulnerability identification}. 
% The workflow is as follows.

% three submodules: 1) \emph{keyword list construction}, 2) \emph{TPC detection}, and 3) \emph{vulnerability identification}.

% \subsubsection{Coarse-grained Analysis}  The workflow is as follows.

\noindent\textbf{TPC Database Construction.} Our analysis strategy requires a TPC database that contains the possible TPCs used in firmware and the detailed vulnerability information for each TPC version. Nevertheless, unlike the Maven Repository~\cite{maven} that indicates the TPCs used in Android apps, there is no such a publicly accessible database for IoT firmware. The main challenge here is to collect as many TPCs as possible from reliable sources and map the vulnerabilities to the TPC versions. To overcome this challenge, we first
% collect the TPCs from four sources and leverage \textit{cve-search} and CVE Detail~\cite{cvedetail} to gather the corresponding vulnerabilities. More specifically, 
collect the possible TPCs used in IoT firmware from four sources: 1) linked libraries extracted from the firmware; 2) open-source IoT projects; 3) SDKs from multiple IoT platforms, e.g., AWS IoT; and 4) a shortlist of TPCs from TSmart, as shown in Table~\ref{Table: common TPCs}. 
% (Appendix~\ref{section: TPCs from TSmart.}). 
Second, we leverage the \textit{cve-search}~\cite{cvesearch}, a professional CVE search tool, to query the TPCs from the CVE database~\cite{CVE}, and implement a script to query the NVD~\cite{nvd} and CVE Detail~\cite{cvedetail} to collect the TPC CVEs. With these two methods, we can get the CVEs that correspond to different versions of TPCs. The constructed TPC database for IoT firmware has the following fields: TPCs, licenses, versions, CVEs, CVSS score~\cite{CVSS}, and CVE description. We finally collect 1,261 TPCs. 
%1,191

\noindent\textbf{TPC Detection.} To precisely detect the TPCs used in firmware at version-level is vital in our analysis since we will use the exact version of the detected TPC to confirm its vulnerabilities. Nevertheless, it is difficult to distinguish the same TPCs at version-level since we can only obtain limited information from firmware, and the code difference of different versions might be tiny. To address this challenge, we implement a novel TPC detection method based on two kinds of features: syntactical features and CFG features, which are hardly changed between the source files and binaries. We first extract the above two features from TPCs and firmware. Next, we leverage the edit distance~\cite{zhan2021atvhunter,editdistance}, a widely used method to measure the similarity
between two strings, and ratio-based matching to calculate the similarity of syntactical features and use customized \textit{Gemini}~\cite{xu2017neural} to compare the CFG features. Based on the comparison results, we finally confirm the usage of TPCs at version-level. The workflow is as follows. 

\noindent$\bullet$\textit{TPC feature extraction.} \textbf{First}, we implement a parser to extract the syntactical features from the C/C++ source files of TPCs. The syntactical features include the string literals (e.g., unique string), function information (e.g., function names and function parameter types). For each kind of TPC, we summarize the common syntactical features in its all versions, which are regarded as sharing syntactical features. We then identify the specific syntactical features in each version of the TPC, which are regarded as unique syntactical features. \textbf{Second}, we extract the attributed control-flow graphs (ACFGs)~\cite{feng2016scalable,xu2017neural}
% , which is first introduced by \textit{Genius}~\cite{feng2016scalable}, 
from each version of TPCs. Each vertex in an ACFG is a basic block labeled with a set of attributes. Except for the block-level attributes used in \textit{Gemini}~\cite{xu2017neural}, we also use three function-level attributes, as shown in Table~\ref{Table: Function-level and Block-level Attributes}. The function-level attributes provide more detailed information on the structure of CFGs, which are ignored by \textit{Gemini}. 
% Based on the attributes used in \textit{Genius}, we use more fine-grained attributes, as shown in . The categories of API calls have been used by prior work for malware detection. We also adopt these features since they can represent the high level TPC operations.
% We leverage the same attributes as \textit{Genius}: string and numeric constants, the number of transfer instructions, calls, instructions, and arithmetic instructions at block-level; betweenness and the number of offspring at inter-block-level.

% \usepackage{multirow}
\begin{table}[]
 \setlength{\abovecaptionskip}{0.08cm}
\centering
\renewcommand\arraystretch{0.8}
\caption{Block-level and Function-level Attributes.}
\setlength\tabcolsep{2pt}
\footnotesize
\label{Table: Function-level and Block-level Attributes}
\begin{tabular}{clcc}
\toprule[1.5pt]
Type                            & Attribute Name                 & \system & \textit{Gemini}~\cite{xu2017neural} \\ \midrule[1pt]
\multirow{8}{*}{Block-level}    & String Constants               &    \textcolor{green}{\checkmark}     &     \textcolor{green}{\checkmark}   \\
                                & Numeric Constants              &   \textcolor{green}{\checkmark}      &   \textcolor{green}{\checkmark}     \\
                                & No. of Transfer Instructions   &  \textcolor{green}{\checkmark}       &      \textcolor{green}{\checkmark}  \\
                                & No. of Calls                   &    \textcolor{green}{\checkmark}     &     \textcolor{green}{\checkmark}   \\
                                & No. of Instructions            &   \textcolor{green}{\checkmark}      &    \textcolor{green}{\checkmark}    \\
                                & No. of Arithmetic Instructions &   \textcolor{green}{\checkmark}      &     \textcolor{green}{\checkmark}   \\
                                & No. of Offspring               &     \textcolor{green}{\checkmark}    &    \textcolor{green}{\checkmark}    \\ \midrule[1pt]
\multirow{3}{*}{Function-level} & No. of Basic Blocks            & \textcolor{green}{\checkmark}         & \textcolor{red}{\XSolidBrush}        \\
                                & No. of Edges                   &  \textcolor{green}{\checkmark}       & \textcolor{red}{\XSolidBrush}        \\
                                & No. of Variables               &   \textcolor{green}{\checkmark}      & \textcolor{red}{\XSolidBrush}        \\ \bottomrule[1.5pt]
\end{tabular}
\vspace{-1em}
\end{table}

\noindent$\bullet$\textit{Firmware feature extraction.} \textbf{First}, we extract the same types of syntactical features as the previous step from firmware. Though we successfully disassemble the Linux-based firmware and monolithic firmware, many function names are unrecognized by \textit{IDA}, especially in monolithic firmware. To address this problem, we equip \textit{IDA} with a large number of signature files of TPCs to identify the function names in disassembled files. The signature files are mainly collected from the Internet, e.g., GitHub projects, and we also manually generate a part of signature files. \textbf{Second}, we extract the ACFGs from the disassembled firmware with the same attributes as those extracted from TPCs. However, the original extraction tool used by \textit{Gemini} cannot extract the ACFGs from monolithic firmware since it cannot disassemble monolithic firmware. Therefore, we customize the extraction tool by integrating our firmware disassembly module.

% two syntactical features from the firmware: string literals and exported functions, which contain the important information that can be used for identifying the TPCs. Syntactical features are hardly changed between source files and binary files . We implement an extraction tool around a Python module used in \cite{duan2017identifying}, which is originally designed to extract the similar features from Android apps. Second, we extract 

% In this step, we generate the fingerprints for each TPC in our database. 
% We implement a parser to extract the 

% (1) We first extract the . (2) Next, we extract the function names from the source files of the TPCs. We first identify the function names that are commonly used in each version of the TPC, which are called public functions.   (3) 
% We then identify the same function names in each version of the same TPCs. We then . We give a general fingerprint example of the TPC: regular expression, public functions, specific functions, and CFG features.

\noindent$\bullet$\textit{Syntactical feature matching.} In this step, we conduct syntactical feature matching to identify the TPCs. Unfortunately, the direct features matching, e.g., regex, will cause low precision and recall. Design a new matching method with high precision and recall is the main challenge in this step. To address this challenge, we utilize the edit distance and ratio-based matching to measure the similarity between the TPC and firmware. We use $D(S_{TPC},S_{Firmware})$ to represent the edit distance between the syntactical features from TPCs and firmware. If $D(S_{TPC},S_{Firmware})$ exceeds the given threshold $\alpha$, we regard the features are matched. We then calculate the distance of each extracted syntactical feature and record the number of matched features. Next, we calculate the ratio of matched features to all features extracted from the TPC, which can be represented as $\frac{S_{TPC}\cap S_{Firmware}}{S_{TPC}}$. If $\frac{S_{TPC}\cap S_{Firmware}}{S_{TPC}}$ exceeds the given threshold $\beta$, we regard the TPC is matched. Given some firmware may adopt the partially built TPCs, the ratio-based matching can improve the precision under these circumstances. According to the evaluation results in  Section~\ref{section: System Evaluation}, we finally set the threshold of $D(S_{TPC},S_{Firmware})$ as 0.74 and the threshold of $\frac{S_{TPC}\cap S_{Firmware}}{S_{TPC}}$ as 0.52. Based on the above matching strategy, we then leverage the sharing syntactical features for TPC-level identification and use the unique syntactical features for version-level identification. We mark the matched results by syntactical features as $R_{Syntax}$.

% We first calculate
% Our matching process is twofold. \textbf{First}, we compare the . For TPC-level identification, we leverage the public syntactical features of each TPC to compare with the If the $30\%$
% For version-level identification, we use the private syntactical features to confirm their exact version.
% We leverage the public syntactical features to identify the TPCs used in firmware, 
% and use the private syntactical features to confirm their exact versions.
% \textbf{Second}, we leverage \textit{Gemini}~\cite{xu2017neural} to identify the TPCs through ACFGs.  More specifically, we    The dataset includes the ACFGs we extracted from each TPC. We then apply  For TPC-identification

% The dataset includes the ACFGs from 8 popular TPCs: \textit{OpenSSL}, \textit{BusyBox}, \textit{uClibc}, \textit{glibc}, \textit{Curl}, \textit{IPTables}, \textit{Wpa supplicant}, and \textit{Lighttpd}.  We combine the matching results from above two steps as our final results. If there are conflicts between

\noindent$\bullet$\textit{CFG feature matching.} \bin{In this step, we leverage the customized \textit{Gemini}~\cite{xu2017neural} to conduct the CFG feature matching. The original \textit{Gemini} is designed to detect specific vulnerabilities, e.g., \textit{OpenSSL} Heartbleed vulnerability, in firmware. It first extracts CFGs from the vulnerability and firmware respectively. Then, \textit{Gemini} converts all CFGs to ACFGs, which are noted as $Vul_{ACFG}$ and $Firm_{ACFG}$. The ACFG is represented as a seven-dimensional vector, which consists of seven block-level attributes as shown in Table ~\ref{Table: Function-level and Block-level Attributes}. Next, it utilizes an embedding network, which is based on \textit{struct2vec}~\cite{dai2016discriminative}, to calculate the cosine similarity between these ACFGs, which is noted as  $Cosine(Vul_{ACFG},Firm_{ACFG})$. Finally, it lists the top-K similar functions in firmware according to $Cosine(Vul_{ACFG},Firm_{ACFG})$. }

Nevertheless, the original \textit{Gemini} cannot directly apply to TPC detection. In our task, we want to confirm the similarity between the TPC and firmware rather than the similarity of individual functions. Given each TPC has many ACFGs, the similarity of a single ACFG cannot determine the similarity between the TPC and firmware. The main challenge here is to design a method to aggregate the similarity of each ACFG in the TPC to represent the final similarity between the TPC and firmware. To address this challenge, we normalize and aggregate the similarity of each ACFG based on the weight of the corresponding CFG. We consider that if the CFG is more complicated, its internal logic will also be more complicated. Therefore, the high-complexity CFGs in the TPC are more likely to have a larger difference from the CFGs in other TPCs than low-complexity CFGs, which are more suitable as features to identify TPCs. Based on the above analysis, we give a high weight to the complex CFGs.
We use cyclomatic complexity (CC)~\cite{cyclomaticcomplexity} to evaluate the complexity of a CFG.
CC is calculated as follows:
\vspace{-1mm}
\begin{small}
\begin{equation}
    CC = e - b + 2
    \vspace{-1mm}
\end{equation}
\end{small}where $e$ is the number of edges in the CFG, and b is the number of basic blocks in the CFG. Next, we calculate the ratio of the complexity of each CFG to the complexity of all CFGs in the TPC as the weight of each CFG, which is defined as follows:
\vspace{-1mm}
\begin{small}
\begin{equation}
% \small
Weight(CFG) = \frac{CC(CFG)}{\sum_{i=0}^{N}CC(CFG_{i})}
\vspace{-1mm}
% Sim(TPC, Firmware) = \sum_{i=0}^{N}Sim(ACFG_{i})*Weight(ACFG_{i})
\end{equation}
\end{small}where $N$ is the number of CFGs in the TPC.
Finally, we aggregate the similarity of each ACFG to represent the similarity of the TPC and firmware, which is denoted as follows: 
\vspace{-1mm}
\begin{small}
\begin{equation}
Sim(TPC, Firmware) = \sum_{i=0}^{N}Sim(ACFG_{i})*Weight(CFG_{i})
% \vspace{-1mm}
\end{equation}
\end{small}where $Sim(ACFG_{i})$ is the similarity between each ACFG in the TPC and the most similar function in firmware. With the above strategy, we successfully apply the customized \textit{Gemini} to our task. We regard the TPC as matched if $Sim(TPC, Firmware)$ exceeds the threshold $\gamma$.
We train the customized \textit{Gemini} on Dataset \uppercase\expandafter{\romannumeral1} and then set threshold as 0.64, which are described in  Section~\ref{section: System Evaluation}.
% Given \textit{Gemini} supports retraining the model to adapt to specific tasks, we prepare an extra training dataset for \textit{Gemini}. The dataset includes the ACFGs we extracted from each TPC. 
% We retrain the pre-trained model used in \textit{Gemini} with our dataset for 100 epochs.
% Based on our evaluation results in  Section~\ref{section: System Evaluation}, we set the threshold as 0.64. 
We mark the matched results by CFG features as $R_{CFG}$. 

We finally take the union of $R_{Syntax}$ and $R_{CFG}$ as our final results, which can be represented as $R_{Syntax} \cup R_{CFG}$. For instance, the identification result of sample.bin in $R_{Syntax}$ is: \textit{OpenSSL unknown, uClibc 0.9.32.1}, and in $R_{CFG}$ is: \textit{OpenSSL 0.9.8, uClibc 0.9.32.1}. Then, $R_{Syntax} \cup R_{CFG}$ should be \textit{OpenSSL 0.9.8, uClibc 0.9.32.1}.
% We combine the results of sample.bin in $R_{Syntax}$ and $R_{CFG}$, thus the final results of sample.bin should be: 

% We then apply  the retrained model, which utilizes an embedding network, to calculate the cosine similarity between the two kinds of ACFGs from firmware and TPCs.   \textit{Gemini} will return the top-N candidate TPCs that used in firmware. In this work, we select the top-1 candidate TPC at version-level as the result.

% We use the $Sim(ACFG_{TPC}, ACFG_{Firmware})$ to represent the similarity between the TPC with the firmware. 

% \begin{algorithm}
% \caption{Pseudo code for the matching strategy}
% \label{alg:A}
% \begin{algorithmic}
% \Procedure{Roy}{$a,b$}
% \STATE {set $r(t)=x(t)$} 
% \REPEAT 
% \STATE set $h(t)=r(t)$ 
% \REPEAT
% \STATE set $h(t)=r(t)$ 
% \UNTIL{B} 
% \UNTIL{B}
% \end{algorithmic}
% \end{algorithm}

% \textit{Gemini}  extracted from  TPCs and firmware.   The retrained model

% adapting to specific  
% retraining the pre-trained model to adapt to specific tasks, in this paper, we prepare a training dataset to apply on detecting TPCs used in firmware.
% In this paper, we transfer by retraining its pre-trained model with our training dataset. 

% More specifically, we adopt its pre-trained model, and retrain the model with our training dataset to enhance the robustness of model. 

\noindent\textbf{Vulnerability Identification.}  In this step, we leverage versions check to identify the vulnerabilities of the detected TPCs in firmware based on our TPC database. The database includes the CVEs that correspond to different versions of TPCs. Therefore, we implement a script to automatically query the database with the TPCs and the corresponding versions (e.g., \textit{OpenSSL 0.9.8}), and record the returned vulnerability information. We need to clarify that some vulnerabilities may cannot be exploited since it is possible that some of the vulnerable code cannot be reached due to other remedies, such as disabling certain configuration options or performing some checks to prevent its use. Therefore, we regard identified vulnerabilities as potential vulnerabilities. We finally generate a report for the tested firmware, which indicates its potential risks and provides a series of suggestions for fixing the vulnerabilities.

\vspace{-1mm}
% \section{\textcolor{red}{System Evaluation}}
\section{System Evaluation}
\label{section: System Evaluation}
% In this section, we mainly evaluate the performance of \system. 
% Though the goal of this paper is not to propose a novel method, we want to convince people that the measurement results based on \system are reliable through the evaluation. % 
In this section, we first introduce the composition of our dataset.
% ~\Footnotes\footnote{We provide some information about the firmware in our dataset on https://github.com/FirmSecDataset/FirmSecDataset.}
%  which will be used in system evaluation and further  analysis.
 Next, we evaluate the performance of \system, and compare it with three state-of-the-art  works from  academia  and  commercial  systems  from  industry:  \textit{Gemini}~\cite{xu2017neural}, \textit{BAT}~\cite{hemel2011finding},  \textit{OSSPolice}~\cite{duan2017identifying}, \textit{binaré}~\cite{binare}, \textit{360 FirmwareTotal}~\cite{360iot}, and \textit{Alibaba FSS}~\cite{aliiot}.

% : \textit{Gemini}~\cite{xu2017neural}, \textit{BAT}~\cite{hemel2011finding},  \textit{OSSPolice}~\cite{duan2017identifying}, \textit{binaré}~\cite{binare}, \textit{360 FirmwareTotal}~\cite{360iot}, and \textit{Alibaba FSS}~\cite{aliiot}.
% evaluate the accuracy of \system which achieves 
% % $100\%$ accuracy in identifying TPCs and 
% $96.3\%$  in discovering corresponding vulnerabilities. 
% % Then, we perform the \system on our dataset which identifies $584$ TPCs and discovers a total of $128,757$ security vulnerabilities that are caused by $429$ CVEs. 
% Then,  we compare \system with one academic system developed by Costin et al.~\cite{costin2014large}, and two industry-leading systems from \textit{360}~\cite{360iot} and \textit{Alibaba}~\cite{aliiot}. 
% \system detects an average of $15.8$ TPCs and $114.2$ vulnerabilities per firmware image while the corresponding results of \cite{costin2014large}, \textit{360}, and \textit{Alibaba} are $7.7$ and $82.9$, $4.8$ and $62.3$, and $3$ and $55.5$, respectively. In addition, \system spends an average of $4.2s$ processing each firmware while \cite{costin2014large} spends $303s$, \textit{360} spends $60s$, and \textit{Alibaba} spends $174.7s$.
% \system is eligible to identify more TPCs and discover more corresponding vulnerabilities based on the same vulnerability database that faster than 360~\cite{360iot} and \textit{Alibaba}~\cite{aliiot}. 

\begin{table*}[]
  \setlength{\abovecaptionskip}{0.08cm}
 %\centering
 \caption{Dataset Composition.}
 \label{Table: Dataset Composition}
 % \tiny
 % \renewcommand\arraystretch{0.85}
 % \renewcommand\arraystretch{1.2}
 \resizebox{\textwidth}{!}{%
 
 \begin{tabular}{cccccccccccccc}
 \toprule[1.5pt]
 Vendor &
   Category &
   \begin{tabular}[c]{@{}c@{}}\# Download\\ Firmware\end{tabular} &
   \begin{tabular}[c]{@{}c@{}}\# Valid\\ Firmware\end{tabular} &
   SquashFS &
   CramFS &
   JFFS2 &
   \begin{tabular}[c]{@{}c@{}}Other \\ Filesystems\end{tabular} &
   ARM &
   MIPS &
   X86 &
 \begin{tabular}[c]{@{}c@{}}Other \\ ARCH\end{tabular} &
   Linux-based &
   Non-Linux based \\ \midrule[1pt]
 Xiongmai                  & Camera          & 1,038 & 520   & -     & 45  & -  & 475   & 326   & -   & -  & 194   & 520   & -  \\ \hline
 Tomato-shibby             & Router          & 642   & 230   & 230   & -   & -  & -     & 168   & -   & -  & 62    & 230   & -  \\ \hline
 Phicomm                   & Router          & 107   & 107   & 73    & -   & -  & 34    & 30    & 25  & -  & 52    & 107   & -  \\ \hline
 \multirow{2}{*}{Fastcom}  & Router          & 200   & 149   & 13    & -   & -  & 136   & 53    & 34  & -  & 62    & 149   & -  \\ \cline{2-14} 
                           & Unknown         & 10    & 10    & 6     & -   & -  & 4     & -     & 4   & -  & 6     & 10    & -  \\ \hline
 \multirow{4}{*}{Trendnet} & Camera          & 492   & 477   & 8     & 44  & -  & 425   & 274   & 100 &    & 103   & 477   & -  \\ \cline{2-14} 
                           & Router          & 463   & 336   & 261   & -   & -  & 75    & 119   & 91  & -  & 126   & 336   & - \\ \cline{2-14} 
                           & Switch          & 162   & 162   & 93    & -   & 18 & 51    & 116   & 17  & -  & 29    & 162   & - \\ \cline{2-14} 
                           & Unknown         & 168   & 106   & 12    & 18  & 4  & 72    & 81    & 2   & -  & 23    & 106   & -  \\ \hline
 Xiaomi                    & Router          & 21    & 21    & 21    & -   & -  & -     & 8     & 9   & -  & 4     & 21    & -  \\ \hline
 \multirow{4}{*}{TP-Link}  & Camera          & 384   & 319   & 242   & -   & 35 & -     & 170   & 145 & -  & 4     & 319   & -  \\ \cline{2-14} 
                           & Router          & 996   & 820   & 709   & -   & 19 & 92    & 143   & 638 & -  & 39    & 820   & - \\ \cline{2-14} 
                           & Switch          & 270   & 270   & 22    & -   & -  & 248   & 110   & 43  & -  & 117   & 270   & - \\ \cline{2-14} 
                           & Unknown         & 48    & 48    & 20    & -   & 1  & 27    & 19    & 16  & -  & 13    & 48    & -  \\ \hline
 \multirow{4}{*}{D-Link}   & Camera          & 360   & 360   & 17    & 11  & -  & 332   & 192   & 106 & -  & 62    & 360   & -  \\ \cline{2-14} 
                           & Router          & 555   & 552   & 401   & -   & 2  & 149   & 268   & 109 & 6  & 169   & 552   & - \\ \cline{2-14} 
                           & Switch          & 695   & 545   & 198   & -   & -  & 347   & 312   & 18  & 8  & 207   & 545   & - \\ \cline{2-14} 
                           & Unknown         & 91    & 91    & 4     & 8   & 5  & 74    & 25    & 7   & 1  & 58    & 91    & - \\ \hline
 Hikvision                 & Camera          & 158   & 139   & -     & 47  & 1  & 91    & 131   & -   & -  & 8     & 139   & -  \\ \hline
 Foscam                    & Camera          & 113   & 113   & -     & -   & -  & 113   & 53    & -   & -  & 60    & 113   & -  \\ \hline
 Dahua                     & Camera          & 419   & 419   & 9     & 19  & -  & 341   & 260   & 2   & 1  & 156   & 419   & -  \\ \hline
 \multirow{20}{*}{TSmart}  & Camera          & 326   & 326   & 88    & 123 & 43 & 72    & 191   & 45  & -  & 90    & 296   & 30 \\ \cline{2-14} 
                           & Smart Switch    & 2,053 & 2,053 & 1     & -   & -  & 2,052 & 267   & 2   & -  & 1,784 & 29    & 2,024  \\ \cline{2-14} 
                           & Sweeper         & 33    & 33    & -     & -   & -  & 33    & 10    & 2   & -  & 21    & 25    & 8  \\ \cline{2-14} 
                           & Light           & 8,089 & 8,089 & -     & -   & 1  & 8,088 & 617   & -   & -  & 7,472 & 12    & 8,077  \\ \cline{2-14} 
                           & General         & 871   & 856   & 2     & 9   & 44 & 801   & 212   & 2   & -  & 642   & 776   & 80  \\ \cline{2-14} 
                           & Plugin          & 8,294 & 8,294 & -     & -   & -  & 8,294 & 478   & -   & -  & 7,816 & 14    & 8,280  \\ \cline{2-14} 
                           & Heater          & 361   & 361   & -     & -   & -  & 361   & -     & -   & -  & 361   & -     & 361  \\ \cline{2-14} 
                           & Blueteeth Light & 1,000 & 1,000 & -     & -   & -  & 1,000 & 637   & -   & -  & 363   & -     & 1,000  \\ \cline{2-14} 
                           & Air             & 517   & 517   & -     & -   & -  & 517   & 96    & -   & -  & 421   & -     & 517  \\ \cline{2-14} 
                           & Curtain         & 195   & 195   & -     & -   & -  & 195   & 8     & -   & -  & 187   & -     & 195  \\ \cline{2-14} 
                           & Lock            & 68    & 68    & -     & -   & -  & 68    & -     & -   & -  & 68    & -     & 68  \\ \cline{2-14} 
                           & Freezer         & 44    & 44    & -     & -   & 4  & 40    & 4     & -   & -  & 40    & 4     & 40  \\ \cline{2-14} 
                           & Air Purifier    & 24    & 24    & -     & -   & -  & 24    & -     & -   & -  & 24    & -     & 24  \\ \cline{2-14} 
                           & Humidifier      & 12    & 12    & -     & -   & -  & 12    & -     & -   & -  & 12    & -     & 12  \\ \cline{2-14} 
                           & Dehumidifier    & 57    & 57    & -     & -   & -  & 57    & -     & -   & -  & 57    & -     & 57  \\ \cline{2-14} 
                           & Heat Controller & 49    & 49    & -     & -   & -  & 49    & -     & -   & -  & 49    & -     & 49  \\ \cline{2-14} 
                           & Fan             & 14    & 14    & -     & -   & -  & 14    & -     & -   & -  & 14    & -     & 14  \\ \cline{2-14} 
                           & Washer          & 4     & 4     & -     & -   & -  & 4     & 4     & -   & -  & -     & -     & 4  \\ \cline{2-14} 
                           & Gateway         & 60    & 60    & 3     & -   & -  & 57    & 14    & 14  & -  & 32    & 3     & 57  \\ \cline{2-14} 
                           & Others          & 994   & 994   & 42    & 21  & 22 & 909   & 337   & 6   & -  & 651   & 97    & 897  \\ \hline
 OpenWrt                   & Router          & 5,585 & 5,292 & 3,903 & -   & 41 & 1,348 & 2,415 & 194 & 10 & 2,673 & 5,292 & - \\ \bottomrule[1.5pt]
 \end{tabular}}
 \end{table*}

\vspace{-2mm}
\subsection{Dataset Composition}
\label{subsection: Dataset Composition}
Based on the firmware collection module, 
we initially collect a total of %$37,266$%  
$35,978$ firmware images varying from $13$ vendors which involve $35$ kinds of devices. More specifically, %$13,328$% 
$12,913$ firmware images are crawled from the Internet and %$23,938$%
$23,065$ firmware images are obtained from TSmart. Our dataset actually involves hundreds of vendors since the private firmware images from TSmart are manufactured by hundreds of vendors. TSmart provides a platform that can enable the devices from various vendors to become smart products. We use TSmart as the vendor of private firmware for convenience. 
% Next, we filter the non-firmware files from our raw dataset and identify the related information of remaining unrecognized firmware, according to the firmware preprocessing module. 
We list the detailed composition of our dataset as shown in Table~\ref{Table: Dataset Composition}. 
% (Appendix~\ref{section: Vulnerability Evaluation}). 
After data filtration, we get rid of %$2,770$%
$1,842$ non-firmware files and finally obtain %$33,833$%
$34,136$ valid firmware images across $13$ vendors including %$13,173$%
$11,086$ publicly accessible firmware images and 
%$23,430$% 
$23,050$ private firmware images from TSmart. 
Our dataset includes  $35$ kinds of known IoT devices and a part of unknown IoT devices.  $2,694$ ($7.9\%$) firmware images are used in camera, $7,293$ ($21.3\%$) firmware images belong to router, $1,191$ ($3.5\%$) firmware images are deployed on switch, $23,050$ ($67.5\%$) firmware images from TSmart have been equipped on smart homes, and $255$ ($0.7\%$) are unknown.
% , and $401$ ($1.2\%$) firmware images are from other kinds of devices.
% However, there are still $1,249$ ($3.7\%$) firmware images whose exact device categories cannot be identified. We classify these unidentified firmware images as unknown. 
Apart from the vendors and devices mentioned above, our dataset also covers several instruction sets, of which ARM ($23.9\%$) takes the majority and MIPS follows ($4.9\%$). SquashFS, CramFS, and JFFS2 are three popular filesystems included in the dataset.  Most of the unknown filesystems actually belong to the above three filesystems based on our further analysis. Vendors customize the above filesystems in firmware, which also changes the magic number of the original filesystems used for identification. Moreover, 12,342 ($36.2\%$) firmware images are Linux-based and 21,794 ($63.8\%$) firmware images are non-Linux based.

\vspace{-2mm}
\subsection{Evaluation}
\label{section: Accuracy Evaluation}
% \subsubsection{End-to-End Accuracy}
% Before we perform \system on our dataset, it is important to obtain the ground truth that whether the results generated by \system are correct at first. 

% in identifying the TPCs and .
% vulnerabilities caused by TPCs in firmware.

% Though there are some systems~\cite{backes2016reliable,duan2017identifying,zhang2019libid} that have been proposed to detect the TPCs used in Android apps, they cannot analyze the IoT firmware. These systems require Android features support. Besides, there is no 
% We finally compare \system with BAT, Both \textit{Gemini}

\noindent\textbf{Evaluation Dataset Construction.} We first build
four datasets: (1) Dataset \uppercase\expandafter{\romannumeral1} for training the customized \textit{Gemini} and evaluating the accuracy of the model;
(2) Dataset \uppercase\expandafter{\romannumeral2} for choosing the appropriate thresholds used in  \system ; (3) Dataset \uppercase\expandafter{\romannumeral3} for evaluating the accuracy of \system in detecting the TPCs at TPC-level and version-level in firmware; (4) Dataset \uppercase\expandafter{\romannumeral4} for comparing the performance of \system with three commercial systems from industry; (5) Dataset \uppercase\expandafter{\romannumeral5} for evaluating the precision of \system in identifying the vulnerabilities caused by TPCs in firmware.

Dataset \uppercase\expandafter{\romannumeral1} includes the ACFGs we extracted from 1,192 TPCs in our TPC database.
Dataset \uppercase\expandafter{\romannumeral2} and Dataset \uppercase\expandafter{\romannumeral3} have a labeled mapping of firmware to TPCs usage for ground truth. 
To the best of our knowledge, there are no such datasets available from previous works. Given it is difficult to know the specific TPCs used in commercial firmware, we finally use  Tomato-shibby and OpenWrt firmware images to build our datasets since they have source code and clearly indicate the adopted TPCs with exact versions in the  configuration files. For Dataset \uppercase\expandafter{\romannumeral2}, we randomly collect $200$ Tomato-shibby firmware images from our dataset, which include $17,918$ TPC-version pairs ($73$ different TPCs with $211$ different versions).  For Dataset \uppercase\expandafter{\romannumeral3}, we randomly select $300$ OpenWrt firmware images from our dataset, which include $19,645$ TPC-version pairs ($92$ different TPCs with $194$ different versions).
 We use the two datasets to perform the threshold selection and evaluation respectively to avoid bias. For Dataset \uppercase\expandafter{\romannumeral4}, we randomly select 24 firmware images include 19 Linux-based firmware images and five monolithic firmware images. For Dataset \uppercase\expandafter{\romannumeral5}, we randomly select 35 firmware images which have 382 vulnerabilities identified by \system.

% \noindent$\bullet$\textit{Dataset \uppercase\expandafter{\romannumeral1}}: We first need to construct a dataset e. 

% \noindent$\bullet$\textit{Dataset \uppercase\expandafter{\romannumeral2}}: This dataset includes the firmware which has the confirmed vulnerabilities. We mainly evaluate \system on two vulnerabilities: OpenSSL Heartbleed vulnerability (CVE-2014-0160) and glibc Ghost  vulnerability (CVE-2015-0235). For each vulnerability, we collect $30$ vulnerable firmware images based on manual analysis.

% We first build a ground-truth dataset to evaluate the accuracy of \system in detecting the TPCs used in firmware. This dataset has a labeled mapping of firmware to TPCs usage for ground truth. To the best of our knowledge, there is no such a dataset available from previous works. Given it is difficult to know the specific TPCs used in commercial firmware, we finally collect $300$ OpenWrt firmware images. These open-source firmware images have source code and clearly indicate the adopted TPCs with exact versions in the  configuration files. More specifically, these $300$ firmware images include $72$ different TPCs with $194$ different versions. 
\noindent\bin{\textbf{Model Accuracy.} We split  Dataset
\uppercase\expandafter{\romannumeral1} into three subsets for training, validation, and testing
respectively according to the ratio of 6:2:2. 
We train the customized \textit{Gemini} for 100 epochs based on the original \textit{Gemini}'s training process. The model that achieves the best AUC (Area Under the
Curve) on the validation set during the 100 epochs is saved. We finally test the model on the testing set and the AUC is 0.953. We also retrain the original \textit{Gemini} with the same process, which AUC is only 0.912.}

\noindent\textbf{Threshold Selection.}   Our final results are the union of the results that matched by syntactical features and CFG features. We do not directly use the thresholds when the respective method achieves the optimal performance since the union results may not reach the best at this time. We take three thresholds as a whole and utilize the true positive rate (TPR) at version-level as the metric to select the appropriate thresholds. We combine the three thresholds and their corresponding TPR as a four-dimensional vector: [$\alpha$, $\beta$, $\gamma$, TPR]. Each threshold ranges from 0.01 to 1.00. We select the thresholds when the TPR reaches the highest. Based on our experiment, \system achieves the highest TPR (91.47\%) when the thresholds of $D(S_{TPC},S_{Firmware})$, $\frac{S_{TPC}\cap S_{Firmware}}{S_{TPC}}$, and $Sim(ACFG_{TPC}, ACFG_{Firmware})$  are 0.74, 0.52, and 0.64, respectively. We leverage the above thresholds for the following evaluation and analysis.

% two datasets: (1) Dataset \uppercase\expandafter{\romannumeral1} for evaluating the accuracy of \system in detecting the TPCs at version level in firmware; (2) Dataset \uppercase\expandafter{\romannumeral2} for evaluating the accuracy of \system in detecting the specific vulnerabilities in firmware.

% \noindent$\bullet$\textit{Dataset \uppercase\expandafter{\romannumeral1}}: We first need to construct a dataset e. 

% \noindent$\bullet$\textit{Dataset \uppercase\expandafter{\romannumeral2}}: This dataset includes the firmware which has the confirmed vulnerabilities. We mainly evaluate \system on two vulnerabilities: OpenSSL Heartbleed vulnerability (CVE-2014-0160) and glibc Ghost  vulnerability (CVE-2015-0235). For each vulnerability, we collect $30$ vulnerable firmware images based on manual analysis.

% \usepackage{multirow}
\begin{table}[]
 \setlength{\abovecaptionskip}{0.08cm}
\caption{Comparison of \system, \textit{Gemini}, \textit{BAT}, and \textit{OSSPolice}.}
\renewcommand\arraystretch{0.85}
\small
\label{Table: Academic Comparison}
\centering
\begin{tabular}{ccccc}
\toprule[1.5pt]
\multirow{2}{*}{Tools} & \multicolumn{2}{c}{TPC-level} & \multicolumn{2}{c}{Version-level} \\ \cmidrule{2-5} 
                       & Precision      & Recall       & Precision        & Recall         \\ \midrule[1pt]
\system              & 92.09\%        & 95.24\%      & 91.03\%          & 92.26\%        \\
Syntax-based              & 92.38\%        & 86.29\%      & 91.47\%          & 81.66\%        \\
CFG-based              & 93.72\%        & 82.76\%      & 94.65\%          & 80.90\%        \\
\textit{Gemini}~\cite{xu2017neural}              & 89.60\%        & 74.19\%      & 90.78\%          & 71.73\%        \\
\textit{BAT}~\cite{hemel2011finding}                    & 70.74\%        & 56.38\%      & NA               & NA             \\
\textit{OSSPolice}~\cite{duan2017identifying}              & 86.63\%        & 71.85\%      & 82.51\%          & 67.05\%        \\ \bottomrule[1.5pt]
% \textit{Asm2Vec}
\end{tabular}
\vspace{-2em}
\end{table}

\noindent\textbf{Evaluation Results.}   We evaluate the detection accuracy of \system on Dataset \uppercase\expandafter{\romannumeral3} at TPC-level and version-level with two evaluation metrics: precision ($\frac{TP}{TP+FP}$) and recall ($\frac{TP}{TP+FN}$). The false positives represent the TPCs and versions that we wrongly identified, and the false negatives represent the TPCs and versions that we do not find. As shown in Table~\ref{Table: Academic Comparison}, \system achieves $92.09\%$ precision, $95.24\%$ recall at TPC-level, and $91.03\%$ precision, $92.26\%$ recall at version-level. We also list the syntax-based matching results and the CFG-based matching results when the TPR of each method reaches the highest. We find that the precision of the union results is very close to the respective methods, but the recall rate is much higher. We explore the reasons that why the union results have a great performance than the separate methods. First, the true positives of the two methods have non-overlapping parts, thus reducing the union results' false negatives. For instance, the syntax-based method is hard to detect the TPCs which have limited syntactical features (e.g., without unique string literals), but they can be identified through CFG features. Second, their false positives have overlapping parts, which will not greatly increase the false positives of the union results. We further analyze the false positives and false negatives. First, the false positives are mainly caused by two reasons: 1) TPCs overlapping. Several TPCs reuse the code of other TPCs with minor changes. Under these circumstances, \system will report all matched TPCs. 2) High similarity between different versions. Some versions of the same TPC have little difference or even no difference in their code. Second, the false negatives are due to two reasons: 1) Uncommon TPCs used in firmware. \system cannot detect the TPCs that are not included in our database. 2) Insufficient features. Some firmware only use partially built TPCs which have limited features that can be used for identification.

\begin{table*}[]
\centering
% \large
\tiny

\renewcommand\arraystretch{1.0}
\setlength\tabcolsep{4.5pt}
\caption{Comparison of \textit{\system}, \textit{binaré}, \textit{360 FirmwareTotal}, and \textit{Alibaba FSS}.}
\label{Table: Comparison}
% \vspace{-5mm}
\resizebox{\textwidth}{!}{
% \Rotatebox{90}{%

\begin{tabular}{ccccccclccclccclccc}
\toprule[1.5pt]
\multirow{2}{*}{Vendor} &
  \multirow{2}{*}{Category} &
  \multirow{2}{*}{Firmware} &
  \multirow{2}{*}{OS Support} &
  \multicolumn{3}{c}{\system} &
  &
  \multicolumn{3}{c}{
\textit{binaré}\cite{binare}} &
  &
  \multicolumn{3}{c}{\textit{360 FirmwareTotal}~\cite{360iot}} &
  &
  \multicolumn{3}{c}{\textit{Alibaba FSS}~\cite{aliiot}} \\ \cmidrule{5-7} \cmidrule{9-11} \cmidrule{13-15} \cmidrule{17-19} 
 &
  &
  &
  &
  \# TPC &
  \# Vul. &
  Time (s) &
  &
  \# TPC &
  \# Vul. &
  Time (s) &
  &
  \# TPC &
  \# Vul. &
  Time (s) &
  &
  \# TPC &
  \# Vul. &
  Time (s) \\ \midrule[1pt]
D-Link   & Router     & 3001\_DIR885LA1.bin & \textcolor{green}{\checkmark}   & \textbf{17} & \textbf{124} & \textbf{9} &  & 12 & 69 & 420 &  & 4  & 7            & 60 &  & 7  & 90           & 47    \\
D-Link   & Camera     & dcs-935l\.bin  & \textcolor{green}{\checkmark}        & \textbf{11}  & 65           & \textbf{7} &  & 5  & 56  & 300 &  & 1  & 1            & 60 &  & 5  & \textbf{78}  & 44    \\
D-Link   & Router     & dir895la1.bin  & \textcolor{green}{\checkmark}        & \textbf{17} & 114         & \textbf{9} &  & 11 & 113 & 300 &  & 6  & \textbf{173} & 60 &  & 7  & 96           & 20    \\
TP-Link  & Router     & c7v5.bin    & \textcolor{green}{\checkmark}           & \textbf{19} & \textbf{146} & \textbf{6} &  & 12 & 119 & 420 &  & 7  & 10           & 60 &  & NA & NA           & 47    \\
TP-Link  & Router     & c2600v1.bin   & \textcolor{green}{\checkmark}         & \textbf{31} & \textbf{148} & \textbf{9} &  & 12 & 119 & 480 &  & 10 & 99           & 60 &  & 1  & 1            & 209   \\
TP-Link  & Router     & archer\_a7(eu).bin  & \textcolor{green}{\checkmark}    & \textbf{18} & \textbf{123} & \textbf{6} &  & 11 & 96 & 240 &  & 7  & 88           & 60 &  & NA & NA           & 53    \\
TP-Link  & Router     & ad7200v2.bin  & \textcolor{green}{\checkmark}         & \textbf{30} & \textbf{140} & \textbf{8} &  & 12 & 114 & 360 &  & 10 & 99           & 60 &  & 1  & 1            & 220   \\
TP-Link  & Router     & e4r.bin  & \textcolor{green}{\checkmark}              & \textbf{15} & 106           & \textbf{8} &  & 11 & 89 & 300 &  & 5  & \textbf{228} & 60 &  & 1  & 0            & 31    \\
TP-Link  & Router     & p7 1.0\_en\_1.2.0.bin & \textcolor{green}{\checkmark} & \textbf{14}  & \textbf{101}  & \textbf{5} &  & 8  & 72  & 300 &  & 4  & 69           & 60 &  & NA & NA           & 68    \\
TP-Link  & Camera     & tl-ipc423(p).bin  & \textcolor{green}{\checkmark}     & \textbf{16}  & 125           & \textbf{5} &  & 4  & 114 & 300 &  & 5  & 9            & 60 &  & 6  & \textbf{135} & 176   \\
TP-Link  & Camera     & tl-ipc543k(p).bin  & \textcolor{green}{\checkmark}    & \textbf{14}  & 106           & \textbf{7} &  & 4  & 105 & 240 &  & 4  & 7            & 60 &  & 6  & \textbf{135} & 60    \\
Trendnet & Router     & luci-ar71xx.bin   & \textcolor{green}{\checkmark}     & \textbf{15} & \textbf{107} & \textbf{9} &  & 8  & 85  & 300 &  & NA & NA           & 60 &  & NA & NA           & 46    \\
Trendnet & Router     & tew755ap.bin    & \textcolor{green}{\checkmark}       & \textbf{24} & 141          & \textbf{5} &  & 8  & 114 & 360 &  & 4  & 78           & 60 &  & 13 & \textbf{330} & 32    \\
Trendnet & Switch     & tl2-g244-all.hex  & \textcolor{green}{\checkmark}     & \textbf{4}  & \textbf{48}  & \textbf{8} &  & 4  & 27  & 240 &  & 1  & 6            & 60 &  & 1  & 42           & 51    \\
Dahua    & Camera     & wifimac\_upall.bin  & \textcolor{green}{\checkmark}   & \textbf{15} & \textbf{111}  & \textbf{6} &  & 8  & 84  & 240 &  & 7  & 40           & 60 &  & 1  & 1            & 28    \\
Xiaomi   & Router     & miwifi\_r1cm\_all.bin & \textcolor{green}{\checkmark} & \textbf{16} & 128          & \textbf{7} &  & 7  & 100  & 360 &  & 7  & \textbf{176} & 60 &  & NA & NA           & 66    \\
Xiaomi   & Router     & miwifi\_all\_bd6da.bin & \textcolor{green}{\checkmark} & \textbf{16} & \textbf{116} & \textbf{5} &  & 8  & 88  & 360 &  & 7  & 84           & 60 &  & NA & NA           & 32    \\
TSmart   & Camera & ymsmart-2.0.2.bin  & \textcolor{green}{\checkmark}    & \textbf{8}  & \textbf{133} & \textbf{5} &  & NA & NA  & 120 &  & NA & NA           & 60 &  & 6  & 131          & 1,977 \\
TSmart   & Camera & lcipc-consumer.bin  & \textcolor{green}{\checkmark}   & \textbf{5}  & \textbf{102}  & \textbf{3} &  & 2  & 11  & 300 &  & 2  & 9            & 60 &  & 2  & 14           & 78    \\
TSmart   & Light      & XR871\_QIO\_1.1.1.bin & \textcolor{red}{\XSolidBrush} & \textbf{1}          & \textbf{27}           & \textbf{8} &  & NA & NA  & 120 &  & NA & NA           & 60 &  & NA & NA           & 34    \\

TSmart   & Light      & esp\_warm\_light\_1.1.0.bin & \textcolor{red}{\XSolidBrush} & \textbf{1}          & \textbf{2}           & \textbf{7} &  & NA & NA  & 120 &  & NA & NA           & 60 &  & NA & NA           & 55    \\ 
TSmart   & Doorbell    & DOORBELL-2.0.3.20180425.bin & \textcolor{red}{\XSolidBrush} & \textbf{1}          & \textbf{16}           & \textbf{8} &  & NA & NA  & 120 &  & NA & NA           & 60 &  & NA & NA           & 49    \\  
TSmart   & Plug      & esp\_12F\_xw\_plug\_1.0.5.bin & \textcolor{red}{\XSolidBrush} & \textbf{1}          & \textbf{36}           & \textbf{9} &  & NA & NA  & 120 &  & NA & NA           & 60 &  & NA & NA           & 21    \\ 
TSmart   & Plug      & SIG-WITHOUTSYS\_0.0.16.bin & \textcolor{red}{\XSolidBrush} & \textbf{1}          & \textbf{36}           & \textbf{7} &  & NA & NA  & 120 &  & NA & NA           & 60 &  & NA & NA           & 42    \\
\bottomrule[1.5pt]
% & & & \# $\overline{3rd \ P.C.}$ & \# $\overline{Vul.}$ &  $\overline{Time(s)}$ & & \# $\overline{3rd \ P.C.}$ & \# $\overline{Vul.}$ &  $\overline{Time(s)}$ & & \# $\overline{3rd \ P.C.}$ & \# $\overline{Vul.}$ &  $\overline{Time(s)}$ & & \# $\overline{3rd \ P.C.}$ & \# $\overline{Vul.}$ & \# $\overline{Time(s)}$ &
%  & Average Results of 19 Linux-based Firmware Images & &  & \textbf{15.8} & \textbf{114.2} & \textbf{4.2} & & 7.7 & 82.9 & 303 & & 4.8 & 62.3 & 60 & & 3 & 55.5 & 174.7 & \hline 

\end{tabular}
}
\vspace{-1em}
\end{table*}

% \vspace{-0.3cm}
% \subsection{Comparison With  State of the Arts}
% \label{section: performance measurement}
% The \system achieves great results in matching the TPCs and identifying the N-days vulnerabilities from $34,430$ firmwares.
% Though multiple static analysis methods are proposed to detect the vulnerabilities in the firmware, we cannot compare \system with all of them. Most of these methods require two inputs: . In this paper, we compare which are still working now.
% Besides, there are several companies also provide commercial tools that claim to detect the N-days vulnerabilities caused by TPCs from the firmware and most of these tools are still in test.

\noindent\textbf{Comparative Analysis With Academic Works.}
% \noindent\textbf{Comparison Results.} 
 We compare \system with three state of the arts: \textit{Gemini}~\cite{xu2017neural}, \textit{BAT}~\cite{hemel2011finding} and \textit{OSSPolice}~\cite{duan2017identifying}. The original \textit{Gemini} uses 8 block-level attributes to conduct code similarity comparison. We enhance it with our CFG weight-based method to detect the TPCs in firmware. 
\textit{BAT} leverages string literals to identify the TPCs in binaries, but cannot detect the exact versions. \textit{OSSPolice} is originally designed to detect the open-source software usage at version-level in Android apps. Given it supports finding TPCs in C/C++ native libraries, we successfully apply it on IoT firmware and compare it with \system. Before we conduct comparison, we generate the corresponding TPC database for \textit{BAT} and \textit{OSSPolice} based on their instructions respectively.
Table~\ref{Table: Academic Comparison} presents the comparison results of \system with \textit{Gemini}, \textit{BAT}, and \textit{OSSPolice}. We do not report the results of \textit{BAT} at version-level since it does not support version-level identification. For TPC-level identification, \system is far better than other tools. \system reports more TPCs at a higher precision and recall rate. For version-level identification, \system outperforms \textit{Gemini} and \textit{OSSPolice} on both metrics. We further explore the reasons that why \system is better than other tools. First, \textit{Gemini} ignores the function-level attributes which offer much useful information on the structure of CFGs. Second, \textit{BAT} mainly utilizes string literals to identify the TPCs. It uses the direct feature matching to compare the string literals extracted from TPCs and firmware, which causes a low precision and recall rate. Finally, \textit{OSSPolice} utilizes a hierarchical matching strategy, which relies on the package structures of TPCs, to identify TPCs. Nevertheless, the package structures of the same TPC in different versions can be easily changed. 
Besides, its feature extraction tool does not perform well on IoT firmware since it is not optimized for firmware.

\noindent\textbf{Comparative Analysis With Industry Systems.} We compare \system with
% one academic incubated system:  \textit{binaré}~\cite{costin2014large, binare} and
three industry systems: \textit{binaré}~\cite{binare}, \textit{360 FirmwareTotal}~\cite{360iot}, and \textit{Alibaba FSS}~\cite{aliiot}. \textit{binaré} is actually developed from a state-of-the-art work~\cite{costin2014large}. The prototype system proposed in \cite{costin2014large} did not support finding the TPCs and the corresponding vulnerabilities in firmware while its industry version supports this function. \textit{360} and \textit{Alibaba} are two well-known companies worldwide, and both have much experience in IoT security. They are the few that offer publicly accessible systems that can detect the vulnerability caused by TPCs in firmware.  \textit{binaré} and \textit{FirmwareTotal} offer free trials and \textit{FSS} costs \$300 for each usage. 

\bin{We do not perform the comparative analysis with industry systems on Dataset \uppercase\expandafter{\romannumeral3} due to two reasons. First, two industry systems limit the number of free trials. We are not allowed to perform them on hundreds of firmware images. Second, during our analysis, we find two industry systems directly analyze the configuration files in OpenWrt firmware to identify the TPCs, hindering the fairness of the experiment. Therefore, we finally perform the comparative analysis on Dataset \uppercase\expandafter{\romannumeral4}, including 24 closed-source firmware images. The analysis results are as follows.} 

\textbf{First}, \system can analyze all test firmware images and is the only one that supports analyzing monolithic firmware. For the 19 Linux-based firmware images,  \textit{binaré} is unable to process $1$ firmware images, \textit{FirmwareTotal} cannot deal with $2$ firmware images, and \textit{FSS} fails to analyze $7$ firmware images.
% Those Linux-based firmware images that cannot be analyzed by \textit{360} and \textit{Alibaba} do not adopt the traditional filesystems which increase the difficulty of analysis.
% In addition, to measure the performance of \system comprehensively, we specially select a firmware image from TSmart which cannot be analyzed by \system, as shown in the last line of Table~\ref{Table: Comparison}. \cite{costin2014large}, \textit{360}, and \textit{Alibaba} also fail to analyze this firmware. 
% To explore the reasons behind, we use the \textit{IDA Pro} to conduct static analysis on this firmware. Based on our result, we find this firmware is just composed of pieces of function codes. We do not detect any TPCs from this smart home device.
\textbf{Second}, 
% compared with the other three systems, 
\system detects the most TPCs in each test case. The TPCs discovered by \system have covered all the results of \textit{binaré}, \textit{FirmwareTotal}, and \textit{FSS}.
% We conclude the excellent performance of \system to three reasons: 1) \system conducts a more thorough extraction of firmware. Since \system contains a custom extraction tool, it extracts more objects from firmware that keep more information.  2) \system maintains a large-scale TPC list. 
We notice that the results from \textit{binaré}, \textit{FirmwareTotal} and \textit{FSS} mainly contain several popular TPCs, e.g., BusyBox and OpenSSL,  while \system can detect some uncommon TPCs, e.g., libogg. 
% 3) \system adopts an effective matching algorithm. 
% \system maintains a large-scale TPC list and adopts a custom matching algorithm, . The system 
% \textit{Alibaba} does not achieve a great performance in detecting the TPCs since it 
% % The \system maintains a large scale vulnerability list to match the TPCs in firmwares.
% % \system has discovered all the TPCs with the same version that reported by \textit{360} and \textit{Alibaba}. Since \system maintains a large-scale TPC list and adopts a custom matching algorithm, it discovers 
% %  it discovers more TPCs which contain confirmed N-days vulnerabilities than 360 and \textit{Alibaba}.
\textbf{Third}, \system identifies more N-days vulnerabilities in $12$ of the $19$ Linux-based firmware images in comparison with the other three systems. We conclude two reasons that why \system detects fewer vulnerabilities than other systems in the other 7 test cases. (1) \textit{binaré}, \textit{FirmwareTotal}, and  \textit{FSS} incorrectly report N-days vulnerabilities not belonging to the corresponding firmware. (2) For several cases, they report multiple versions of the same TPCs. For example, \textit{FSS} reports two different versions of OpenSSL in tl-ipc423(p).bin, and calculates the vulnerabilities according to these two versions. We further remove the above false positives of other systems in these 7 test cases and find \system can detect more N-days vulnerabilities than them in all cases.
\textbf{Finally}, \system spends less time when analyzing firmware. We implement \system on an Ubuntu server equipped with the i7-9700K CPU, RTX 2080 GPU, and 32 GB memory. 
\system analyzes each firmware image in less than 10 seconds which is much quicker than other systems in all test cases. 

% For each firmware image, \textit{360} spends 60 seconds to analyze it. \textit{binaré} spends an average of  303 seconds processing each firmware image.
% \textit{Alibaba} has an unstable performance on analyzing firmware where the best case is 20 seconds and the worst case is 33 minutes.

Overall, our results show that \system has a much better performance
than state-of-the-art tools from academia and commercial tools from industry. \system has a higher precision and recall than \textit{Gemini}, \textit{BAT}, and \textit{OSSPolice} at TPC-level and version-level.
In comparison with \textit{binaré}, \textit{FirmwareTotal}, and \textit{FSS}, \system achieves a higher success rate in analyzing firmware with less time, and can detect more TPCs, as well as identify more vulnerabilities.

\subsection{Identified Vulnerability Precision Evaluation}
\label{section: Vulnerability Evaluation}
% \subsubsection{End-to-End Accuracy}
% Before we perform \system on our dataset, it is important to obtain the ground truth that whether the results generated by \system are correct at first. 
In this subsection, we evaluate the precision of \system in identifying the vulnerabilities caused by TPCs in firmware.

\noindent\textbf{Experiment Setup.}  In our evaluation, we employ three methods to verify the vulnerabilities: 1) source code audit; 2) static analysis; and 3)  emulating-based dynamic analysis.
% 1) use source code audit to check the vulnerabilities if the source code of the firmware is available; 2) utilize static analysis to verify the vulnerabilities which have disclosed the vulnerable code snippets; and 3) adopt emulating-based dynamic analysis to verify the vulnerabilities which have public Proof-of-Checks (PoCs). 
We 
choose these methods for three reasons. First of all, a part of firmware  from OpenWrt has source code. Therefore, we can check its source code to verify the vulnerabilities. Second, since a large number of firmware images do not have source code,  static analysis is an acceptable method to verify the vulnerabilities in this case. Third, static analysis may fail to check several vulnerabilities that do not disclose the corresponding insecure codes. In this case, dynamic analysis can be used to make up for the shortcomings of static analysis. We here employ  \textit{FIRMADYNE}~\cite{chen2016towards}  to conduct dynamic analysis on firmware. 
% \yuan{\color{red}How do you combine the dynamic and static analysis?}
\noindent\textit{FIRMADYNE} supports emulating the Linux-based firmware on desktop. It overcomes the general challenges in emulating firmware, such as the presence of hardware-specific peripherals. 
Our evaluation method can confirm the true positives and false positives of our results, but hard to tell the false negatives since we do not have the ground truth of all TPC-related vulnerabilities in firmware. 

In consideration of our large-scale dataset, it is unrealistic to evaluate \system's precision on all firmware images. Therefore, we select some representative firmware according to two standards: 1) each identified vulnerability of the firmware has practical verification methods, whether through source code audit, static analysis, or dynamic analysis; and 2) include as many vendors and categories as possible.
% We thus choose five firmwares, which are convenient to conduct static analysis and dynamic analysis, for this test.
% \yuan{\color{red}How many images did you choose? Need to be specific. How do you choose them? If you are not doing a random sampling, people might say you are cherry picking results.} 
We finally perform the evaluation on Dataset \uppercase\expandafter{\romannumeral5}, including $35$ randomly selected  firmware images which have a total of $382$ vulnerabilities that identified by \system, as shown in Table~\ref{Table: Precision Evaluation}.

% $5$ out of the $35$ firmware images are from OpenWrt which have the corresponding source code. For these firmware images, we adopt source code audit to verify the vulnerabilities.
% The other $30$ firmware images are without source codes. We check the vulnerabilities in them either through static analysis or dynamic analysis.

\noindent\textbf{Vulnerabilities Verification.} \textbf{First}, since $5$ firmware images from OpenWrt have source code, we adopt source code audit to verify the vulnerabilities. 
% According to the result, \system reaches $100\%$ precision in identifying the $47$ vulnerabilities from these firmware images. 
\textbf{Second}, we combine static analysis and dynamic analysis to evaluate the remaining 30 firmware images. We manually review the disclosure reports of the $335$ vulnerabilities in them to determine the appropriate verification methods for each vulnerability. Then, we divide the $335$ vulnerabilities into two parts according to their corresponding verification method. $192$ vulnerabilities in the first part can be verified through static analysis and $143$ vulnerabilities in the second part can be validated by dynamic analysis. 

In summary, we finally confirm $368$ of $382$ vulnerabilities really exist. We fail to verify $14$ vulnerabilities in four firmware images, with one from TP-Link, and the other three from TSmart. Overall, \system achieves an average of $96.3\%$ precision in identifying the vulnerabilities caused by TPCs in firmware.

\begin{table}[]
\centering
\caption{Precision Evaluation of \system.}
\vspace{-1mm}
% \tiny
\renewcommand\arraystretch{0.9}
\resizebox{0.5\textwidth}{!}{
\label{Table: Precision Evaluation}
\begin{tabular}{ccccc}
\toprule[1.5pt]
Vendor   & Category     & Firmware                          & \# Vul. & \# Verified  Vul. \\ \midrule[1pt]
Trendnet & Router       & openwrt-ar71xx-generic-823dru.bin & 21      & 21                \\
Trendnet & Router       & tew-654tr\_a1\_fw100b19.bin       & 19      & 19                \\
Trendnet & Router       & TEW-638APBV2\_V1.1.10.BIN         & 5       & 5                 \\
Trendnet & Camera       & FW\_344345\_V1.2.2\_20170207.BIN  & 1       & 1                 \\
Trendnet & Switch       & TEG-082WS-1.00.12-ALL.HEX         & 23      & 23                \\
TP-Link  & Router       & wr810nv2\_eu\_boot(160509).bin    & 24      & 24                \\
TP-Link  & Router       & TL-R4149GV1.BIN                   & 56      & 51                \\
TP-Link  & Router       & ARCHER\_D20V1\_0.8.0.BIN          & 7       & 7                 \\
TP-Link  & Router       & TL-WR802NV4\_EU\_0.9.1\_3.16.BIN  & 5       & 5                 \\
TP-Link  & Router       & MR10UV1.BIN                       & 2       & 2                 \\
TP-Link  & Camera       & NC230\_1.2.1.BIN                  & 9       & 9                 \\
D-Link   & Router       & DIR-605L\_206B01.BIN              & 4       & 4                 \\
D-Link   & Router       & DIR100A1\_FW114CNB01.BIX          & 4       & 4                 \\
D-Link   & Camera       & DCS-932L\_REVB\_V2.12.01.BIN      & 5       & 5                 \\
D-Link   & Camera       & DCS-800L\_A1\_V1.06.10.BIN        & 16      & 16                \\
Xiongmai & Camera       & GENERAL\_HZXM\_IPC.BIN            & 2       & 2                 \\
Xiongmai & Camera       & xmjp\_liteos\_v1.01.bin           & 24      & 24                \\
Dahua    & Camera       & UPALL\_IPC.BIN                    & 2       & 2                 \\
Phicomm  & Camera       & PHICOMM\_ROUTER.BIN               & 8       & 8                 \\
OpenWrt  & Router       & OpenWrt-Router-1.bin              & 9       & 9                 \\
OpenWrt  & Router       & OpenWrt-Router-2.bin              & 11       & 11                 \\
OpenWrt  & Router       & OpenWrt-Router-3.bin              & 10       & 10                 \\
OpenWrt  & Router       & OpenWrt-Router-4.bin              & 10       & 10                 \\
OpenWrt  & Router       & OpenWrt-Router-5.bin              & 7       & 7                 \\
TSmart   & Camera       & ipcam.20190314.v02.bin            & 20      & 14                \\
TSmart   & Camera       & PPSTRONG-2.1.2.BIN                & 6       & 6                 \\
TSmart   & Camera       & ppstrong-c2-atlantic.bin          & 6       & 6                 \\
TSmart   & Camera       & PPSTRONG-C2-TSMART.bin            & 7       & 7                 \\
TSmart   & Camera       & CHMI-CAMERA.BIN                   & 2       & 2                 \\
TSmart   & Camera       & DIFF-OTA-4.0.7.2017112719.BIN     & 2       & 2                 \\
TSmart   & Gateway      & TSMART\_RTL8196E\_GW.BIN          & 5       & 5                 \\
TSmart   & Doorbell     & HANKE-DOORBELL.BIN                & 5       & 5                 \\
TSmart   & Story Bot    & XR871\_STORYBOT\_QIO.BIN          & 14      & 13                \\
TSmart   & Light        & YMSMART-2.0.2.BIN                 & 21      & 19                \\
TSmart   & Sweeper & YY5EV2TP6C\_0.BIN                 & 10      & 10 \\ \bottomrule[1.5pt]
\end{tabular}
}
% \vspace{-7em}
\end{table}

% \begin{table*}[]
% \centering
% \caption{Accuracy Evaluation of \system}
% \renewcommand\arraystretch{0.9}
% \label{Table: Accuracy Evaluation}
% \begin{tabular}{ccccccc}
% \hline \hline
% Vendor &
%   Category &
%   firmware &
% %   \begin{tabular}[c]{@{}c@{}}\#  Third-party\\ Components\end{tabular} &
%   \# Vul. &
% %   \begin{tabular}[c]{@{}c@{}}\# Verified\\ TPCs\end{tabular} &
%   \begin{tabular}[c]{@{}c@{}}\# Verified\\ Vul.\end{tabular} \\ \hline
% Trendnet & Router & openwrt-ar71xx-generic-tew-823dru-squashfs-factory.bin             & 21  & 21 \\
% Trendnet & Router & tew-654tr\_a1\_fw100b19.bin                                        & 19  & 19 \\
% TP-Link  & Router & wr810nv2\_eu\_3\_16\_9\_up\_boot(160509).bin                       & 24  & 24 \\
% TSmart     & Camera & ipcam.20190314.v02.bin                                             & 20  & 14 \\
% Xiongmai & Camera & xmjp\_liteos\_hi3518ev200\_8188eu\_v1.01.liteos.20170630\_all.bin  & 24  & 24 \\ \hline \hline
% \end{tabular}
% \end{table*}

% \vspace{-0.4cm}
\section{DATA CHARACTERIZATION}

\begin{table}[]
 \setlength{\abovecaptionskip}{0.08cm}
\caption{Analysis Results of Our Dataset. 
% \# TPC and \# Vul. represent the number of TPCs and vulnerabilities respectively. 
\# $\overline{TPC}$ and \# $\overline{Vul.}$ represent the average number of TPCs and vulnerabilities of each firmware image respectively.}
\label{Table: Vulnerability}
% \centering
\tiny
\resizebox{0.5\textwidth}{!}{
\begin{tabular}{ccccccc}
\toprule[1.5pt]
Vendor &
  Category &
    \#  Firmware &
  \# TPC &
  \# $\overline{TPC}$ &
  \# Vul. &
  \# $\overline{Vul.}$ \\ \midrule[1pt]
Xiongmai                  & Camera      & 520    & 232   & 0.45    & 313    & 0.60  \\ \hline
Tomato-shibby             & Router      & 230    & 2,088   & 9.08  & 11,948 & 51.95 \\ \hline
Phicomm                   & Router      & 107    & 405     & 3.79  & 1,818  & 16.99 \\ \hline
\multirow{2}{*}{Fastcom}  & Router      & 149    & 274     & 1.83  & 1,849  & 12.41  \\ \cline{2-7} 
                          & Unknown     & 10     & 0       & 0     & 0      & 0     \\ \hline
\multirow{4}{*}{Trendnet} & Camera      & 477    & 136     & 0.28  & 1,395   & 4.32 \\ \cline{2-7} 
                          & Router      & 336    & 1,762   & 5.24  & 7,903  & 23.52 \\ \cline{2-7} 
                          & Switch      & 162    & 366     & 2.26  & 3,157  & 19.49 \\ \cline{2-7} 
                          & Unknown     & 106    & 164     & 1.54  & 158    & 1.52 \\ \hline
Xiaomi                    & Router      & 21     & 251     & 11.95 & 2,440  & 116.19   \\ \hline
\multirow{4}{*}{TP-Link}  & Camera      & 319    & 1,981   & 6.21  & 27,001 & 84.64 \\ \cline{2-7} 
                          & Router      & 606    & 4,222   & 6.97  & 30,612 & 50.51 \\ \cline{2-7} 
                          & Switch      & 484    & 77      & 0.16  & 795    & 1.64 \\ \cline{2-7}
                          & Unknown     & 48     & 67      & 1.40  & 639    & 13.31 \\ \hline
\multirow{4}{*}{D-Link}   & Camera      & 360    & 113     & 0.31  & 737    & 2.04 \\ \cline{2-7} 
                          & Router      & 552    & 2,823   & 5.11  & 14,495 & 26.26 \\ \cline{2-7} 
                          & Switch      & 545    & 80      & 0.15  & 1062   & 1.95     \\ \cline{2-7}
                          & Unknown     & 91     & 30      & 0.33  & 266     & 2.92 \\ \hline
Hikvision                 & Camera      & 139    & 8       & 0.06  & 127    & 0.91  \\ \hline
Foscam                    & Camera      & 113    & 0       & 0     & 0      & 0     \\ \hline
Dahua                     & Camera      & 419    & 43      & 0.10  & 430    & 1.03 \\ \hline
TSmart                    & Smart Homes & 23,050 & 856     & 0.04  & 4,353  & 0.19 \\ \hline
OpenWrt                   & Router     & 5,292   & 300,020 & 56.69 & 13,486 & 2.55  \\ \bottomrule[1.5pt]
\end{tabular}}
\vspace{-5mm}
\end{table}

\begin{figure*}[] 
 \setlength{\abovecaptionskip}{0.05cm}
\centering
\includegraphics[width=\textwidth]{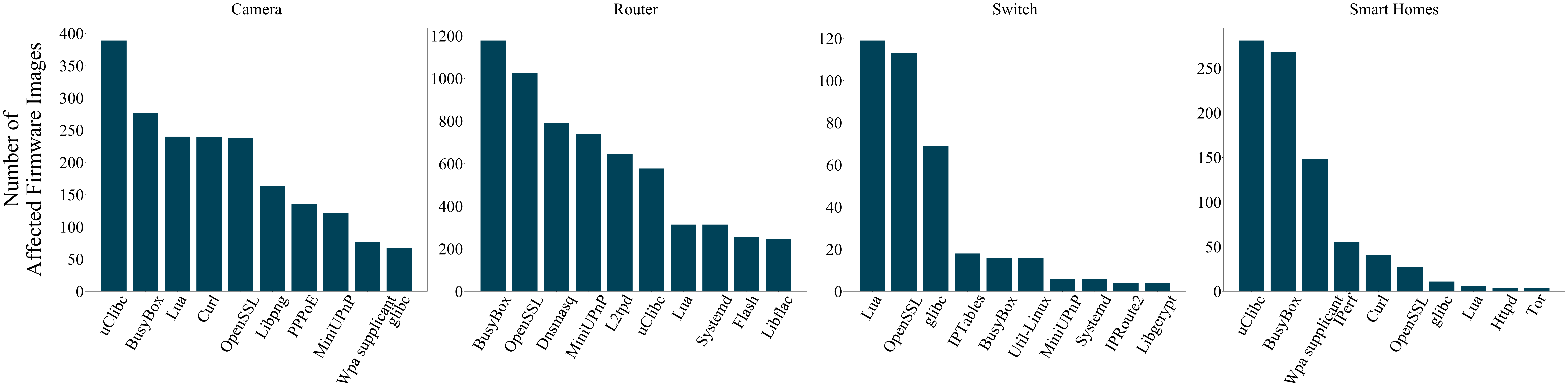}
\vspace{-2em}
\caption{Top 10 TPCs Used in Each Kind of Firmware.} 
\label{Figure: Top 10 TPCs Used in each kind of Firmware} 
\vspace{-1em}
\end{figure*}

% \vspace{-3mm}
\subsection{TPC Usage}
\label{section: TPCs Identification}
In this subsection, we deploy \system on the dataset to first identify the TPCs used in firmware. 
% \textbf{First},  
\system successfully unpacks and disassembles $96\%$ firmware images in the whole dataset. We summarize the reasons why the remaining $4\%$ firmware images cannot be analyzed: 1) $123$ firmware images are encrypted; 2) $972$ firmware images from TSmart have unknown processors; thus, we fail to implement plugins for \textit{IDA} to process them; and 3) We do not support the unknown filesystems used in $269$ Linux-based firmware images. 
We have excluded 
the remaining $4\%$ firmware images in further analysis  in Section~\ref{Section: Analysis Results}. 

As shown in Table~\ref{Table: Vulnerability}, \system identifies $584$ different TPCs used in $34,136$ firmware images. Since there are lots of  identified TPCs, we decide to present the results of top 10 TPCs used in each kind of firmware, as shown in Figure~\ref{Figure: Top 10 TPCs Used in each kind of Firmware}. Based on the results, we have the following findings. (1) Routers from OpenWrt contain the most TPCs, reaching an average of $56.69$ per firmware image. It is reasonable that OpenWrt utilizes many TPCs since OpenWrt is an open-source project for embedded systems. (2) Smart homes from TSmart contain fewer TPCs which have $0.04$ TPCs per firmware image. Most of the smart homes leverage the monolithic firmware. According to our further analysis, we find the monolithic firmware used in smart homes usually adopts their own implementations to replace the TPCs. Besides, some monolithic firmware is composed of a piece of code and data that achieves simple logistic functions. 
(3) 10 unknown firmware images from Fastcom and 113 cameras from Foscam are encrypted. For these encrypted firmware images, \system cannot obtain useful information to identify the TPCs used in them. Currently, there is no effective method that can analyze the encrypted firmware automatically.  (4) The same kind of firmware from different vendors adopts similar TPCs. For instance, all routers adopt  \textit{BusyBox}, \textit{OpenSSL}, \textit{Dnsmasq}, \textit{MiniUPnP}, \textit{uClibc}, \textit{L2tpd} and \textit{Util-linux} in their firmware.
(5) Different kinds of firmware have commonalities in adopting TPCs. For instance,  \textit{BusyBox}, \textit{Lua}, \textit{OpenSSL} are all in the top 10 TPCs usage list of each kind of firmware.
\bin{\textbf{In addition}, we have a counterintuitive finding. We find that only \textless 1\% firmware images are using \textit{MbedTLS}, which is a popular lightweight TPC designed to replace \textit{OpenSSL} in embedded systems.
% the top 10 TPCs used in each kind of firmware do not include \textit{MbedTLS}, 
We propose two possible reasons why more firmware images are using \textit{OpenSSL} rather than \textit{MbedTLS}. (1) \textit{OpenSSL} has more features compared to \textit{MbedTLS}. When computing power is allowed, there are more reasons for developers to utilize \textit{OpenSSL}. (2) Though \textit{MbedTLS} is used in many popular IoT frameworks, e.g., FreeRTOS, many vendors have their own frameworks to develop the firmware which do not use \textit{MbedTLS}.}

\begin{figure*}
 \setlength{\abovecaptionskip}{0.05cm}
 \setlength{\belowcaptionskip}{-0.3cm}
  \centering
 
  \subfigure[Number of Affected Firmware Images by Top 10 TPCs.]
  {
  \begin{minipage}[b]{0.28\textwidth}
   \setlength{\abovecaptionskip}{0.02cm} 
    \label{Subfig: Number of Affected firmwares by Top 10 TPCs} %% label for first subfigure
    \includegraphics[width=\textwidth]{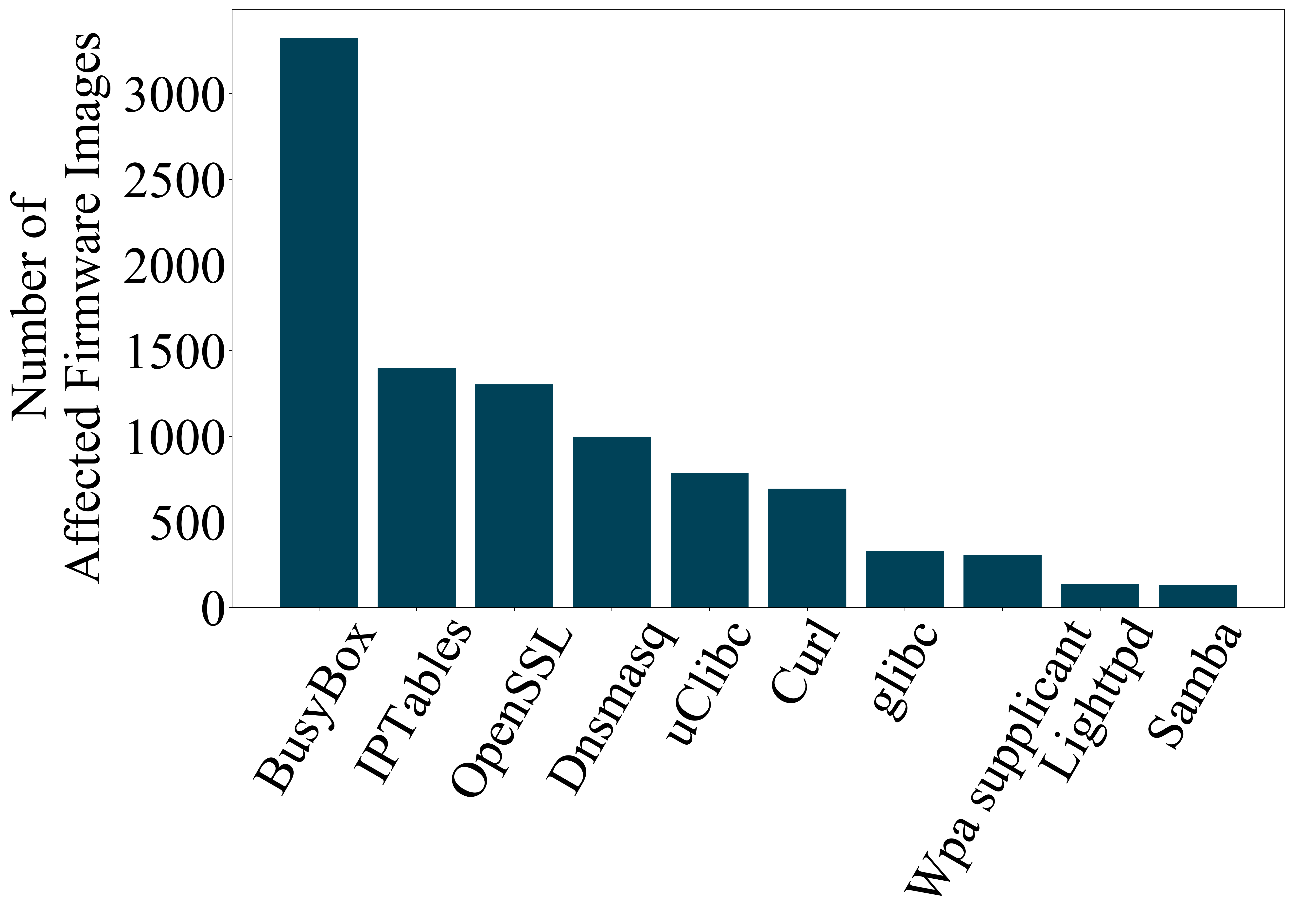}
    \end{minipage}
  }
  \hspace{3mm}
  \subfigure[Number of CVEs Caused by Top 10 TPCs.]{
  \begin{minipage}[b]{0.28\textwidth}
    \label{Subfig: Number of CVEs Caused by Top 10 TPCs} %% label for second subfigure
    \includegraphics[width=\textwidth]{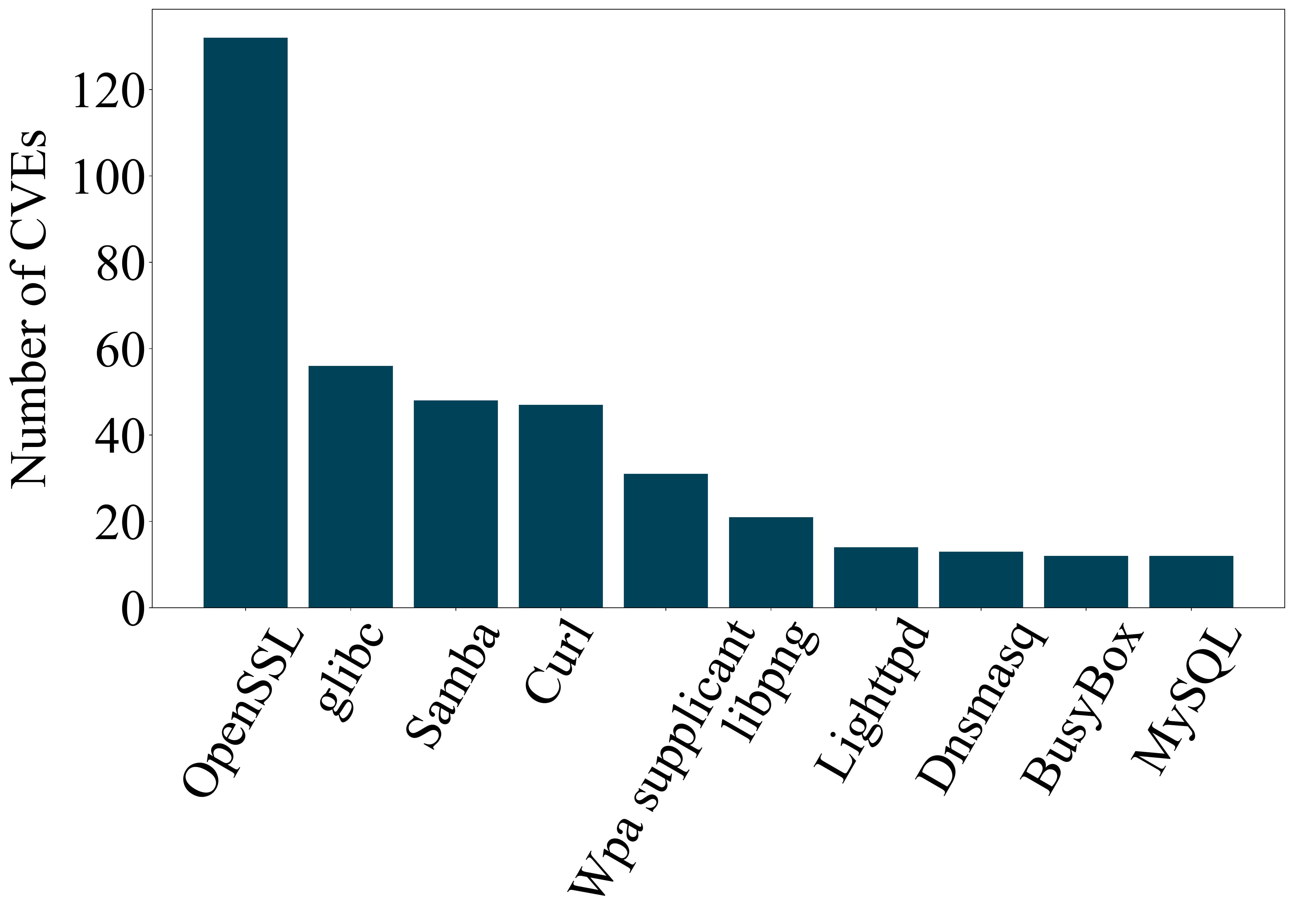}
    \end{minipage}
    }
  \hspace{3mm}
  \subfigure[Number of Vulnerabilities Caused by Top 10 TPCs.]{
  \begin{minipage}[b]{0.28\textwidth}
    \label{Subfig: Number of Vulnerabilities Caused by Top 10 TPCs} %% label for second subfigure
    \includegraphics[width=\textwidth]{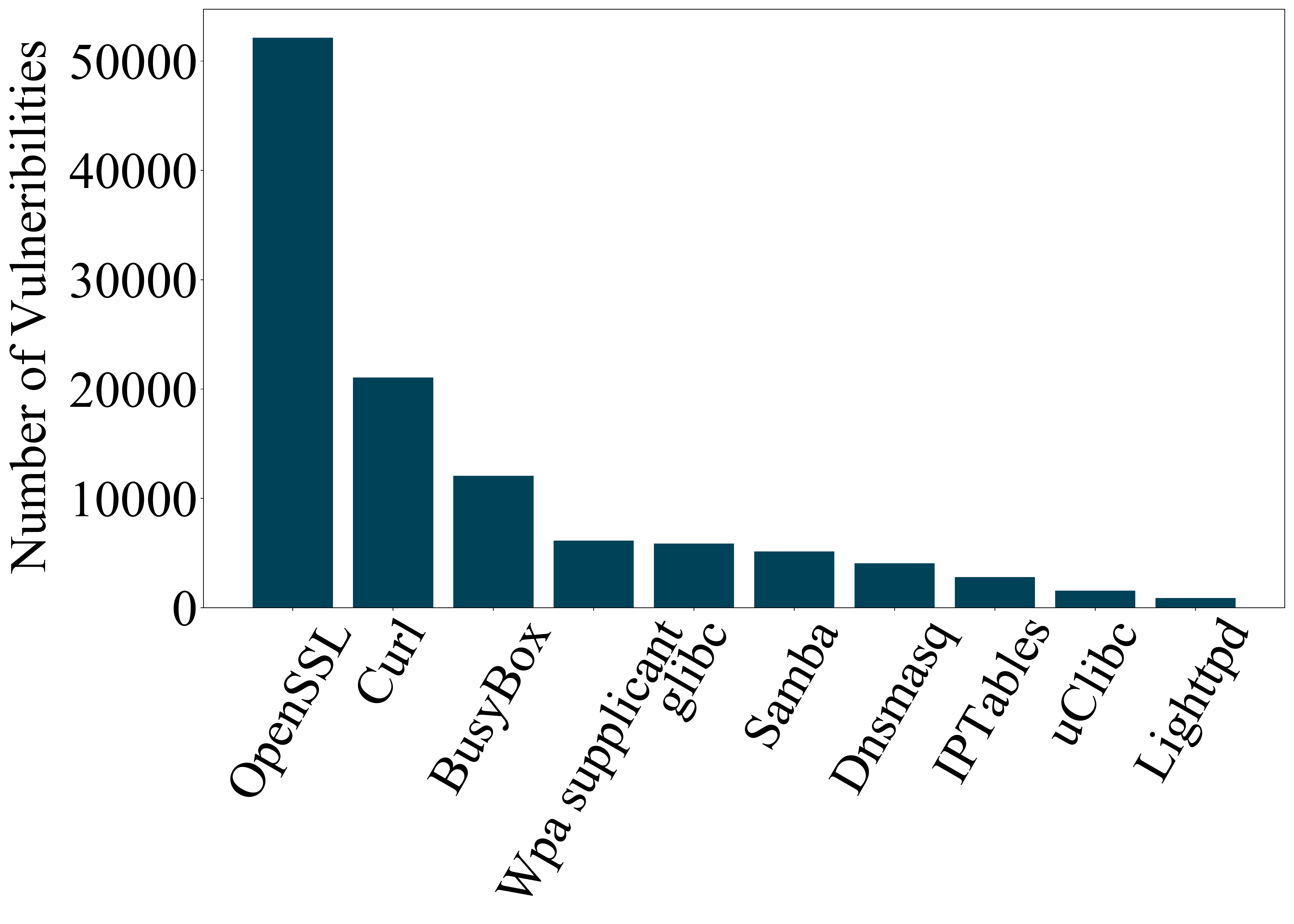}
    \end{minipage}
    }
    \vspace{-2mm}
  \caption{Number of Affected Firmware Images, CVEs and Vulnerabilities Caused by the Top 10 TPCs.}
  \label{Figure: Number of Affected firmwares, CVEs and Vulnerabilities Caused By Top 10 TPCs} %% label for entire figure
%   \vspace{-2em}
\end{figure*}
%%End: Figure: Number of Affected firmwares, CVEs and Vulnerabilities Caused By Top 10 TPCs

\begin{table}[]
 \setlength{\abovecaptionskip}{0.08cm}
% \tiny
\footnotesize
\caption{Top 10 CWE Software Weaknesses.}
% \vspace{-3mm}
\renewcommand\arraystretch{0.9}
\setlength\tabcolsep{10pt}
\label{Table: Top 10 CWE Software Weaknesses}
\begin{tabular}{cccc}
\toprule[1.5pt]
    & CWE ID & Weakness          & \# CVEs \\ \midrule[1pt]
1.  & 399    & Resource Management Error & 87      \\
2.  & 119    & Buffer Overflow           & 84      \\
3.  & 310    & Cryptographic Issues      & 47      \\
4.  & 20     & Improper Input Validation & 39      \\
5.  & 264    & Access Control Error      & 36      \\
6.  & 200    & Information Disclosure    & 31      \\
7.  & 189    & Numeric Errors            & 20      \\
8.  & -      & Insufficient Information  & 18      \\
9.  & 94     & Code Injection            & 8       \\
10. & 362    & Race Condition            & 7       \\ \bottomrule[1.5pt]
\end{tabular}
\vspace{-2em}
\end{table}
% \vspace{-3mm}

\vspace{-2mm}
\subsection{Introduced Vulnerabilities Overview}
% \textbf{Vulnerabilities Detection.} \ \ \ 
After we obtain the TPCs used in firmware, we then search the corresponding vulnerabilities in our vulnerability database. As shown in Table~\ref{Table: Vulnerability}, we detect a total of $128,757$ potential vulnerabilities, which involve $429$ CVEs, in $34,136$ firmware images. In this paper, we count each CVE found in a firmware image as a separate vulnerability.  
Table~\ref{Table: Top 10 CWE Software Weaknesses} lists the top 10 CWE software weaknesses by the number of CVEs, accounting for $88\%$ of all the CVEs we detected. The results indicate that the vulnerabilities in firmware
involve a wide variety of security issues. Based on the results, we find IoT devices are suffering from substantial security risks since most of them contain a significant number of vulnerabilities. 
The routers from Xiaomi are in a very critical situation with $116.19$ mean vulnerabilities, the most of all vendors.
Moreover, routers produced by other vendors also have a great number of vulnerabilities.
TP-Link, D-Link and Trendnet are well-known IoT vendors and all have multiple kinds of IoT devices. However, their products all have lots of vulnerabilities.
The cameras from Xiongmai, Hikvision and Dahua all have nearly one vulnerability per firmware image.
What's more, though OpenWrt contains the highest average number of TPCs per firmware image, we find comparatively few vulnerabilities in it, which has an average of $2.55$ vulnerabilities for each firmware image. 
Next, \system detects a comparative few vulnerabilities from TSmart's firmware in consideration of its largest scale. TSmart has the best performance among the involved vendors, which only contains $0.19$ vulnerabilities per firmware image.
%\yuan{\color{red}Do we know which category of devices are more vulnerable? Also, can we also report the list of most vulnerable vendors (by average number of vulnerabilities per image)? These would be interesting results.}

Besides, though we detect a large number of TPCs, most of the vulnerabilities are concentrated on a few TPCs. As shown in Figure~\ref{Figure: Number of Affected firmwares, CVEs and Vulnerabilities Caused By Top 10 TPCs}, we list the top 10 TPCs from three aspects: the number of affected firmware images, the number of caused CVEs and the number of caused vulnerabilities.
\textit{OpenSSL} contains the most CVEs, totaling $132$, that affects $1,304$ firmware images and causes  $52,135$ vulnerabilities totally.
\textit{Busybox} is the most widely used TPC that has been identified in $3,326$ firmware images. We detect  $12$ CVEs of \textit{Busybox} that cause $12,072$ vulnerabilities, which reaches an average of $1,006$ vulnerabilities per CVE.
Though \textit{IPTables} has been identified in a great number of firmware images, it just causes a few vulnerabilities.
Moreover, as shown in Figure~\ref{Subfig: Number of CVEs Caused by Top 10 TPCs}, we have identified a total of $386$ CVEs from these $10$ TPCs, accounting for $90\%$ of all the CVEs we detected.

% In addition, Busybox is the most widely used TPCs that has been identified in $3,326$ firmware images according to Figure~\ref{Figure: the Number of Affected firmware images}. 
% We detect a total of $12$ corresponding CVEs of Busybox that cause $12,072$ vulnerabilities, which reaches an average of $1,006$ vulnerabilities  per CVE.

% \vspace{-1mm}
\section{Analysis Results}
\label{Section: Analysis Results}
In this section, we aim to answer the following research questions.
% we analyze the data from four different perspectives. 

% \noindent$\bullet$ \textbf{RQ1:} What are the risk rankings of different firmware and vendors?

\noindent$\bullet$ \textbf{RQ1:} How vulnerable are firmware images of different kinds and from different vendors?

% \noindent$\bullet$ \textbf{RQ1:} What are the differences in the security of different kinds of firmware?

% \noindent$\bullet$ \textbf{RQ2:} What are the differences in the security of firmware from different vendors?

\noindent$\bullet$ \textbf{RQ2:}  What is the trend of using TPCs in firmware from different vendors over time?

\noindent$\bullet$ \textbf{RQ3:} What is the geographical distribution of the devices using vulnerable firmware?

\noindent$\bullet$ \textbf{RQ4:}  Does the firmware adopt the latest TPCs at the time when it was released?

\noindent$\bullet$ \textbf{RQ5:} What happens to TPCs when the firmware is updated?

\noindent$\bullet$ \textbf{RQ6:} Are there any TPC license violations?

% \begin{table}[]
% \caption{Vulnerability of Different Kinds of Firmware. $\overline{Critical}$, $\overline{High}$, $\overline{Medium}$ and $\overline{Low}$ represent the average number of critical, high, medium and low vulnerabilities.}
% % \vspace{-1em}
% \centering
% \footnotesize
% \label{Table: risk firmware}
% \renewcommand\arraystretch{0.85}
% \begin{tabular}{cccccc}
% \hline \hline
% Category    & $\overline{Vul.}$ & $\overline{Critical}$ & $\overline{High}$ & $\overline{Medium}$ & $\overline{Low}$ \\ \hline
% Router& 22.92& 1.48& 2.73 & 17.59 & 1.12 \\
% Camera&13.68& 0.41& 2.18& 10.64 & 0.45\\
% Switch& 4.34& 0.38& 0.60& 2.92& 0.44\\
% Smart Homes & 0.22 & 0.03& 0.06 & 0.11 & 0.02 \\ 
% \hline \hline
% \vspace{-2em}
% \end{tabular}

\begin{table}[]
 \setlength{\abovecaptionskip}{0.06cm}
\caption{Vulnerability of Different Kinds of Firmware. $\overline{Critical}$, $\overline{High}$, $\overline{Medium}$ and $\overline{Low}$ represent the average number of critical, high, medium and low vulnerabilities.}
% \vspace{-1em}
\centering
\footnotesize
\label{Table: risk firmware}
\renewcommand\arraystretch{0.85}
\begin{tabular}{cccccc}
\toprule[1.5pt]
Category    & $\overline{Vul.}$ & $\overline{Critical}$ & $\overline{High}$ & $\overline{Medium}$ & $\overline{Low}$ \\ \midrule[1pt]
Router& 22.92& 1.48& 2.73 & 17.59 & 1.12 \\
% Camera&13.68& 0.41& 2.18& 10.64 & 0.45\\
% Switch& 4.34& 0.38& 0.60& 2.92& 0.44\\
% Smart Homes & 0.22 & 0.03& 0.06 & 0.11 & 0.02 \\ 
Camera&9.81& 0.32& 1.92& 7.20 & 0.37\\
Switch& 5.29& 0.22& 0.62& 3.98& 0.47\\
Smart Homes & 0.19 & 0.01& 0.05 & 0.11 & 0.02 \\ 
\bottomrule[1.5pt]
% \vspace{-3em}
\end{tabular}
% 
%   \footnotesize{$^*$ $\overline{Critical}$, $\overline{High}$, $\overline{Medium}$ and $\overline{Low}$ represent the average number of critical, high, medium and low vulnerabilities respectively.}
\end{table}

\begin{table}[]
 \setlength{\abovecaptionskip}{0.05cm}
\caption{Vulnerability of Firmware From Different Vendors.}
% \vspace{-1em}
\centering
\footnotesize
\renewcommand\arraystretch{0.9}
\label{Table: risk vendors}
\begin{tabular}{cccccc}
\toprule[1.5pt]
Vendor & $\overline{Vul.}$ & $\overline{Critical}$ & $\overline{High}$ & $\overline{Medium}$ & $\overline{Low}$ \\ \midrule[1pt]
    %   Xiaomi&116.48&2.94&18.53&78.63&16.38\\
       
    %   Tomato-shibby&51.95&2.77&8.46&35.49&5.27\\
       
    %   TP-Link&42.06&1.45&7.26&30.34&3.01\\
       
    %   Phicomm&16.99&0.41&3.88&12.16&0.54\\
       
    %   D-link&11.64&0.52&1.85&8.01&1.26\\
       
    %   Trendnet&12.09&0.31&1.92&8.90&0.96\\
       
    %   Fastcom&12.41&0.44&1.75&8.72&1.50\\
       
    %   OpenWrt&2.82&0.00&0.47&2.35&0.00\\
       
    %   Dahua&1.03&0.03&0.14&0.66&0.20\\
       
    %   Hikvision&0.91&0.06&0.18&0.62&0.05\\
       
    %   Xiongmai&0.63&0.00&0.22&0.34&0.07\\
       
    %   TSmart&0.22&0.03&0.06&0.11&0.02\\
    %   Foscam&0.00&0.00&0.00&0.00&0.00\\
           Xiaomi&116.19&2.86&18.52&78.43&10.67\\
       Tomato-shibby&51.95&2.77&8.46&35.49&1.84\\
       
       TP-Link&39.20&1.37&6.97&28.59&2.26\\
       
       Phicomm&16.99&0.41&3.88&11.28&0.54\\
       
       D-link&11.87&0.55&1.95&8.03&1.34\\
       
       Trendnet&11.02&0.29&1.85&7.98&0.90\\
       
       Fastcom&9.13&0.44&1.35&6.81&0.53\\
       
       OpenWrt&2.55&0.00&0.46&1.58&0.00\\
       Dahua&1.03&0.03&0.14&0.66&0.16\\
       Hikvision&0.91&0.05&0.17&0.60&0.03\\
       Xiongmai&0.60&0.00&0.21&0.32&0.07\\
       TSmart&0.19&0.01&0.05&0.11&0.02\\
       \bottomrule[1.5pt]
\end{tabular}
\vspace{-1em}
\end{table}

\begin{table}[]
 \setlength{\abovecaptionskip}{0.08cm}
\centering
\caption{The Firmware Affected by Two Vulnerabilities.}
\label{Table: Firmware images affected by two vulnerabilities}
\setlength\tabcolsep{12pt}
\footnotesize
\begin{tabular}{ccc}
\toprule[1.5pt]
Vendors & \begin{tabular}[c]{@{}c@{}}Heartbleed\\ (CVE-2014-0160)\end{tabular} & \begin{tabular}[c]{@{}c@{}}GHOST\\ (CVE-2015-0235)\end{tabular} \\ \midrule[1pt]
Fastcom       & 2   & 1  \\
Trendnet      & 36  & 87 \\
Tomato-shibby & 24  & -  \\
TP-Link       & 301 & 91 \\
D-Link        & 3   & 45 \\
Hikvision     & 1   & -  \\
Dahua         & 5   & -  \\
TSmart        & 8   & -  \\ \bottomrule[1.5pt]
\end{tabular}
\vspace{-5mm}
\end{table}

% \subsection{\textcolor{red}{Firmware Vulnerability}}
% \vspace{-0.2cm}
\subsection{Firmware Vulnerability}
This subsection answers \textbf{RQ1}. 
% We study these two questions based on two metrics: the vulnerabilities of different and the delay time of the TPCs used in firmware.
Though we have identified many vulnerabilities in firmware, we still lack an understanding of how vulnerable are firmware images of different kinds and from different vendors. To answer this question, we evaluate the security of firmware based on the average number of vulnerabilities of different severity in firmware. We also discover some critical vulnerabilities still threaten the security of firmware.
% the firmware is, and the impact of these vulnerabilities on firmware. We first evaluate the vulnerability of the firmware based on the average number of vulnerabilities of different severity in firmware. Next, we study the impact of representative critical vulnerabilities on firmware.
% the impact of these vulnerabilities on firmware and how vulnerable the firmware is.
% In this subsection, we first present the average number of vulnerabilities of different severity in different kinds of firmware  
% Next, we present the impact of two well-known vulnerabilities: \textit{OpenSSL} Heartbleed vulnerability  (CVE-2014-0160) and \textit{glibc}  GHOST  vulnerability  (CVE-2015-0235) on our dataset.
% introduce the risk score for measuring the risk of firmware, which is based on the severity of the firmware vulnerabilities. The risk score is calculated as follows.
% Through the vulnerabilities we identified in the firmware images, we can figure out which kinds of firmware images and which vendors are most vulnerable. 
% In this subsection, we calculate the risk score for each kind of firmware and each vendor. 

\textbf{First}, we explore the vulnerability of different kinds of firmware. As shown in Table~\ref{Table: risk firmware}, we list the average number of different severity vulnerabilities in each kind of firmware. Since the number of routers from OpenWrt is much higher than other vendors, which may bring bias to our results, we finally randomly select $500$ OpenWrt routers for analysis. 
% We notice that router and camera have a similar number of vulnerabilities per firmware image, which is significantly . Router is the only category that has more than one critical vulnerability averagely per firmware image. Smart homes has very few vulnerabilities of different severity since it rarely uses. 
We notice that 
the router is more vulnerable to attacks than the other three kinds of firmware since it has the most vulnerabilities of different severity. The router is also the only category that has more than one critical vulnerability averagely per firmware image. The camera and switch have a similar number of critical vulnerabilities and low vulnerabilities per firmware image, while the camera has more high vulnerabilities and medium vulnerabilities than the switch. Smart home is the least vulnerable category since it has few vulnerabilities of different severity. \textbf{Second}, we study the vulnerability of firmware from different vendors. As shown in Table~\ref{Table: risk vendors}, Xiaomi has the most vulnerabilities of different severity, which causes Xiaomi more vulnerable to attacks. Both Xiaomi and Tomato-shibby have nearly three critical vulnerabilities averagely per firmware image, which is significantly higher than the other vendors. The vulnerabilities detected in TP-Link, Phicomm, D-Link, Trendnet, and Fastcom are mainly at the medium severity level. 
OpenWrt, Dahua, Hikvision, Xiongmai, and TSmart have very few vulnerabilities per firmware image. Besides, both OpenWrt and Xiongmai have no critical vulnerability in each firmware image.

% During the above analysis, we surprisingly find that some critical vulnerabilities are still having severe impact on the firmware in our dataset. We take two representative critical vulnerabilities: \textit{OpenSSL} Heartbleed vulnerability  (CVE-2014-0160) and \textit{glibc}  GHOST  vulnerability  (CVE-2015-0235) for instance. We find these two  vulnerabilities have affected 604 firmware images which account for $1.8\%$ of the dataset.
% % , as shown in Table~\ref{Table: Firmware images affected by two vulnerabilities}.  
% More specifically, 380 firmware images
% % from 8 vendors 
% are vulnerable to the Heartbleed vulnerability and 224 firmware images 
% % from 4 vendors 
% contain the GHOST vulnerability. 
% % Fastcom, Trendnet, TP-Link, and D-Link are all vulnerable to both of these two vulnerabilities. Moreover, TP-Link has the most vulnerable firmware images that affected by these two vulnerabilities. 

During the above analysis, we surprisingly find that some critical vulnerabilities are still having severe impact on the firmware in our dataset. We take two representative critical vulnerabilities: \textit{OpenSSL} Heartbleed   and \textit{glibc}  GHOST  for instance. We find these two vulnerabilities have affected 604 firmware images which account for $1.8\%$ of the dataset, as shown in Table~\ref{Table: Firmware images affected by two vulnerabilities}.  More specifically, 380 firmware images from 8 vendors are vulnerable to the Heartbleed, and 224 firmware images from 4 vendors contain the GHOST.
\begin{figure}
  \centering
  \subfigure[Changes in the Number of TPCs Over Time.]
  {
  \begin{minipage}[b]{0.42\textwidth}
    \label{Subfig: Changes in the Number of TPCs Over Time} %% label for first subfigure
    \includegraphics[width=\textwidth, height=150pt]{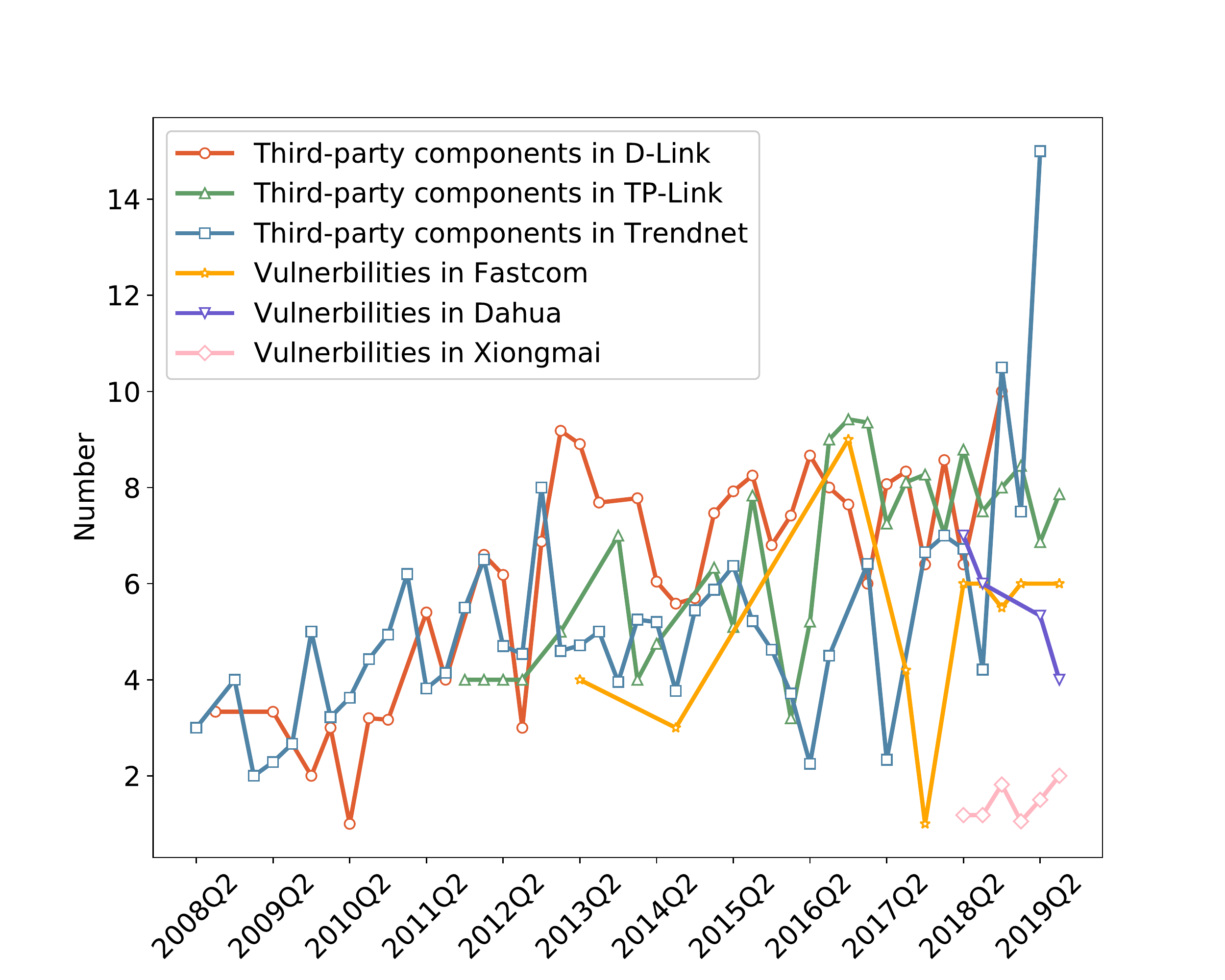}
    \end{minipage}
  } \vspace{-3mm}

  \subfigure[Changes in the Number of Vulnerabilities Over Time.]{
  \begin{minipage}[b]{0.42\textwidth}
    \label{Subfig: Changes in the Number of Vulnerabilities Over Time} %% label for second subfigure
    \includegraphics[width=\textwidth, height=150pt]{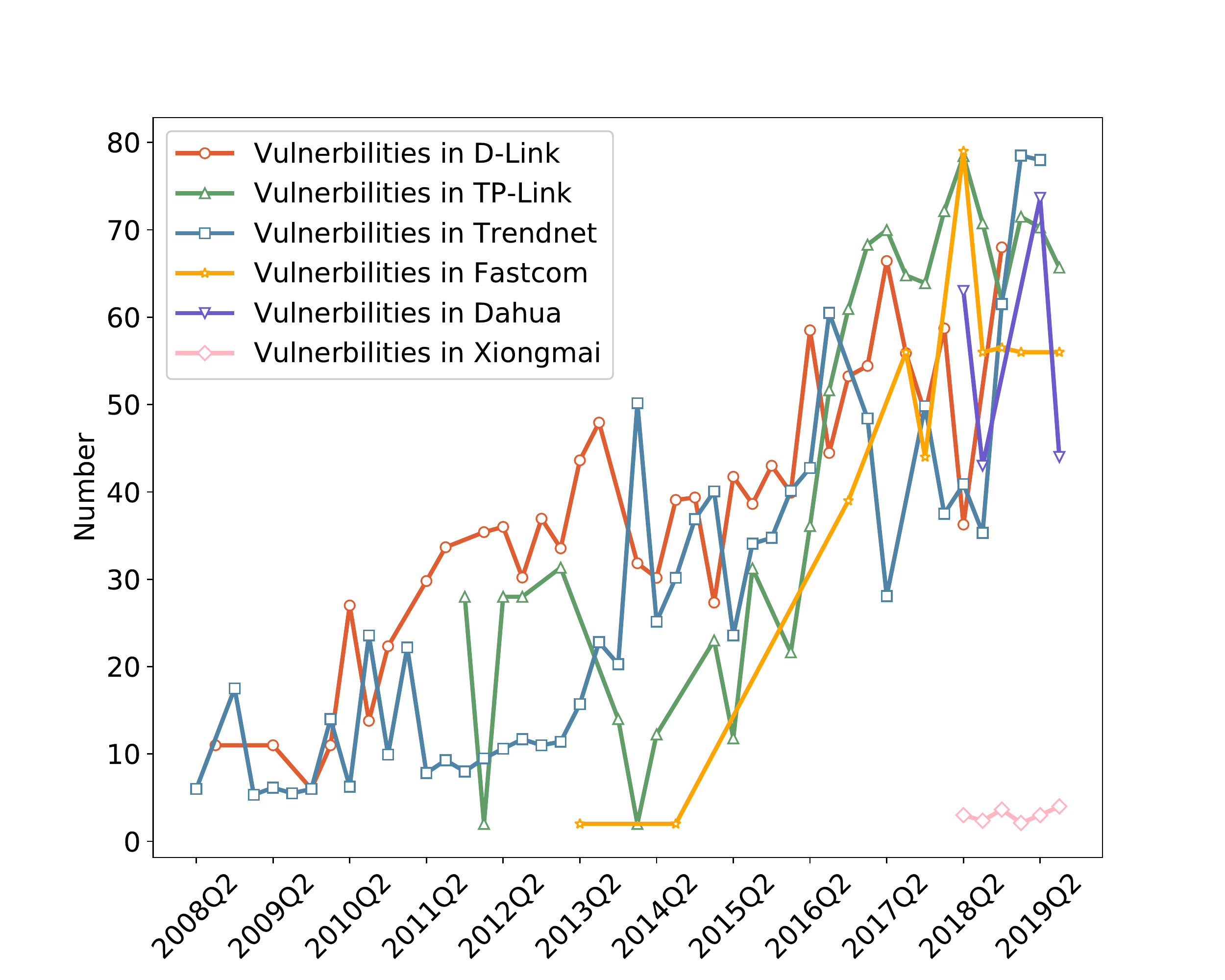}
    \end{minipage}
    }
%  \vspace{-1em}
  \caption{Changes in the Usage of TPCs and Their Corresponding Vulnerabilities Over Time.}
  \label{Figure: Changes in the Usage of TPCs and Their Corresponding Vulnerabilities Over Time} %% label for entire figure
  \vspace{-2em}
\end{figure}

%Begin: Figure: Changes in the Vulnerability Composition of TPCs Over Time
% \vspace{-2cm}
\begin{figure*}
    % \centering
    % \setlength{\abovecaptionskip}{-5.2cm}
    
    \begin{center} 
    \includegraphics[width=\textwidth]{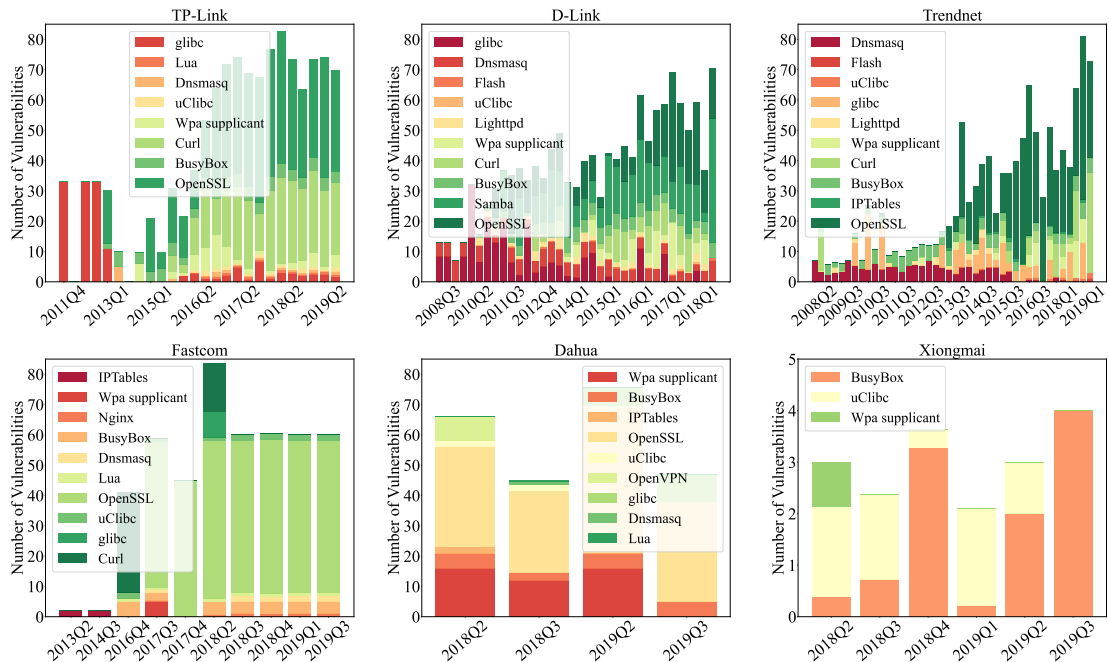}
    \vspace{-1.2cm}
    \caption{Changes in the Vulnerability Composition of TPCs Over Time.}
    \label{Figure: Changes in the Vulnerability Composition of TPCs Over Time}
    \setlength{\belowcaptionskip}{-3pt}
    \end{center} 
    \vspace{-2em}
\end{figure*}

% \vspace{-0.4cm}
\subsection{Trend in the Usage of TPCs Over Time}
\label{section: TPCs and vulnerabilities change over time}
This subsection answers \textbf{RQ2}. With the booming of the IoT ecosystem, IoT devices may adopt new TPCs which will cause different vulnerabilities over time.
In this subsection, we explore the changes in the usage of TPCs and the corresponding vulnerabilities of IoT devices over time.

We first classify firmware images from each vendor by their release time. We find the firmware images from five vendors: Xiaomi, TSmart, Tomato-shibby, Phicomm and OpenWrt are all released in the same period, respectively. For example, all the firmware images from TSmart are released in the first quarter of 2020. Therefore, we cannot obtain the  usage changes of TPCs and corresponding vulnerabilities on the firmware from these vendors.
% Thus, we cannot conduct our analysis on the firmware images from these five vendors. 
Our experiment is then conducted on the firmware from the other six vendors: D-Link, TP-Link, Trendnet, Fastcom, Xiongmai, and Dahua.
% Before we conduct the experiments, we need to classify the firmware images by their release time at first. Nevertheless, we find only a few vendors have sufficient firmware images released at different times which can be used for analysis. We then conduct our experiment on the firmware images from three vendors: TP-Link, Trendnet and D-Link. 
% \yuan{\color{red}Are these three vendors the only ones that have released firmware images at different times. Showing the trend this way seems to only cover a small range of the dataset. Can we compute the average amount of the vulnerabilities of each firmware as well?}
%For each selected vendor, we choose multiple firmware images that released at different times. 
We calculate the average number of the TPCs and corresponding vulnerabilities for the firmware released in the same period. We also list the exact number of vulnerabilities caused by each TPC quarterly.
% Currently, 
% As show in Figure~\ref{Figure: D-Link}, Figure~\ref{Figure: D-Link}, Figure~\ref{Figure: Trendnet}, Figure~\ref{Figure: Trendnet},Figure~\ref{Figure: Fastcom},Figure~\ref{Figure: Xiongmai}

Figure~\ref{Figure: Changes in the Usage of TPCs and Their Corresponding Vulnerabilities Over Time} shows the firmware from TP-Link, D-Link, Trendnet, and Fastcom has similar trends both in the usage of TPCs and their corresponding vulnerabilities. For each vendor, the number of the adopted TPCs is slightly increased over time. Also, the number of corresponding vulnerabilities has increased a lot with time. For Dahua and Xiongmai, since we only collected the firmware images from the recent two years, their changes are relatively small.

% First of all,  
% the number of TPCs used in D-Link has increased by 6 from 2008 to 2019. Nevertheless, the number of corresponding vulnerabilities has increased by nearly 60 per firmware. Second, 
% Finally, we find th

The firmware of TP-Link contains less than $10$ TPCs from the fourth quarter of 2011 to the third quarter of 2019. However, the number of corresponding vulnerabilities has increased dramatically since the second quarter of 2015 and reaches a peak in the second quarter of 2018. Next, our records indicate that the oldest firmware images of D-Link in our dataset were released in 2008 and the newest 
ones were released in 2019. The number of TPCs has not changed a lot which remains less than 10 for 11 years. The number of vulnerabilities of D-Link maintains an upward trend which reaches a peak in the first quarter of 2019. Third, we find the firmware of Trendnet utilizes less than 10 TPCs for most quarters. For the recent quarters, Trendnet begins to adopt more TPCs and the number of corresponding vulnerabilities also reaches a peak at the same time. Then, for Fastcom, the number of TPCs is also slightly changed but the number of corresponding vulnerabilities is rapidly growing from 2014 to 2018 which reaches the peak at the same time with TP-Link. For Dahua, the number of TPCs is decreased in the recent two years and the number of corresponding vulnerabilities fluctuates between 40 to 75. Finally, for  Xiongmai, the number of TPCs and corresponding vulnerabilities maintain a very low level.

We then conduct a further study to explore the specific number of vulnerabilities caused by various TPCs in different periods. Figure~\ref{Figure: Changes in the Vulnerability Composition of TPCs Over Time}  presents the changes in the vulnerability composition of TPCs of the six vendors over time. \textbf{First}, for TP-Link, before the first quarter of 2013, all of the vulnerabilities are caused by \textit{glibc}. Then, \textit{OpenSSL} and \textit{curl} are introduced to TP-Link which bring the most vulnerabilities in recent years. 
\textbf{Second}, for D-Link, we notice that the newly introduced TPCs, are the main reasons for the increase in the number of vulnerabilities. \textbf{Third}, the reason for the rapid increase in the number of vulnerabilities in Trendnet is that they started to use \textit{OpenSSL}. \textbf{Moreover}, we notice that \textit{OpenSSL} also contributes the most vulnerabilities in Fastcom's firmware. \textbf{Furthermore}, \textit{OpenSSL} and \textit{Wpa supplicant} are the main sources of vulnerabilities in Dahua in recent years.
\textbf{Finally}, the vulnerabilities in Xiongmai are mainly caused by two TPCs: \textit{BusyBox} and \textit{uClibc}.
These vendors have overlaps in using the TPCs, e.g., \textit{BusyBox} is adopted in each vendor's firmware.
Though they adopt several same TPCs, they do not use the same version. The \textit{curl} utilized by TP-Link has more vulnerabilities than that of D-Link and Trendnet. The \textit{OpenSSL} used by TP-Link and Trendnet has a similar number of vulnerabilities, while that used in D-Link contains fewer vulnerabilities.

% The routers of D-Link and Trendnet have a similar tendency with TP-Link, which both have a steady trend on the increment of TPCs and have a dramatic increment of corresponding vulnerabilities as shown in Figure~\ref{Subfig:D-Linka} and Figure~\ref{Subfig:Trendneta}. The number of vulnerabilities of D-Link reaches a peak in the first quarter of 2019 while Trendnet reaches a peak in the third quarter of 2019. The D-Link and Trendnet utilize the similar TPCs in their firmware images. As shown in Figure~\ref{Subfig:D-Linkb}, D-Link begins to utilize OpenSSL and Samba since the second quarter of 2013 which also bring most of the vulnerabilities in the later. The number of vulnerabilities of Trendnet begins to increase since it adopts the OpenSSL in the third quarter of 2013 as shown in Figure~\ref{Subfig:Trendnetb}. 

% In addition, the routers from camera also adopt less than $10$ TPCs since the second quarter of 2013. The number of corresponding vulnerabilities reaches nearly $80$ at the second quarter of 2018 and decreases to $52$ at the third quarter of 2018 which hardly changed in the next quarters. Similar to other vendors, most vulnerabilities are also caused by OpenSSL as shown in Figure~\ref{Figure: Fastcom}.

Overall, for the firmware from these vendors except for Xiongmai, we all find that the number of adopted TPCs is slightly increased while the number of corresponding vulnerabilities increases a lot. The newly introduced TPCs, e.g.,  \textit{OpenSSL}, bring more vulnerabilities. In addition, though different vendors may use the same TPCs, they may adopt different versions which will bring different vulnerabilities.

\vspace{-3mm}
\subsection{Geographical Distribution}
\begin{figure}
    \centering
    \includegraphics[scale=0.17]{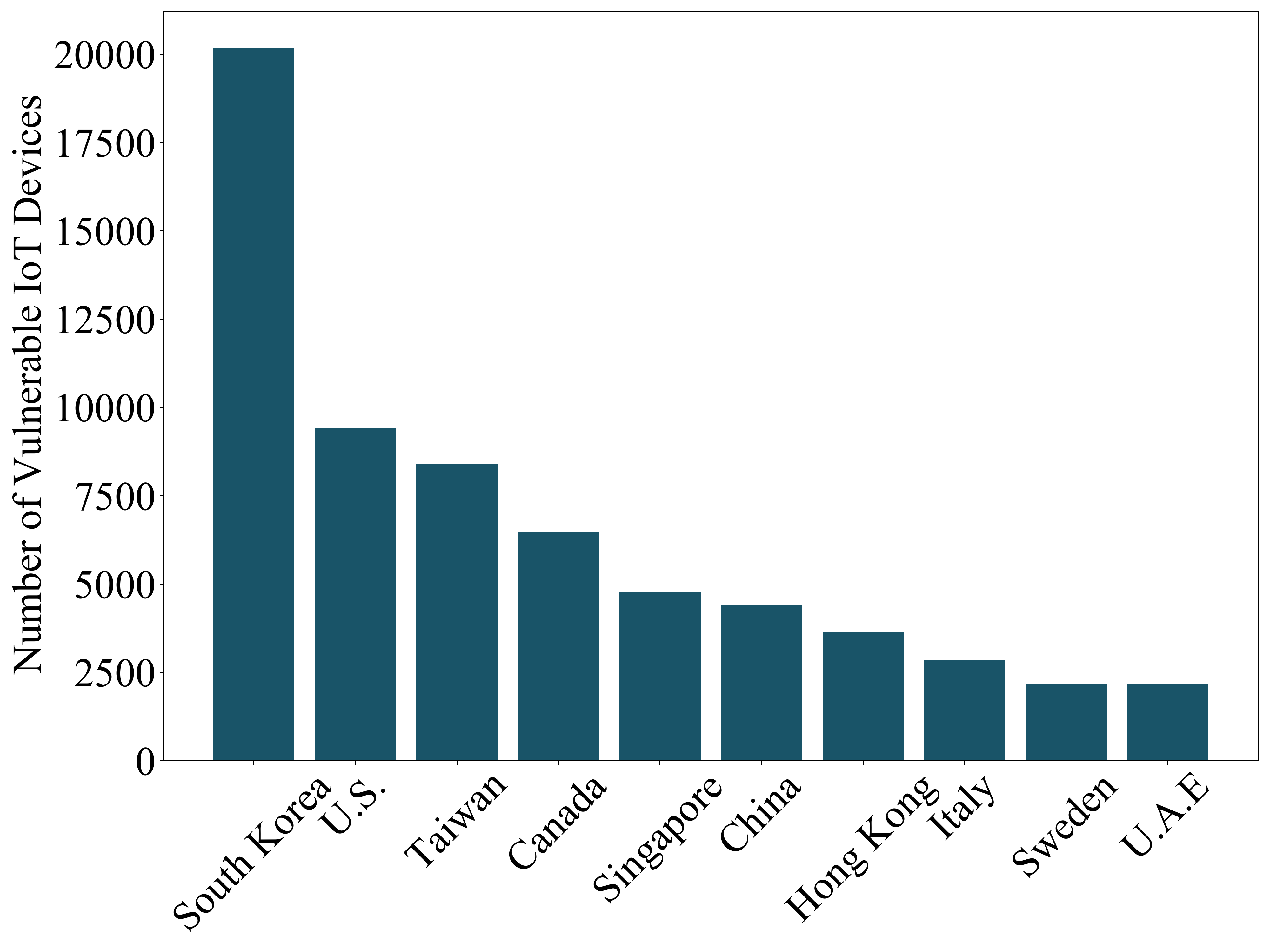}
       \setlength{\abovecaptionskip}{0.02cm}
    \caption{Top 10 Regions with the Most Vulnerable Devices.}
 
    \label{Figure: Top 10 Regions}
    % \vspace{-8mm}
\end{figure}

This subsection answers \textbf{RQ3}. We regard that
geographical distribution of vulnerable devices can reflect the potential imbalance between regions in terms of IoT devices’ threat level.
% We
% % conduct an experiment to 
% explore the geographical distribution of the devices using vulnerable firmware.
To answer this question, 
\textbf{first}, we map firmware to its exact device model, which is necessary for searching IoT devices in the IoT search engine. We successfully find the corresponding device models with exact firmware versions for %455 
$1,247$
firmware images. For instance, we confirm the firmware DIR827A1\_FW103.bin is actually used in D-Link Amplifi HD Media Router 2000 (DIR-827), whose firmware version is 1.0.3.
% Though we obtain a great number of firmware images, we only know a small portion of their exact corresponding device models. Therefore, we only conduct this test on $455$ vulnerable firmware images.
%\yuan{\color{red}With the firmware, how difficult is it to know which country it comes from? Do we have to first map it to the device model first? How do we select these 455 firmware images? Why are they representative?}
\textbf{Second}, we utilize  \textit{Shodan}~\cite{shodan} to find the distribution of these possible vulnerable IoT devices, and choose the devices whose version is the same as the vulnerable firmware.
% \textit{Shodan} is a well-known IoT search engine that can search the IoT devices exposed to the Internet. 
For example, we use \textbf{DIR-827} as the keyword to search the DIR-827 routers. Then, we select the routers whose version is 1.0.3 that is shown in their banner.
% Then, we crawl the banner of the searched devices, and then search with a version of $1.0.3$ on the Internet. 
Based on the searching results, we list the top 10 regions with the most vulnerable IoT devices, as shown in Figure~\ref{Figure: Top 10 Regions}.
% create a world distribution map of vulnerable IoT devices, as shown in Figure~\ref{Figure: Risk Heat Map}. For 
% the top 10 regions with the most vulnerable IoT devices,
There are six regions: South Korea, Taiwan, Singapore, China, Hongkong, and the United Arab Emirates, located in Asia. South Korea contains the most vulnerable IoT devices all over the world, which reaches 20,191. 
The two countries in North America - the United States and Canada both have a large number of vulnerable IoT devices. 
Europe contains relatively few vulnerable IoT devices.%, though Sweden and Italy are two of the top 10 regions with the most vulnerable IoT devices in the world.

% \begin{figure}
%     \centering
%     \includegraphics[scale=0.27]{Figures/VulneraableIoTDeviceRegion.eps}
%       \setlength{\abovecaptionskip}{0.02cm}
%     \caption{Top 10 Regions with the Most Vulnerable Devices.}
 
%     \label{Figure: Top 10 Regions}
%     \vspace{-2.0em}
% \end{figure}

We propose three possible reasons for the difference in the distribution of vulnerable devices. (1) The number of devices sold by vendors varies from region to region. (2) The security of the devices on factory mode is different in different regions. The same devices sold in some regions may have already been equipped with the updated firmware on factory mode. (3) Vendors have different firmware update mechanisms in different regions. Some regions may get firmware updates preferentially.

\begin{figure}
\begin{center}
\includegraphics[scale=0.15]{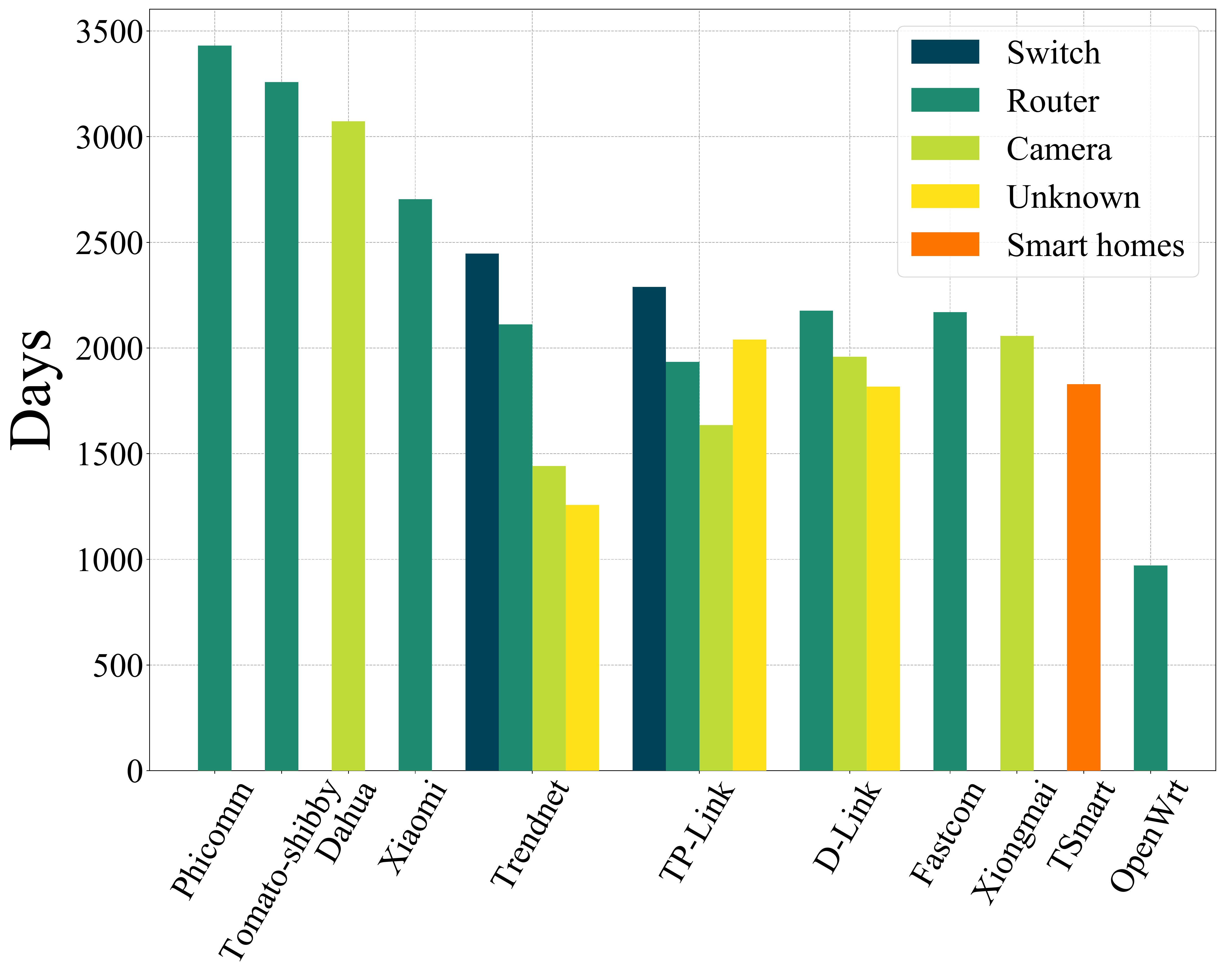}
% \vspace{-1em}
\setlength{\abovecaptionskip}{0.02cm}
\caption{Delay Time of TPCs.}
\label{Figure: the Delay Time of TPCs}
\end{center}
\vspace{-0.65cm}
\end{figure}
%%End: Figure: Delay Time of TPCs

% \subsection{Vulnerability Ranking of Different Vendors}
\vspace{-1mm}
\subsection{Delay Time of TPCs}
This subsection answers question \textbf{RQ4}. Most of the firmware vulnerabilities caused by TPCs are due to that firmware is still utilizing outdated TPCs. However, the latest TPCs may not be suitable for firmware, since they may lower the performance and influence the stability. Vendors may ignore the vulnerabilities and still use the old version of TPCs. Considering this situation, we want to explore the delay time of TPCs used in firmware. The delay time represents the days from the release date of the TPC that the firmware was using to the latest version of the TPC when the firmware was released.
%\yuan{\color{red}Why do we only analyzed a few vendors here?}

% We first obtain the release dates of firmware images. Based on which, we can know the 
We first obtain the release dates of TPCs based on their versions.
Then, we obtain the release dates of the corresponding firmware. Based on this, we infer the latest versions of the TPCs used in firmware at the time when the firmware was released. 
Finally, we calculate the delay time by the difference between these two release dates. Figure~\ref{Figure: the Delay Time of TPCs} shows the average delay time of the used TPCs for each vendor. The TPCs used in Phicomm have the longest delay time, which reaches $3457.2$ days. In other words, Phicomm is still using the TPCs that were released ten years ago.
OpenWrt has the shortest delay time, which is less than two years. It could be a reason that OpenWrt has few vulnerabilities, as shown in Table~\ref{Table: Vulnerability}.
Xiaomi also has a long delay time,  nearly seven years. What's more, we also notice that Xiaomi contains the most number of vulnerabilities per firmware image, which also confirms the relationship between the delay time and the number of vulnerabilities.
D-Link, TP-Link, and Trendnet all have a couple of different kinds of IoT devices. These vendors have a common phenomenon that the router has a longer delay time than the camera. %In addition, the switches from TP-Link and Trendnet have the longest delay time, which is over $2,000$ days.
The average delay time of TPCs for all involved firmware images is $1948.2$ days, which shows that the TPCs they used have fallen behind by five years. Our results reveal the widespread usage of outdated TPCs in IoT firmware.

\begin{table}[]

  \setlength{\tabcolsep}{3mm}
  \setlength{\abovecaptionskip}{0.08cm}
  \renewcommand\arraystretch{1.18}
  % \captionsetup{width=.8\linewidth}
  \caption{Consecutive Firmware Set.} 
  % $^{\rm a}$ indicates the total number of CVEs of the TPC. $^{\rm b}$ indicates the total number of CVEs of the TPC disclosed before the corresponding firmware.}
  \label{Table: Consecutive Firmware Set}
  \resizebox{0.5\textwidth}{!}{
  \begin{tabular}{cccccccc}
  \toprule[1.5pt]
  \multicolumn{1}{c}{\textbf{\# Set}} &
  \multicolumn{1}{c}{\textbf{Vendor}} &
    \multicolumn{1}{c}{\textbf{Category}} &
    \textbf{Firmware} &
    \textbf{TPC} &
    \textbf{Version} &
    \textbf{\# CVE} \\ \midrule[1pt]
  \multicolumn{1}{c}{\multirow{4}{*}{1}} &
  \multicolumn{1}{c}{\multirow{4}{*}{Trendnet}} &
    \multicolumn{1}{c}{\multirow{4}{*}{Router}} &
    TEW-637AP\_V1.2.0.26 &
    IPv6 & -
    & -
    \\ \cmidrule{4-7} 
  \multicolumn{1}{c}{} &
    \multicolumn{1}{c}{} &
    \multicolumn{1}{c}{} &
    TEW-637APV2\_V1.3.0.X (Total 3) &
    BusyBox &
    1.8.2 &
    5 (0) \\ \cmidrule{4-7} 
    \multicolumn{1}{c}{} &
  \multicolumn{1}{c}{} &
    \multicolumn{1}{c}{} &
    \multirow{2}{*}{TEW-637APV3\_V3.1.X (Total 2)} &
    BusyBox &
    1.13.4 &
    5 (0) \\
    \multicolumn{1}{c}{} &
  \multicolumn{1}{c}{} &
    \multicolumn{1}{c}{} &
    &
    OpenSSL &
    0.9.8b &
    46 (32) \\ \midrule
  \multicolumn{1}{c}{\multirow{4}{*}{2}} &
  \multicolumn{1}{c}{\multirow{4}{*}{Trendnet}} &
    \multicolumn{1}{c}{\multirow{4}{*}{Router}} &
    TEW-638APBV2\_V1.1.X (Total 3) &
    BusyBox &
    1.8.2 &
    5 (0) \\ \cmidrule{4-7} 
  \multicolumn{1}{c}{} &
  \multicolumn{1}{c}{} &
    \multicolumn{1}{c}{} &
    TEW-638APBV2\_V1.2.X (Total 2) &
    BusyBox &
    1.8.2 &
    5 (0) \\ \cmidrule{4-7} 
  \multicolumn{1}{c}{} &
  \multicolumn{1}{c}{} &
    \multicolumn{1}{c}{} &
    \multirow{2}{*}{TEW-638APBV3\_V3.1.4.0} &
    BusyBox &
    1.13.4 &
    5 (0) \\
  \multicolumn{1}{c}{} &
  \multicolumn{1}{c}{} &
    \multicolumn{1}{c}{} &
    &
    OpenSSL &
    0.9.8b &
    46 (46) \\ \midrule
  \multicolumn{1}{c}{\multirow{6}{*}{3}} &
  \multicolumn{1}{c}{\multirow{6}{*}{Trendnet}} &
    \multicolumn{1}{c}{\multirow{6}{*}{Router}} &
    TEW-639GR\_V1.0.X (Total 2) &
    IPv6 & -
    & -
    \\ \cmidrule{4-7} 
  \multicolumn{1}{c}{} &
    \multicolumn{1}{c}{} &
    \multicolumn{1}{c}{} &
    \multirow{3}{*}{TEW-639GR\_V2.0.X  (Total 6)} &
    BusyBox &
    1.12.1 &
    5 (0) \\
  \multicolumn{1}{c}{} &
    \multicolumn{1}{c}{} &
    \multicolumn{1}{c}{} &
    &
    TOR &
    0.1.0.16 &
    20 (5) \\
  \multicolumn{1}{c}{} &
    \multicolumn{1}{c}{} &
    \multicolumn{1}{c}{} &
    &
    Dnsmasq &
    2.4 &
    7 (2) \\ \cmidrule{4-7} 
  \multicolumn{1}{c}{} &
    \multicolumn{1}{c}{} &
    \multicolumn{1}{c}{} &
    \multirow{2}{*}{TEW-639GR\_V3.X (Total 3)} &
    BusyBox &
    13.4 &
    5 (1) \\
  \multicolumn{1}{c}{} &
    \multicolumn{1}{c}{} &
    \multicolumn{1}{c}{} &
    &
    Dnsmasq &
    2.4 &
    7 (2) \\ \midrule
  \multicolumn{1}{c}{\multirow{3}{*}{4}} &
  \multicolumn{1}{c}{\multirow{3}{*}{Trendnet}} &
    \multicolumn{1}{c}{\multirow{3}{*}{Router}} &
    \multirow{3}{*}{\begin{tabular}[c]{@{}c@{}}TEW-691GR\_V1.0.X (Total 4) \\ TEW-691GR\_V1.1.X (Total 2)\end{tabular}} &
    BusyBox &
    1.12.1 &
    5 (0) \\
  \multicolumn{1}{c}{} &
    \multicolumn{1}{c}{} &
    \multicolumn{1}{c}{} &
    &
    Dnsmasq &
    2.4 &
    7 (2) \\
  \multicolumn{1}{c}{} &
    \multicolumn{1}{c}{} &
    \multicolumn{1}{c}{} &
    &
    TOR &
    0.1.0.16 &
    20 (12) \\ \midrule
    \multicolumn{1}{c}{\multirow{7}{*}{5}} &
  \multicolumn{1}{c}{\multirow{7}{*}{Trendnet}} &
    \multicolumn{1}{c}{\multirow{7}{*}{Router}} &
    \multirow{3}{*}{TEW-812DRU\_V1.0.X (Total 7)} &
    BusyBox &
    1.7.2 &
    5 (1) \\
  \multicolumn{1}{c}{} &
    \multicolumn{1}{c}{} &
    \multicolumn{1}{c}{} &
    &
    Dnamasq &
    2.4 &
    7 (4) \\
  \multicolumn{1}{c}{} &
    \multicolumn{1}{c}{} &
    \multicolumn{1}{c}{} &
    &
    TOR &
    0.1.0.16 &
    20 (16) \\ \cmidrule{4-7} 
  \multicolumn{1}{c}{} &
    \multicolumn{1}{c}{} &
    \multicolumn{1}{c}{} &
    \multirow{4}{*}{\begin{tabular}[c]{@{}c@{}}TEW-812DRUV2\_V2.0.X (Total 5)\\ TEW-812DRUV2\_V2.1.X (Total 3)\end{tabular}} &
    BusyBox &
    1.7.2 &
    5 (1) \\
  \multicolumn{1}{c}{} &
    \multicolumn{1}{c}{} &
    \multicolumn{1}{c}{} &
    &
    IPTables &
    1.4.12 &
    1 (1) \\
  \multicolumn{1}{c}{} &
    \multicolumn{1}{c}{} &
    \multicolumn{1}{c}{} &
    &
    Dnsmasq &
    2.4 &
    7 (6) \\
  \multicolumn{1}{c}{} &
    \multicolumn{1}{c}{} &
    \multicolumn{1}{c}{} &
    &
    TOR &
    0.1.0.16 &
    20 (16) \\ \midrule
    \multicolumn{1}{c}{6} &
  \multicolumn{1}{c}{Trendnet} &
    \multicolumn{1}{c}{Camera} &
    FW\_TV\_V1.3.X (Total 4) &
    OpenSSL &
    1.0.2d &
    54 (0)\\ \midrule
    \multicolumn{1}{c}{\multirow{3}{*}{7}} &
  \multicolumn{1}{c}{\multirow{3}{*}{Trendnet}} &
    \multicolumn{1}{c}{\multirow{3}{*}{Switch}} &
    \multirow{3}{*}{TEG-448WS-ALL\_v1.0.X (Total 6)} &
    Glibc &
    2.8 &
    23 (20) \\
  \multicolumn{1}{c}{} &
    \multicolumn{1}{c}{} &
    \multicolumn{1}{c}{} &
    &
    Lua &
    5.1 &
    1 (0) \\
  \multicolumn{1}{c}{} &
    \multicolumn{1}{c}{} &
    \multicolumn{1}{c}{} &
    &
    OpenSSL &
    0.9.8h &
    45 (35) \\ \midrule
    \multicolumn{1}{c}{\multirow{5}{*}{8}} &
  \multicolumn{1}{c}{\multirow{5}{*}{TP-Link}} &
    \multicolumn{1}{c}{\multirow{5}{*}{Router}} &
    \multirow{5}{*}{LTE\_GATEWAYV3\_1.X (Total 6)} &
    BusyBox &
    1.23.2 &
    3 (2) \\
  \multicolumn{1}{c}{} &
    \multicolumn{1}{c}{} &
    \multicolumn{1}{c}{} &
    &
    Glibc &
    2.21 &
    9 (9) \\
  \multicolumn{1}{c}{} &
  \multicolumn{1}{c}{} &
    \multicolumn{1}{c}{} &
    &
    OpeSSL &
    0.9.8z &
    4 (4) \\
  \multicolumn{1}{c}{} &
    \multicolumn{1}{c}{} &
    \multicolumn{1}{c}{} &
    &
    Samba &
    3.6.22 &
    21 (17) \\
  \multicolumn{1}{c}{} &
    \multicolumn{1}{c}{} &
    \multicolumn{1}{c}{} &
    &
    curl &
    7.29.0 &
    31 (24) \\ \midrule
    \multicolumn{1}{c}{\multirow{6}{*}{9}} &
  \multicolumn{1}{c}{\multirow{6}{*}{TP-Link}} &
    \multicolumn{1}{c}{\multirow{6}{*}{Router}} &
    \multirow{6}{*}{TL-R480T+V9\_UN\_9.0.X (Total 3)} &
    BusyBox &
    1.22.1 &
    4 (3) \\
  \multicolumn{1}{c}{} &
    \multicolumn{1}{c}{} &
    \multicolumn{1}{c}{} &
    &
    curl &
    7.38.0 &
    31 (19) \\
  \multicolumn{1}{c}{} &
    \multicolumn{1}{c}{} &
    \multicolumn{1}{c}{} &
    &
    Dnsmasq &
    2.71 &
    2 (0) \\
  \multicolumn{1}{c}{} &
    \multicolumn{1}{c}{} &
    \multicolumn{1}{c}{} &
    &
    Lua &
    5.1.5 &
    1 (1) \\
  \multicolumn{1}{c}{} &
    \multicolumn{1}{c}{} &
    \multicolumn{1}{c}{} &
    &
    Nginx &
    1.8.1 &
    1 (0) \\
  \multicolumn{1}{c}{} &
    \multicolumn{1}{c}{} &
    \multicolumn{1}{c}{} &
    &
    OpenSSL &
    1.0.1j &
    50 (49) \\ \midrule
    \multicolumn{1}{c}{\multirow{2}{*}{10}} &
  \multicolumn{1}{c}{\multirow{2}{*}{TP-Link}} &
    \multicolumn{1}{c}{\multirow{2}{*}{Camera}} &
    NC250\_1.0.X (Total 2) &
    \multirow{2}{*}{OpenSSL} &
    \multirow{2}{*}{0.9.8z} &
    \multirow{2}{*}{4 (3)} \\
  \multicolumn{1}{c}{} &
    \multicolumn{1}{c}{} &
    \multicolumn{1}{c}{} &
    NC250\_1.2.1 &
    &
    &
    \\ \midrule
    \multicolumn{1}{c}{\multirow{2}{*}{11}} &
  \multicolumn{1}{c}{\multirow{2}{*}{TP-Link}} &
    \multicolumn{1}{c}{\multirow{2}{*}{Camera}} &
    NC450\_1.2.4 &
    \multirow{2}{*}{OpenSSL} &
    \multirow{2}{*}{0.9.8z} &
    \multirow{2}{*}{4 (3)} \\
  \multicolumn{1}{c}{} &
    \multicolumn{1}{c}{} &
    \multicolumn{1}{c}{} &
    NC450\_1.3.X (Total 2) &
    &
    &
    \\ \midrule
    \multicolumn{1}{c}{\multirow{5}{*}{12}} &
  \multicolumn{1}{c}{\multirow{5}{*}{TP-Link}} &
    \multicolumn{1}{c}{\multirow{5}{*}{Camera}} &
    \multirow{5}{*}{TL-IPC42A-4\_V.X.0 (Total 5)} &
    BusyBox &
    1.19.4 &
    5 (5) \\
  \multicolumn{1}{c}{} &
    \multicolumn{1}{c}{} &
    \multicolumn{1}{c}{} &
    &
    curl &
    7.29.0 &
    33 (32) \\
  \multicolumn{1}{c}{} &
    \multicolumn{1}{c}{} &
    \multicolumn{1}{c}{} &
    &
    Lua &
    5.1.4 &
    1 (1) \\
  \multicolumn{1}{c}{} &
    \multicolumn{1}{c}{} &
    \multicolumn{1}{c}{} &
    &
    OpenSSl &
    1.0.1e &
    54 (54) \\
  \multicolumn{1}{c}{} &
    \multicolumn{1}{c}{} &
    \multicolumn{1}{c}{} &
    &
    libpng &
    1.2.37 &
    16 (16) \\ \midrule
    \multicolumn{1}{c}{\multirow{4}{*}{13}} &
  \multicolumn{1}{c}{\multirow{4}{*}{D-Link}} &
    \multicolumn{1}{c}{\multirow{4}{*}{Router}} &
    \multirow{4}{*}{DIR505\_FW\_1.0.X (Total 5)} &
    Dnsmasq &
    2.41 &
    7 (5) \\
  \multicolumn{1}{c}{} &
    \multicolumn{1}{c}{} &
    \multicolumn{1}{c}{} &
    &
    Flash &
    1 &
    1 (0) \\
  \multicolumn{1}{c}{} &
    \multicolumn{1}{c}{} &
    \multicolumn{1}{c}{} &
    &
    Lighttpd &
    1.4.28 &
    6 (1) \\
  \multicolumn{1}{c}{} &
    \multicolumn{1}{c}{} &
    \multicolumn{1}{c}{} &
    &
    OpenSSL &
    1.0.0d &
    53 (0) \\ \midrule
    \multicolumn{1}{c}{\multirow{4}{*}{14}} &
  \multicolumn{1}{c}{\multirow{4}{*}{D-Link}} &
    \multicolumn{1}{c}{\multirow{4}{*}{Router}} &
    \multirow{4}{*}{DIR826L\_FW\_1.0.X (Total 5)} &
    BusyBox &
    1.12.1 &
    5 (1) \\
  \multicolumn{1}{c}{} &
    \multicolumn{1}{c}{} &
    \multicolumn{1}{c}{} &
    &
    Dnsmasq &
    2.41 &
    7 (2) \\
  \multicolumn{1}{c}{} &
    \multicolumn{1}{c}{} &
    \multicolumn{1}{c}{} &
    &
    Samba &
    3.0.37 &
    32 (9) \\
  \multicolumn{1}{c}{} &
    \multicolumn{1}{c}{} &
    \multicolumn{1}{c}{} &
    &
    OpenSSL &
    0.9.8e &
    44 (31) \\ \midrule
    \multicolumn{1}{c}{\multirow{4}{*}{15}} &
  \multicolumn{1}{c}{\multirow{4}{*}{D-Link}} &
    \multicolumn{1}{c}{\multirow{4}{*}{Router}} &
    \multirow{4}{*}{DGL5500A1\_FW1.1.X (Total 5)} &
    BusyBox &
    1.6.1 &
    5 (1) \\
  \multicolumn{1}{c}{} &
    \multicolumn{1}{c}{} &
    \multicolumn{1}{c}{} &
    &
    curl &
    7.29.0 &
    33 (0) \\
  \multicolumn{1}{c}{} &
    \multicolumn{1}{c}{} &
    \multicolumn{1}{c}{} &
    &
    Dnsmasq &
    2.41 &
    7 (4) \\
  \multicolumn{1}{c}{} &
    \multicolumn{1}{c}{} &
    \multicolumn{1}{c}{} &
    &
    Lighttpd &
    1.4.28 &
    6 (1) \\ \midrule
    \multicolumn{1}{c}{\multirow{4}{*}{16}} &
  \multicolumn{1}{c}{\multirow{4}{*}{D-Link}} &
    \multicolumn{1}{c}{\multirow{4}{*}{Camera}} &
    \multirow{4}{*}{DCS-960L\_A1\_FW\_1.X (Total 4)} &
    BusyBox &
    1.22.1 &
    4 (2) \\
  \multicolumn{1}{c}{} &
    \multicolumn{1}{c}{} &
    \multicolumn{1}{c}{} &
    &
    curl &
    7.37.0 &
    31 (5) \\
  \multicolumn{1}{c}{} &
    \multicolumn{1}{c}{} &
    \multicolumn{1}{c}{} &
    &
    Lua &
    5.1.4 &
    1 (1) \\
  \multicolumn{1}{c}{} &
    \multicolumn{1}{c}{} &
    \multicolumn{1}{c}{} &
    &
    OpenSSL &
    1.0.1p &
    21 (3) \\ \midrule
    \multicolumn{1}{c}{\multirow{6}{*}{17}} &
  \multicolumn{1}{c}{\multirow{6}{*}{Fastcom}} &
    \multicolumn{1}{c}{\multirow{6}{*}{Router}} &
    \multirow{6}{*}{FER1200G V1.0.2.X (Total 7)} &
    BusyBox &
    1.22.1 &
    4 (3) \\
  \multicolumn{1}{c}{} &
    \multicolumn{1}{c}{} &
    \multicolumn{1}{c}{} &
    &
    Dnsmasq &
    2.71 &
    2 (0) \\
  \multicolumn{1}{c}{} &
    \multicolumn{1}{c}{} &
    \multicolumn{1}{c}{} &
    &
    Lua &
    5.1.5 &
    1 (1) \\
  \multicolumn{1}{c}{} &
    \multicolumn{1}{c}{} &
    \multicolumn{1}{c}{} &
    &
    Nginx &
    1.8.1 &
    1 (0) \\
  \multicolumn{1}{c}{} &
    \multicolumn{1}{c}{} &
    \multicolumn{1}{c}{} &
    &
    OpenSSL &
    1.0.1j &
    50 (49) \\
  \multicolumn{1}{c}{} &
    \multicolumn{1}{c}{} &
    \multicolumn{1}{c}{} &
    &
    uClibc &
    0.9.33.2 &
    2 (0) \\ \midrule
  \multicolumn{1}{c}{18} &
  \multicolumn{1}{c}{Xiongmai} &
    \multicolumn{1}{c}{Camera} &
    WIFI\_NVR\_IPC\_V4.2.X (Total 4) &
    uClibc &
    0.9.33.2 &
    2 (0) \\ \midrule
    \multicolumn{1}{c}{\multirow{5}{*}{19}} &
  \multicolumn{1}{c}{\multirow{5}{*}{TSmart}} &
    \multicolumn{1}{c}{\multirow{5}{*}{Camera}} &
    \multirow{3}{*}{cmcc\_sc002\_QIO\_1.0.0.X (Total 2)} &
    BusyBox &
    1.26.2 &
    2 (2) \\
  \multicolumn{1}{c}{} &
    \multicolumn{1}{c}{} &
    \multicolumn{1}{c}{} &
    &
    curl &
    7.49.1 &
    26 (25) \\
  \multicolumn{1}{c}{} &
    \multicolumn{1}{c}{} &
    \multicolumn{1}{c}{} &
    &
    OpenSSL &
    1.0.1u &
    1 (1) \\ \cmidrule{4-7} 
  \multicolumn{1}{c}{} &
    \multicolumn{1}{c}{} &
    \multicolumn{1}{c}{} &
    \multirow{2}{*}{cmcc\_sc003\_OTA\_1.0.2.4.bin} &
    BusyBox &
    1.26.2 &
    2 (2) \\
  \multicolumn{1}{c}{} &
    \multicolumn{1}{c}{} &
    \multicolumn{1}{c}{} &
    &
    curl &
    7.49.1 &
    26 (25) \\ \midrule
    \multicolumn{1}{c}{\multirow{7}{*}{20}} &
  \multicolumn{1}{c}{\multirow{7}{*}{TSmart}} &
    \multicolumn{1}{c}{\multirow{7}{*}{Camera}} &
    \multirow{3}{*}{\begin{tabular}[c]{@{}c@{}}PPSTRONG-C2-1.8.0\\ PPSTRONG-C2-1.8.1\end{tabular}} &
    BusyBox &
    1.20.2 &
    4 (1) \\
  \multicolumn{1}{c}{} &
    \multicolumn{1}{c}{} &
    \multicolumn{1}{c}{} &
    &
    uClibc &
    0.9.33.2 &
    2 (0) \\
  \multicolumn{1}{c}{} &
    \multicolumn{1}{c}{} &
    \multicolumn{1}{c}{} &
    &
    Wpa supplicant &
    2.6 &
    16 (0) \\ \cmidrule{4-7} 
  \multicolumn{1}{c}{} &
    \multicolumn{1}{c}{} &
    \multicolumn{1}{c}{} &
    PPSTRONG-C2-1.8.7 &
    \multirow{2}{*}{BusyBox} &
    \multirow{2}{*}{1.20.2} &
    \multirow{2}{*}{4 (0)} \\
  \multicolumn{1}{c}{} &
    \multicolumn{1}{c}{} &
    \multicolumn{1}{c}{} &
    PPSTRONG-C2-1.9.X (Total 27) &
    &
    &
    \\
  \multicolumn{1}{c}{} &
    \multicolumn{1}{c}{} &
    \multicolumn{1}{c}{} &
    PPSTRONG-C2-1.10.X   (Total 3) &
    \multirow{2}{*}{uClibc} &
    \multirow{2}{*}{0.9.33.2} &
    \multirow{2}{*}{2 (0)} \\
  \multicolumn{1}{c}{} &
    \multicolumn{1}{c}{} &
    \multicolumn{1}{c}{} &
    PPSTRONG-C2-2.0.X (Total 10) &
    &
    &
    \\ \midrule
    \multicolumn{1}{c}{\multirow{2}{*}{21}} &
  \multicolumn{1}{c}{\multirow{2}{*}{TSmart}} &
    \multicolumn{1}{c}{\multirow{2}{*}{Camera}} &
    \multirow{2}{*}{CHMI-CAMERA-OTA-3.X (Total 10)} &
    \multirow{2}{*}{uClibc} &
    \multirow{2}{*}{0.9.33.2} &
    \multirow{2}{*}{2 (0)} \\
  \multicolumn{1}{c}{} &
    \multicolumn{1}{c}{} &
    \multicolumn{1}{c}{} &
    &
    &
    &
    \\ \midrule
    \multicolumn{1}{c}{\multirow{5}{*}{22}} &
  \multicolumn{1}{c}{\multirow{5}{*}{TSmart}} &
    \multicolumn{1}{c}{\multirow{5}{*}{Camera}} &
    \multirow{5}{*}{FIRMWARE-D621-2-3.0.X   (Total 9)} &
    BusyBox &
    1.22.1 &
    4 (3) \\
  \multicolumn{1}{c}{} &
    \multicolumn{1}{c}{} &
    \multicolumn{1}{c}{} &
    &
    curl &
    7.51.0 &
    16 (8) \\
  \multicolumn{1}{c}{} &
    \multicolumn{1}{c}{} &
    \multicolumn{1}{c}{} &
    &
    OpenSSL &
    1.0.2m &
    11 (2) \\
  \multicolumn{1}{c}{} &
    \multicolumn{1}{c}{} &
    \multicolumn{1}{c}{} &
    &
    uClibc &
    0.9.33.2 &
    2 (2) \\
  \multicolumn{1}{c}{} &
  \multicolumn{1}{c}{} &
    \multicolumn{1}{c}{} &
    &
    Wpa supplicant &
    2.7 &
    9 (0) \\ \bottomrule[1.5pt]
  \end{tabular}
  }
  \end{table}

% \vspace{-3mm}

\subsection{\bin{Consecutive Firmware Set Analysis}}
This subsection answers question \textbf{RQ5}. 
We pick all consecutive firmware sets from our dataset, a series of historical firmware images belonging to the same device. We arrange the order of firmware images in each set according to their release time and list the TPCs used in firmware with the corresponding number of CVEs. As shown in Table~\ref{Table: Consecutive Firmware Set}, we have a total of $22$ consecutive firmware sets involving $162$ firmware images. Based on the results, we have the following two observations.

% As shown in Table~\ref{Table: Consecutive Firmware Set} (Appendix~\ref{section: Vulnerability Evaluation}), we have a total of $22$ consecutive firmware sets involving $162$ firmware images. Based on the results, we have the following two observations.

% We find that vendors rarely update TPCs at the same time as firmware updates, even if the TPCs have corresponding vulnerabilities disclosed before firmware updates.

%  We have X new findings.
$\bullet$
Vendors rarely change the TPCs along with firmware updates. The change of  TPCs only occurs during major firmware updates. Our consecutive firmware sets have $6$ sets involving major firmware updates and $16$ sets merely involving minor firmware updates. The results show that only $6$ sets 
(Set 1, 2, 3, 5, 19, 20) 
that involve major firmware updates have changed the TPCs while the remaining $16$ sets do not. Besides, we notice that not all sets will update the version of TPCs during major updates.
% The firmware may
% Besides, though 
% firmware has changed the TPCs during the period of major firmware updates, it may still use the vulnerable TPCs rather than update them. 
Set 20 is a representative case where 43 consecutive firmware images that involve three major updates still use two same TPCs: \textit{BusyBox 1.20.2} and \textit{uClibc 0.9.33.2}. Nevertheless, these two specific versions of TPCs have multiple known CVEs.

% According to Set 1 to Set 22, we find the consecutive firmware in each set widely adopts TPCs of the same version as the previous firmware.  Another representative case is Set X. In Set X, it has been six years from the release of the first firmware image to the release of the last firmware image while the TPCs adopted are still the same.

% $\bullet$
% The replacement or update of the TPCs only occurs during the period of major firmware updates. 

$\bullet$
A great number of firmware images adopt the TPCs which have corresponding CVEs disclosed before the firmware is released. Although we have listed CVEs that caused by TPCs in  firmware, there is a possibility that the CVE may be released later than the firmware. In this case, it is understandable that the vendor does not fix the vulnerabilities or update TPCs. To further explore it,  we list the number of CVEs of TPCs, i.e., the number in (), that have been disclosed before the corresponding firmware is released. The results turn out that except for the firmware from Set 6 and Set 18, the TPCs used by the firmware in other sets have already known CVEs before the corresponding firmware is released.
In subsequent updated versions, firmware is still using these TPCs which have publicly disclosed CVEs.

% Overall, we sum up two conclusions according to our findings:  1) Vendors hardly check the known vulnerabilities of TPCs before they adopt them. 2) Vulnerabilities caused by TPCs may affect a series of same-source firmware since vendors rarely update TPCs during firmware updates.

% %%Begin: Figure: the Architecture of XXX
% \begin{figure}
% % \centering
% \includegraphics[width=0.50\textwidth]{Figures/CVEMap.eps}
% \caption{World Map of Vulnerable IoT Devices}
% \label{Figure: Risk Heat Map}
% \end{figure}
% %%End: Figure: Vulnerable Rate Trending of Route

\vspace{-2mm}
\subsection{License Violations}
This subsection answers question \textbf{RQ6}. The use of TPCs in firmware can lead to complex license compliance issues. For instance, Cisco has involved in a lawsuit since it did not adhere to the license requirements~\cite{cisco}. We mainly study the license violations caused by two highly restrictive licenses: General Public License (GPL) and Affero General Public License (AGPL), since they are widely used licensing terms and have a basic requirement that developers should provide the source code if they distribute the programs that use the TPCs licensed under GPL/AGPL. According to our in-depth study, we first discover $2,478$ commercial firmware images that have potentially violated GPL/AGPL licensing terms, as shown in Table~\ref{Table: Potential License Violations}. We then study the open-source policy of the involved vendors. We notice that four vendors (TP-Link, D-Link, Trendnet, and Hikvision) have provided distribution sites for downloading the source code of firmware. Nevertheless, we find the source code of some firmware, which utilizes the GPL/AGPL-licensed TPCs, cannot be found on their sites. We further contact the involved vendors, except for the TSmart, to request the source code for some firmware but do not get any response yet. 

We summarize three possible reasons why vendors do not open-source the firmware according to license requirements. \textbf{First}, vendors disregard the restrictions of licenses. Currently, there are no strong measures to enforce the GPL/AGPL compliance since the lawsuit is complicated and may not apply to some countries. \textbf{Second}, open-sourcing the firmware may lead to new attacks. Attackers can find vulnerabilities through auditing the source code, which may affect more firmware images if vendors reuse the vulnerable code in other firmware. \textbf{Third}, the firmware has license conflicts. Vendors may have some commercial licenses that applied to the firmware simultaneously, which conflicts with the open-source licenses. 

% Besides, TSmart's developers tell us that they usually give priority to utilizing free TPCs and their own implementations to replace the GPL/AGPL-licensed ones to avoid license issues. Nevertheless, there are always some  GPL/AGPL-licensed TPCs that cannot be replaced.

\begin{table}[]
\centering
\caption{Potential License Violations.}
\vspace{-3mm}
\setlength\tabcolsep{15pt}
\footnotesize
\label{Table: Potential License Violations}
\begin{tabular}{ccc}
\toprule[1.5pt]
Vendors & \# Firmware & Source Code Available \\ \midrule[1pt]
Xiongmai  & 195 &  \textcolor{red}{\XSolidBrush} \\
Phicomm   & 96  &  \textcolor{red}{\XSolidBrush} \\
Fastcom   & 17  &  \textcolor{red}{\XSolidBrush} \\
Trendnet  & 433 &  \textcolor{green}{\checkmark} \\
Xiaomi    & 20  &  \textcolor{red}{\XSolidBrush} \\
TP-Link   & 847 & \textcolor{green}{\checkmark} \\
D-Link    & 487 & \textcolor{green}{\checkmark} \\
Hikvision & 2   & \textcolor{green}{\checkmark} \\
Dahua     & 11  & \textcolor{red}{\XSolidBrush}\\
TSmart    & 370 & \textcolor{red}{\XSolidBrush} \\ \bottomrule[1.5pt]
\end{tabular}
\vspace{-4mm}
\end{table}

% \vspace{-2mm}
\section{Discussion}
% \yuan{Need to talk about suggestions/takeaways.}
\noindent\textbf{Ethics.}   Our large-scale vulnerability analysis of IoT firmware may raise serious ethical concerns. To avoid these potential hazards,
% In this work, we conduct so far the largest scale vulnerability analysis of IoT firmware, which may raise serious ethical concerns.
% In consideration of these potential hazards,  
we pay special attention to legal and ethical issues.
First, all firmware images are collected and treated legally. 
% Our dataset includes two parts of firmware; one is collected from the public Internet, the other one is obtained from our TSmart. 
For the publicly accessible firmware, we collect it from legal sources, e.g., official websites, and adhere to the Robots Exclusion Protocol (REP)~\cite{REP}. For the private firmware, we only use it for research purposes. Besides, responsible disclosure is also a basic requirement for us. We have actively contacted the related vendors and reported our results to them as detailed as possible. Finally, we have a legal and ethical issues-free plan to open-source our dataset.  For the publicly accessible firmware images, we will provide their official download links.  For the private firmware images,  we have full authorization from TSmart to open-source them after desensitizing.  
\noindent\textbf{Limitations and Future Work.}   
%\yuan{We should first talk about limitations of the work, justify why these limitations don't matter much, and finally what other researchers can build on top of our work.}
In the future, we plan to improve our work in three aspects.
First, we will continue to collect more firmware images to extend our dataset. Though we have collected $34,136$ firmware images, we still lack the firmware of some current popular devices, e.g., smart assistant devices.
% We also want to extend our dataset to more studies, e.g., serve as a benchmark for evaluating the performance of fuzzers.
% We want to apply our analysis to more different kinds of firmware images from more vendors.
% Second, we will improve our matching strategy to discover much more third-party components from firmware. The current matching strategy is heavily based on our constructed keywords list. However, we cannot cover all the third-party components used in firmware.
Second, we will continually enrich our TPC database. Currently, \system can only detect the TPCs included in the database. During our analysis, \system has some false positives at TPC-level identification since our database does not record some uncommon TPCs used in IoT firmware.
% Our database currently has $1,942$ TPCs
% which will not only based on CVE Detail, to discover more vulnerabilities for third-party components. Currently, \textit{360} and \textit{Alibaba} both have established their vulnerability databases which are more detailed than ours as indicated by the results in Section~\ref{section: performance measurement}. 
% Third, we will introduce more metrics to evaluate the risk levels of various kinds of firmware and various vendors. The current metric based on the average number of different severity vulnerabilities in firmware cannot comprehensively reflect the r vendors.
Finally, we will adopt new techniques to conduct a more in-depth analysis of IoT devices. \system has an outstanding performance in finding N-days vulnerabilities caused by TPCs. However, it is hard to detect unknown vulnerabilities. We plan to combine  fuzzing techniques~\cite{lyu2022ems, liu2021ase} with \system to find new vulnerabilities.

% With the TPCs detected in firmware, we actually know a part of the source code of firmware. Therefore, we can regard the firmware as graybox. We can leverage this kind of information to guide the firmware fuzzing.

% adopt new techniques, such as binary code comparison~\cite{ding2019asm2vec}, to improve \system in finding new vulnerabilities from firmware. %\yuan{should write it in the way that others can build on top of our work.}

% We will improve the ability of \system in matching much more third-party components. Though \system supports to match $1,942$ third-party components at present, it still cannot cover all the third-party components used in IoT devices. Moreover, with the development of IoT devices, there will have more and more third-party components are introduced to IoT devices.

\vspace{-2mm}
\section{Related work}
\noindent\textbf{TPC Detection.} Several works have been proposed to discover the TPCs used in Android apps~\cite{backes2016reliable,zhang2019libid,zhan2021atvhunter,nguyen2020up2dep}. However, these works more or less require Android features support. Therefore, it is difficult to apply them to IoT firmware analysis. Duan et al.~\cite{duan2017identifying} proposed \textit{OSSPolice} to detect the open-source software license violations and identify the open-source software at version-level in Android apps. Though \textit{OSSPolice} supports C/C++ native binaries, its feature extraction tool does not perform well on IoT firmware. Besides, its hierarchical matching strategy heavily relies on the correctness of the package structures of TPCs.
Hemel et al.~\cite{hemel2011finding} proposed  \textit{BAT} to detect the usage of TPCs in binaries based on string literals. Nevertheless, the direct feature matching strategy adopted by \textit{BAT} will cause a low precision and recall rate. 
% Moreover, it cannot detect the TPCs at version-level.

\noindent\textbf{Static Analysis.}   Costin et al.~\cite{costin2014large} conducted the first large-scale analysis of firmware. However, they did not delve into the vulnerabilities introduced by TPCs in firmware.
% since their system is not specifically designed for this purpose.
Other works utilized more robust features from I/O behavior~\cite{pewny2015cross} and control flow graphs~\cite{feng2016scalable,eschweiler2016discovre} of an image to discover vulnerabilities. 
% Corteggiani et al.~\cite{corteggiani2018inception} introduced new techniques for symbolic execution in embedded systems by generating and merging LLVM bitcode~\cite{lattner2004llvm} from high-level source code,  binary libraries, etc. 
Nevertheless, these methods cannot be applied for large-scale firmware vulnerability search since they only support firmware using a few kinds of architectures and require a lot of manual work. Several works focus on using similar code detection to find vulnerabilities in firmware. Xu et al.~\cite{xu2017neural} proposed a neural network-based approach to compare the similarity of binary codes.
% according to their control flow graphs. 
David et al.~\cite{david2018firmup} proposed \textit{FirmUp} to conduct a precise static detection of common vulnerabilities in firmware via matching similar procedures in the context of executables. Ding et al.~\cite{ding2019asm2vec} proposed \textit{Asm2Vec} for assembly clone detection based on  learned representation.  However, these works are designed for comparing the similarity between the individual functions rather than the similarity between the entire TPCs with firmware.

% Though we can transfer by retrainning them with TPC dataset,
% It is a chan
% All of these methods require the assembly code of vulnerabilities for comparison. Nevertheless, it is a challenge to prepare a large-scale vulnerability dataset. Therefore, these methods are hard to detect all potential vulnerabilities caused by TPCs but suitable for precisely detecting specific vulnerabilities, e.g., Heartbleed.

\noindent\textbf{Dynamic Analysis.}   Zaddach et al.~\cite{zaddach2014avatar} designed \textit{Avatar}, a dynamic analysis framework for firmware security analysis by forwarding I/O access between the emulator and real device.
Further, Muench et al. ~\cite{muench2018avatar2} described how to orchestrate the execution among different testing environments. Chen et. al~\cite{chen2016towards} presented \textit{FIRMADYNE} to automatically analyze Linux-based firmware. Costin et al.~\cite{costin2016automated} performed the security analysis of web interfaces within embedded devices leveraging several  off-the-shell analysis tools. Feng et al.~\cite{feng2020P2IM} proposed \textit{$P^{2}IM$} to perform fuzz-testing on firmware from MCU devices in a fully emulated fashion. 
% by modeling peripheral interface automatically. 
Nevertheless, it is hard to apply them in a large-scale analysis since they require real devices or a lot of manual work to configure for each firmware image.
\vspace{-3mm}
\section{conclusion}
In this paper, we conduct the largest-scale analysis of the TPC issues in IoT firmware at present. We propose \system which dedicates to finding the vulnerabilities in firmware caused by TPCs. Based on \system, we identify $584$ TPCs and detect a total of $128,757$ security vulnerabilities caused by $429$ CVEs  in $34,136$ firmware images. \system achieves 91.03\% of precision  and 92.26\% of recall in detecting the TPCs at version-level in firmware, which significantly outperforms state-of-the-art tools from academia and commercial tools from industry. Our analysis reveals the widespread usage of outdated TPCs in IoT firmware. Moreover, firmware has more and more vulnerabilities caused by TPCs over time, and vendors hardly update vulnerable components when the firmware is updated. In addition, we present a global view of the geographical difference in the security of IoT devices. Further analysis discloses the GPL/AGPL license violations widely exist in firmware. We believe our work will shed light on the further study of the security of IoT devices.

% \system achieves 91.03\% of precision  and 92.26\% of recall in detecting the TPCs at version-level in firmware.
% , which  significantly  outperforms state-of-the-art tools from academia and commercial  tools  from  industry.
% outperforms three academic works: \textit{Gemini}, \textit{BAT}, and \textit{OSSPolice}, and three industry state-of-the-arts: \textit{binaré}, \textit{360}, and \textit{Alibaba}. 
% Our analysis reveals the widespread usage of vulnerable and outdated TPCs in IoT firmware. 
% % Moreover, firmware has more and more vulnerabilities caused by TPCs over time, and vendors hardly update vulnerable components when the firmware is updated.
% Moreover, we present a global view of the geographical difference in the security of IoT devices. Further analysis discloses the GPL/AGPL license violations widely exist in firmware.
% % Our work provides a comprehensive overview of the current security situations of IoT devices.
% We believe our work will shed light on the further study of the security of IoT devices.

\section*{Acknowledgement}
This work was partly supported by NSFC under No. U1936215, the Zhejiang Provincial Natural Science Foundation for Distinguished Young Scholars under No. LR19F020003, the State Key Laboratory of Computer Architecture (ICT, CAS) under Grant No. CARCHA202001, the Fundamental Research Funds for the Central Universities (Zhejiang University NGICS Platform), Google Research Scholar Award, Facebook Faculty Award, NSFC under No. 62102363,  and the Zhejiang Provincial Natural Science Foundation under No. LQ21F020010.

% %\clearpage
\bibliographystyle{IEEEtran}
{
% \scriptsize
\begingroup
\bibliography{Reference}
\endgroup
}

\begin{IEEEbiography}[{\includegraphics[width=1in,height=1.25in,clip,keepaspectratio]{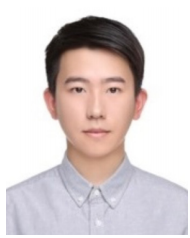}}]{Binbin Zhao}
is currently a Ph.D. candidate in the School of Electrical
and Computer Engineering at Georgia Tech. He received the B.S. degree from the College of
Computer Science at Zhejiang University
in 2018. His research interests include
IoT Security, Fuzzing and Data-driven Security.
\end{IEEEbiography}

\begin{IEEEbiography}[{\includegraphics[width=1in,height=1.25in,clip,keepaspectratio]{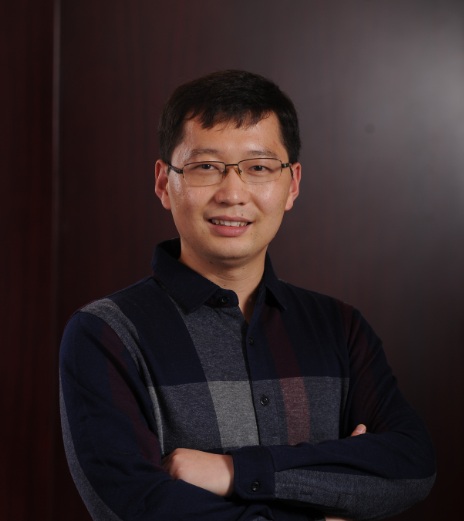}}]{Shouling Ji} is a ZJU 100-Young Professor in the
College of Computer Science and Technology
at Zhejiang University and a Research Faculty
in the School of Electrical and Computer Engineering at Georgia Institute of Technology. He
received a Ph.D. in Electrical and Computer Engineering from Georgia Institute of Technology
and a Ph.D. in Computer Science from Georgia
State University. His current research interests
include AI Security, Data-driven Security and
Data Analytics. He is a member of IEEE and
ACM and was the Membership Chair of the IEEE student Branch at Georgia State (2012-2013).
\end{IEEEbiography}

\begin{IEEEbiography}[{\includegraphics[width=1in,height=1.25in,clip,keepaspectratio]{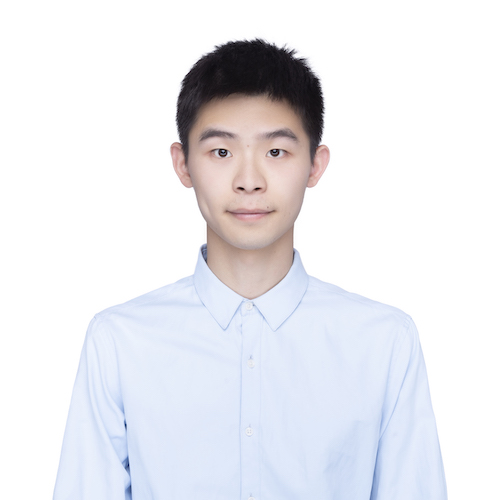}}]{Jiacheng Xu}
is currently a Ph.D. student at Zhejiang University. He received the B.S. degree from the College of Computer Science and Technology at Zhejiang University in 2021. His current research interests include IoT Security, Fuzzing, and Kernel Security.
\end{IEEEbiography}

\begin{IEEEbiography}[{\includegraphics[width=1in,height=1.25in,clip,keepaspectratio]{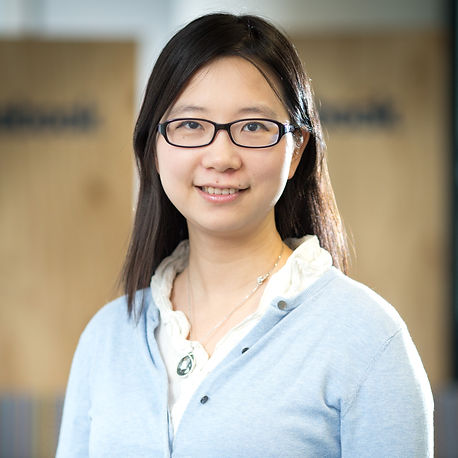}}]{Yuan Tian} is the assistant professor at the University of California, Los Angeles. She received the Ph.D. at Carnegie Mellon University in 2017. Her research focuses on developing novel technologies for the security, privacy, and safety of modern and emerging systems.
\end{IEEEbiography}

\begin{IEEEbiography}[{\includegraphics[width=1in,height=1.25in,clip,keepaspectratio]{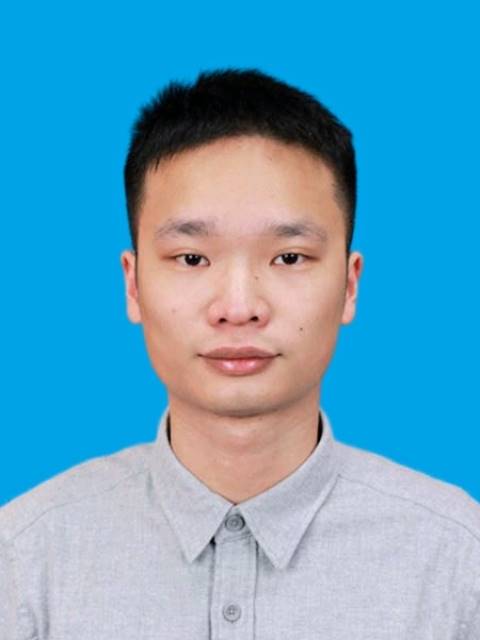}}]{Qiuyang Wei} is currently a master student at the Institute of Software, Chinese Academy of Sciences. He received the B.S. degree from the College of Computer Science and Technology at Zhejiang University in 2021. His research interests include Formal Analysis, Smart Contract and IoT Security.
\end{IEEEbiography}

\begin{IEEEbiography}[{\includegraphics[width=1in,height=1.25in,clip,keepaspectratio]{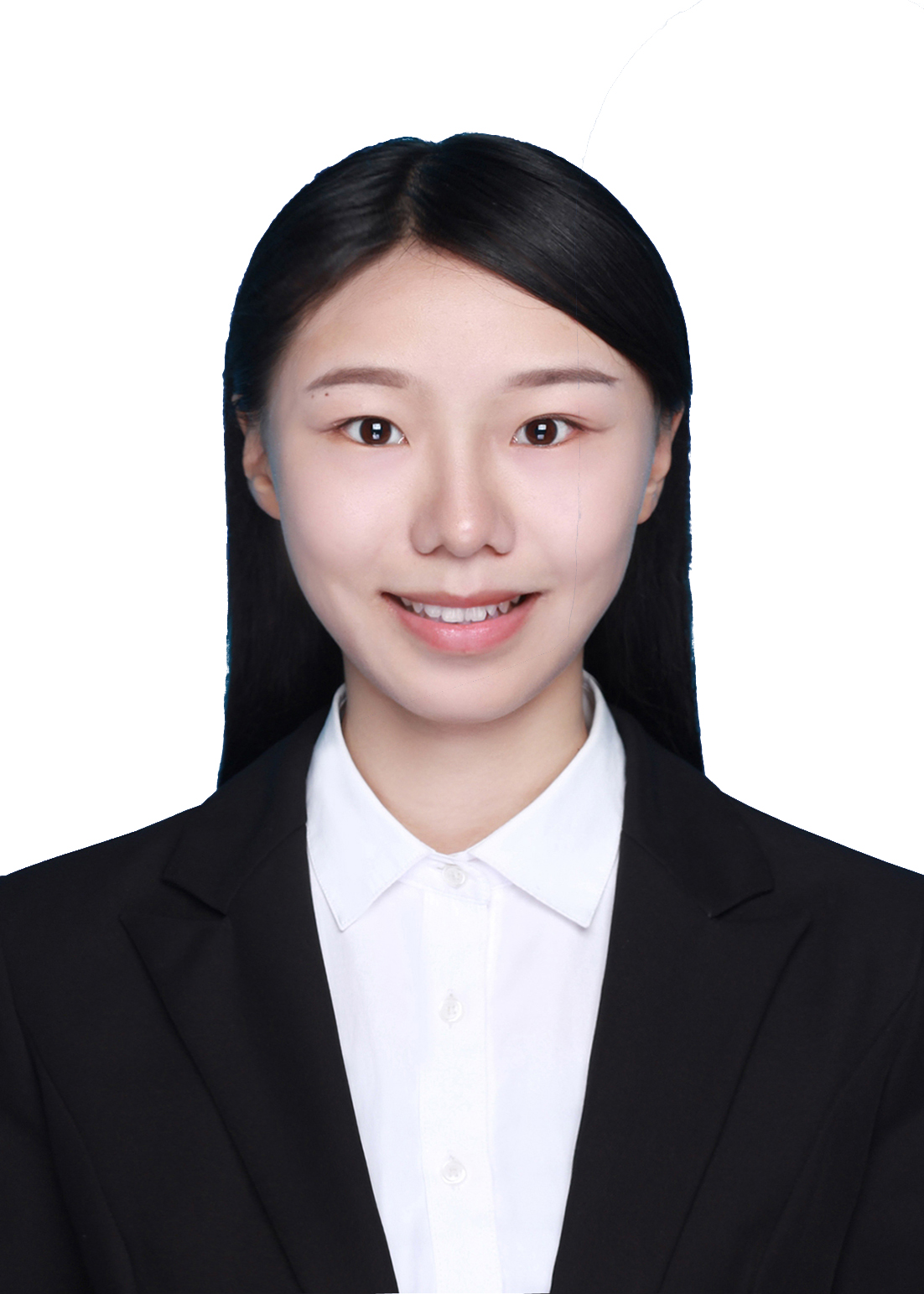}}]{Qinying Wang}
received the B.S. degree from the College of Computer Science and Electronic Engineering at Hunan University in 2018. She is currently working toward the Ph.D. degree with the College of Computer Science and Technology, Zhejiang University. Her research interests include IoT Security, Fuzzing, and AI security.
\end{IEEEbiography}

\begin{IEEEbiography}[{\includegraphics[width=1in,height=1.25in,clip,keepaspectratio]{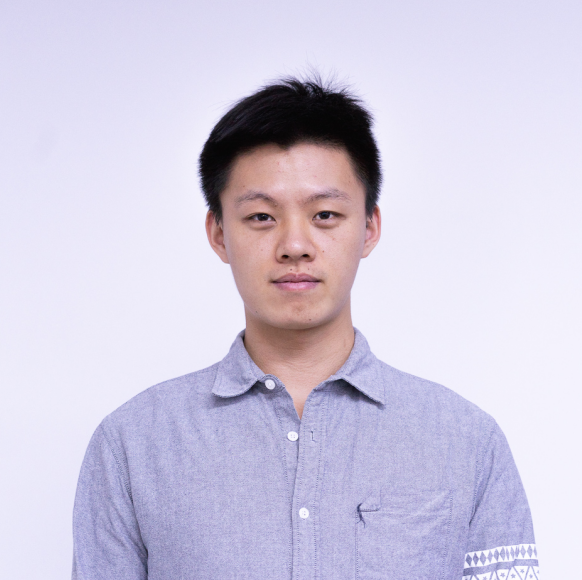}}]{Chenyang Lyu} received his Ph.D. from the College of Computer Science and Technology at Zhejiang University, 2022. He received the B.S. degree from the School of Computer Science and Technology at Huazhong University of Science and Technology. His research interest lies in fuzzing and data-driven technique.
\end{IEEEbiography}

\begin{IEEEbiography}[{\includegraphics[width=1in,height=1.25in,clip,keepaspectratio]{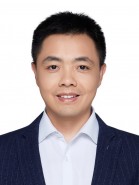}}]{Xuhong Zhang} received his Ph.D. in Computer
Engineering from University of Central Florida in
2017. He received his B.S. Degree in Software
Engineering from Harbin Institute of Technology
in 2011 and received his M.S. degree in Computer Science from Georgia State University in
2013. He is an assistant professor of College
of Control Science and Engineering at Zhejiang
University. His research interests include distributed big data and AI systems, big data mining
and analysis, data-driven security, AI and Security. He has authored over 10 publications in premier journals such
as IEEE Transactions on Parallel and Distributed Systems, and top
conferences such as VLDB, IPDPS.
\end{IEEEbiography}

\begin{IEEEbiography}[{\includegraphics[width=1in,height=1.25in,clip,keepaspectratio]{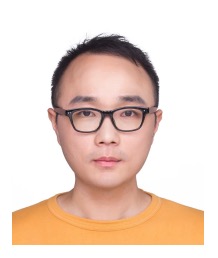}}]{Changting Lin} received the Ph.D degree in computer science from the Zhejiang University in 2018. He is currently a researcher at the Binjiang Institute of Zhejiang University, China. His research interests include IoT security, AI security, and SDN.
\end{IEEEbiography}

\begin{IEEEbiography}[{\includegraphics[width=1in,height=1.25in,clip,keepaspectratio]{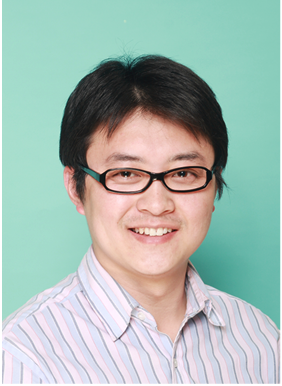}}]{Jingzheng Wu} received his Ph.D. degree in computer software and theory from the Institute of Software, Chinese Academy of Sciences, Beijing, in 2012. He is a research professor at the Institute of Software, Chinese Academy of Sciences, Beijing. His primary research interests include System Security, Vulnerability Detection, Covert Channels.
\end{IEEEbiography}

\begin{IEEEbiography}[{\includegraphics[width=1in,height=1.25in,clip,keepaspectratio]{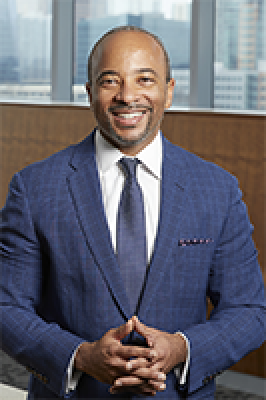}}]{Raheem Beyah} is the Dean of the College of Engineering at Georgia Tech and the Southern Company Chair. He received his Bachelor of Science in Electrical Engineering from
North Carolina A\&T State University in 1998.
He received his Master’s and Ph.D. in Electrical
and Computer Engineering from Georgia Tech
in 1999 and 2003, respectively. His work is at
the intersection of the networking and security
fields His research interests, includes Cyber-Physical Systems Security,
Privacy, Network Security and Network Monitoring and Performance.
\end{IEEEbiography}

% \clearpage
% \input{0x0Appendix}
% \clearpage

\end{document}